\documentclass[a4paper,11pt]{article}
\pdfoutput=1

\usepackage{jheppub} 
\usepackage{epstopdf}
\usepackage{graphicx}
\usepackage{booktabs}
\usepackage{multirow}
\usepackage{subfig}
\usepackage{float}
\usepackage{listings}
\usepackage{placeins}
\usepackage{morefloats}
\usepackage{amsmath}
\usepackage{slashed}

\usepackage{color}

\definecolor{shaded}{RGB}{210,210,210}
\definecolor{bianco}{RGB}{255,255,255}
\definecolor{rosso}{RGB}{210,0,0}
\definecolor{blu}{RGB}{0,0,210}
\definecolor{nero}{RGB}{0,0,0}

\newcommand{\gev}{\,\textrm{GeV}}
\newcommand{\tev}{\,\textrm{TeV}}
\newcommand{\ord}{\mathcal{O}}

\newcommand\aNLO{{\sc\small MadGraph5\_aMC@NLO}}

\definecolor{dkgreen}{rgb}{0,0.6,0}
\definecolor{gray}{rgb}{0.5,0.5,0.5}
\definecolor{mauve}{rgb}{0.58,0,0.82}

\title{On the Impact of Lepton PDFs}

\author[a]{Valerio Bertone,}
\author[b]{Stefano Carrazza,}
\author[c]{Davide Pagani,}
\author[d,e]{Marco Zaro.}

\affiliation[a]{PH Department, TH Unit,\\ CERN, CH-1211 Geneva 23, Switzerland}
\affiliation[b]{Dipartimento di Fisica, Universit\`a di Milano and INFN,
 Sezione di Milano, \\ Via Celoria 16, I-20133 Milano, Italy}
\affiliation[c]{Center for Cosmology, Particle Physics and
  Phenomenology (CP3)\\Universit\'e Catholique de Louvain, B-1348
  Louvain-la-Neuve, Belgium}
\affiliation[d]{Sorbonne Universit\'es, UPMC Univ. Paris 06, UMR 7589, LPTHE, F-75005, Paris, France}
\affiliation[e]{CNRS, UMR 7589, LPTHE, F-75005, Paris, France}

\emailAdd{valerio.bertone@cern.ch}
\emailAdd{stefano.carrazza@mi.infn.it}
\emailAdd{davide.pagani@uclouvain.be}
\emailAdd{marco.zaro@lpthe.jussieu.fr}

\abstract{In this paper we discuss the effect of the complete
  leading-order QED corrections to the DGLAP equations that govern the 
  perturbative evolution of parton distribution functions (PDFs). This
  requires the extension of the purely QCD DGLAP evolution including
  a PDF for the photons and, consistently, also for the charged
  leptons $e^{\pm}$, $\mu^\pm$ and $\tau^\pm$. We present the
  implementation of the QED-corrected DGLAP evolution in the presence
  of photon and lepton PDFs in the {\tt APFEL} program and, by means
  of different assumptions for the initial scale PDFs, we produce for
  the first time PDF sets containing charged lepton distributions. We
  also present phenomenological studies that aim to assess the impact
  of the presence of lepton PDFs in the proton for some relevant SM
  (and BSM) processes at the LHC at 13 TeV and the FCC-hh at 100
  TeV. The impact of the photon PDF is also outlined for those
  processes.}


\begin{document}

\begin{figure}[h]
  \begin{flushleft}
    \includegraphics[width=0.32\textwidth]{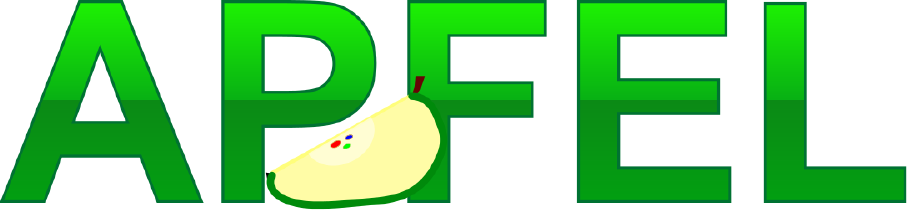}
  \end{flushleft}
\end{figure}
\vspace{-1.0cm}
\begin{flushright}
CERN-PH-TH-2015-202\\
CP3-15-27\\
TIF-UNIMI-2015-5
\end{flushright}

\maketitle

\flushbottom

\section{Introduction}

The Run II at the Large Hadron Collider (LHC) will probe the
interactions of elementary particles at unprecedented scales of energy
and with an outstanding accuracy.  In order to correctly identify
possible Beyond-the-Standard-Model (BSM) effects, the theoretical
predictions for the Standard Model (SM) processes have to match (at
least) the precision of the corresponding experimental measurements.
In other words, the impact of higher-order corrections on
phenomenological predictions has to be known and under control. To
this purpose, the computation of Next-to-Leading-Order (NLO) QCD
corrections is necessary, but often not sufficient.  For example,
at fixed order the inclusion of the second-order (NNLO) corrections
in QCD as well as of the NLO electroweak (EW) corrections is in
general desirable and in particular cases even essential.  In order to
formally achieve this level of accuracy, not only the matrix elements
of the hard processes, but also the so-called parton-distribution
functions (PDFs) have to be known at the same level of precision.
While most of the collaborations already provide PDF sets accurate up
to NNLO in
QCD~\cite{Ball:2014uwa,Harland-Lang:2014zoa,Gao:2013xoa,Abramowicz:2015mha,Alekhin:2013nda},
for the EW corrections the situation is less satisfactory.  Indeed,
NLO EW corrections naturally introduce photon-induced processes and
only two PDF sets including a photon density are presently available:
the MRST2004QED set~\cite{Martin:2004dh}, whose photon PDF is
determined by means of model assumptions, and the NNPDF2.3QED
family~\cite{Ball:2013hta}, where instead the photon PDF is extracted
by means of a fit to data. Contrary to the MRST2004QED set, the photon
PDF of NNPDF2.3QED sets is provided with an uncertainty.  In addition,
the aforementioned PDF sets are accurate up to (N)NLO in QCD but only
at LO in QED evolution. Thus, since NLO EW, and in particular NLO QED
effects, are never included at the level of PDFs, this level of
accuracy cannot be formally claimed at the level of the hadronic cross
section.
 
Before moving to NLO QED accuracy, however, it is worth attempting a
determination of PDFs at (N)NLO QCD + LO QED accuracy for all the
QED-interacting SM particles, $i.e$. by considering also {\it leptons} as
partons of the proton.  In this way predictions for lepton-induced
processes at hadron colliders could be easily calculated by
convoluting partonic cross sections with lepton PDFs. On the one hand,
this can be useful for the study of processes such as same-sign
and/or different-flavour dilepton production, which, with lepton PDFs,
can be directly produced without requiring additional (unresolved)
jets or photons. On the other, lepton-initiated processes are
naturally involved in EW corrections. Indeed, as already mentioned, a
very interesting aspect of EW corrections is the fact that they
naturally require new partons in the proton on top of the usual
(anti)quarks and gluons. Processes with quarks in the initial
state at LO receive NLO EW corrections from photon-initiated
processes. Equivalently, once the photon is considered as a parton,
photon-initiated processes at LO will receive corrections from
processes including leptons in the initial state. However, while
PDF fits that provide an estimate of the photon PDFs are presently
available, this is not the case for leptons.
 
There are (at least) two reasons why lepton PDFs have been neglected
so far in the context of EW corrections. Firstly, they are expected to
be small, since they are suppressed by a factor $\alpha$ with respect
to the photon PDF, which in turn is already suppressed with respect to
quark and gluon PDFs. Secondly, photon-initiated processes are
typically simulated only at LO accuracy, due the smallness of the
photon PDF with respect to quark and gluon ones. Thus, neglecting NLO
EW corrections in photon-initiated processes, the absence of the
lepton PDFs has never been a real issue. On the other hand, the recent
progresses in the automation of NLO EW corrections
\cite{Frixione:2014qaa,Frixione:2015zaa,Kallweit:2014xda,Denner:2014ina,Chiesa:2015mya}
make it possible to take into account these contributions without
additional efforts. Moreover, lepton PDFs can induce not only EW
corrections, but also new LO contributions for many SM and BSM
processes at hadron colliders. In particular, leptons in the initial
state can open new production mechanisms. For instance, at LO the
production of a lepton pair can proceed also through a $t$-channel
photon exchange as it is the case for the elastic scattering at lepton
colliders. This production mechanism is enhanced in the peripheral
region close to the beam pipe, therefore a quantitative estimate of
lepton PDFs is desirable in order not to introduce any systematic
effect in the searches and measurements in the Run II at the LHC and
possibly also at future colliders.

The purpose of this paper is to give the first quantitative estimate
on the leptonic content of the proton and analyse its phenomenological
impact at the LHC at 13 TeV and at a future 100 TeV hadron-hadron
collider (FCC-hh). This will be achieved in two steps. First of all by
implementing the lepton PDF Dokshitzer-Gribov-Altarelli-Lipatov-Parisi
(DGLAP) evolution equations at LO in QED in the so-called
variable-flavour-number (VFN) scheme in the evolution code {\tt
  APFEL}~\cite{Bertone:2013vaa}, and secondly by producing a guess for
the lepton PDFs at the initial scale based on the assumption that
leptons are generated by photon splitting at the respective mass
scales.

The paper is organised as follows. In Section~\ref{sec:PDFs} we
describe the procedure used to generate new PDF sets including
leptons. First, in Section \ref{sec:DGLAPwithLeptons} we show how to
generalise the DGLAP evolution equations in order to include the
complete QED corrections at LO, which involve leptons. Then, in
Section~\ref{sec:model} we discuss how to model the lepton
distributions at the initial scale on the base of different
theoretical assumptions. In Section~\ref{sec:results} we present the
numerical results obtained with {\tt APFEL}, considering different
initial conditions and comparing the results at the level of PDFs and
momentum fractions. In Section~\ref{sec:pheno} we discuss the
phenomenological impact of lepton (and photon) PDFs at the LHC and at
the FCC-hh.  In Section \ref{sec:lumi} we look at the impact of lepton
PDFs on parton luminosities and we analyse their dependence on the
invariant mass and the rapidity of the initial/final state in a
process-independent approach. Then, in Section \ref{sec:processes}, we
turn to the study of the effects of lepton PDFs for some relevant SM
(and BSM) processes at the LHC and at the FCC-hh. Finally, in
Section~\ref{sec:outlook}, we draw our conclusions and discuss
possible phenomenological studies that could be done using PDF sets
with lepton distributions. The PDF grids used in this work have been
made public in the {\tt LHAPDF6} library~\cite{Buckley:2014ana}
format.

\section{PDFs determination} \label{sec:PDFs}

\subsection{DGLAP equations in the presence of photons and leptons}\label{sec:DGLAPwithLeptons}

In this section we will show how to extend the DGLAP equations in
order to include the evolution of photon and lepton PDFs at leading
order (LO) in QED. The inclusion of the photon PDF evolution has
already been treated extensively in
Refs.~\cite{Martin:2004dh,Roth:2004ti,Bertone:2013vaa} and thus we
will not discuss it here. As far as leptons are concerned, at LO in
QED they couple directly only to photons. However, since the photons
couple to quarks that in turn couple to gluons, the lepton PDFs
evolution will indirectly depend on the evolution of all other
partons. Following the notation of Ref.~\cite{Bertone:2013vaa}, where
QCD and QED evolutions are treated separately, the inclusion of
leptons does not imply any change to the pure QCD sector. On the
contrary, with the inclusion of leptons, the QED evolution equations
with respect to the QED factorisation scale $\nu$ read:
\begin{equation}\label{QED_DGLAP}
  \small
  \begin{array}{rcl}
    \displaystyle \nu^{2}\frac{\partial \gamma}{\partial \nu^{2}}
    &=& \displaystyle \frac{\alpha}{4\pi} \left[\left(\sum_{q}^{n_f}N_c e_{q}^{2}\right)
        P_{\gamma\gamma}^{(0)}\otimes\gamma+\sum_q^{n_f}e_{q}^{2}P_{\gamma
        q}^{(0)}\otimes (q+\bar{q}) +\sum_\ell^{n_\ell} P_{\gamma
        \ell}^{(0)}\otimes (\ell^-+\ell^+)
        \right]\,,\\
    \\
    \displaystyle \nu^{2}\frac{\partial q}{\partial
    \nu^{2}} &=&\displaystyle \frac{\alpha}{4\pi}
                 \left[ N_c e_q^{2} P_{q\gamma}^{(0)}\otimes \gamma+e_q^{2}
                 P_{qq}^{(0)}\otimes q\right]\,,\\
    \\
    \displaystyle \nu^{2}\frac{\partial \bar{q}}{\partial
    \nu^{2}} &=&\displaystyle \frac{\alpha}{4\pi}
                 \left[ N_c e_q^{2} P_{q\gamma}^{(0)}\otimes \gamma+e_q^{2}
                 P_{qq}^{(0)}\otimes \bar{q}\right]\,, \\
    \\
    \displaystyle \nu^{2}\frac{\partial \ell^-}{\partial
    \nu^{2}} &=&\displaystyle \frac{\alpha}{4\pi}
                 \left[ P_{\ell\gamma}^{(0)}\otimes \gamma+
                 P_{\ell\ell}^{(0)}\otimes \ell^-\right]\,,\\
    \\
    \displaystyle \nu^{2}\frac{\partial \ell^+}{\partial
    \nu^{2}} &=&\displaystyle \frac{\alpha}{4\pi}
                 \left[ P_{\ell\gamma}^{(0)}\otimes \gamma+
                 P_{\ell\ell}^{(0)}\otimes \ell^+\right]\,, 
  \end{array}
\end{equation}
where $\gamma$, $q(\bar{q})$ and $\ell^-(\ell^+)$ correspond to the
PDFs of the photon, the quark(antiquark) flavour $q$ and the
lepton(antilepton) species $\ell$, respectively. In Eq.~(\ref{QED_DGLAP}),
$e_q$ corresponds to the electric charge of the quark flavour $q$,
$N_c=3$ is the number of QCD colours and $\alpha$ is the running
fine-structure constant. Note also that the indices $q$ and $\ell$ in
the sums in the first line of Eq.~(\ref{QED_DGLAP}) run over the $n_f$
and $n_\ell$ active quarks and leptons at the scale $\nu$,
respectively. The symbol $\otimes$ represents the usual Mellin
convolution operator defined as:
\begin{equation}
  A(x)\otimes B(x) \equiv \int_0^1 dy \int_0^1 dz\,A(y)B(z) \delta(x-yz)\,.
\end{equation}
Finally, the LO QED splitting functions $P_{ij}^{(0)}$ are given by:
\begin{equation}\label{QEDLOsplittingFunctions}
  \begin{array}{rcl}
    \displaystyle P_{\gamma\gamma}^{(0)}(x) &=& \displaystyle
                                                -\frac{4}{3}\delta(1-x)\,,\\
    \\
    \displaystyle P_{q\gamma}^{(0)}(x)&=&P_{\ell\gamma}^{(0)}(x) = \displaystyle 2\left[x^2+(1-x)^2\right],\\
    \\
    \displaystyle P_{\gamma q}^{(0)}(x)&=&P_{\gamma\ell}^{(0)}(x) = \displaystyle
                                           2\left[\frac{1+(1-x)^2}{x}\right],\\
    \\
    \displaystyle P_{qq}^{(0)}(x)&=&P_{\ell\ell}^{(0)}(x) = \displaystyle
                                     2\frac{1+x^2}{(1-x)_+}+3\delta(1-x)\,.
  \end{array}
\end{equation}

Combining the system of differential equations in
Eq.~(\ref{QED_DGLAP}) with the pure-QCD DGLAP equations that govern
the evolution of gluons and quarks, we obtain the full (N)NLO QCD plus
LO QED evolution in the presence of photons and leptons. The solution
of this system is implemented in version 2.4.0, or later, of the {\tt
  APFEL} evolution library~\cite{Bertone:2013vaa}.

It is worth stressing that, contrary to the approach described in the
original reference of {\tt APFEL}~\cite{Bertone:2013vaa}, in the new
implementation of the coupled QCD+QED DGLAP equations the QCD and the
QED evolutions are no longer treated separately and combined \textit{a
  posteriori}. Indeed, thanks to a suitable combination of PDFs that
cleverly diagonalises the full splitting-function matrix in the
presence of QED corrections, we have been able to implement in {\tt
  APFEL} an efficient \textit{unified} solution of the QCD+QED DGLAP
equations where the evolution of PDFs with respect to QCD and QED
factorisation scales takes place at the same time. In other words, QCD
and QED factorisation scales are no longer kept separate and are
instead identified. We present the details of this basis in
Appendix~\ref{sec:appendix}.

As far as the fine-structure constant $\alpha$ is concerned,
consistently with the evolution of PDFs, we only consider the LO
running by solving the renormalisation-group equation:
\begin{equation}
  \nu^2 \frac{d\alpha}{d\nu^2} = \beta_{\rm QED}^{(0)} \alpha^2(\nu)\, ,
\end{equation}
with:
\begin{equation}
  \beta_{\rm QED}^{(0)} = \frac{8}{12\pi}\left(N_c\sum_{q}^{n_f}e_q^2+n_\ell\right)\,,
\end{equation}
where, as a boundary condition for the evolution, we take
$\alpha^{-1}(m_\tau) = 133.4$.

\subsection{Modelling the lepton PDFs}\label{sec:model}

Once the full QCD+QED DGLAP evolution in the presence of photons and
leptons has been implemented, the following step is the determination
of suitable boundary conditions, $i.e.$ the initial scale PDFs, to be
evolved. As far as gluons and quarks are concerned, various solid
determinations are publicly
available~\cite{Ball:2014uwa,Harland-Lang:2014zoa,Gao:2013xoa,Abramowicz:2015mha,Alekhin:2013nda}.
On the other hand, as already mentioned, only two public
determinations of the photon PDF are available at the moment: the
MRST2004QED set~\cite{Martin:2004dh}, where the photon PDF is
determined by means of a model assumption, and the more recent
NNPDF2.3QED family~\cite{Ball:2013hta}, where the photon PDF is
instead determined from a fit to data. No lepton PDF determination has
ever been attempted so far.

In principle, the most reliable way to determine the lepton PDFs is
the extraction from data by means of a fit. However, to our opinion,
this option is currently not viable. The main reason is that lepton
PDFs are expected to be tiny as compared to the quark and gluon ones
and even substantially smaller than the photon PDF. Assuming no
intrinsic leptonic component in the proton, lepton PDFs are expected
to be of the order of $\alpha$ times the photon PDF, where
$\alpha \sim 10^{-2}$. As a consequence, the photon being already
suppressed as compared to quark and gluon PDFs, the contribution of
leptons is likely to be so small that the determination of the
corresponding PDFs from experimental data is hardly achievable.

As an alternative to the fit, one could try to guess the functional
form of the lepton PDFs just by assuming that light leptons, $i.e.$
electrons and muons, are generated by photon splitting at the
respective mass scales. Using the leading-logarithmic (LL)
approximation for the photon splitting into leptons, one can then
infer their distributions at the initial scale $Q_0\simeq1$ GeV as:
\begin{equation}\label{eq:ansatz}
  \ell^-(x,Q_0) =\ell^+(x,Q_0) =
  \frac{\alpha(Q_0)}{4\pi} \ln\left(\frac{Q_0^2}{m_\ell^2}\right)
  \int_x^1\frac{dy}{y} P_{\ell\gamma}^{(0)}\left(\frac{x}{y}\right)
  \gamma(y,Q_0)\,,
\end{equation}
with $\ell=e,\mu$ \footnote{For the light lepton masses, we take
  $m_e = 0.510998928$ MeV and $m_\mu = 105.6583715$ MeV, as quoted in
  the PDG~\cite{Agashe:2014kda}.}. In other words,
Eq.~(\ref{eq:ansatz}) comes from the linearisation of the LO QED DGLAP
evolution for the lepton PDFs $\ell^\pm$ from the mass scale $m_\ell$,
where they are assumed to appear as dynamical partons, to the initial
scale $Q_0$. Indeed, in the situation where both QED and QCD
evolutions are considered, the divergence of the strong coupling
$\alpha_s$ in correspondence of $\Lambda_{\rm QCD}$ (Landau pole),
with $\Lambda_{\rm QCD}>m_\ell$, makes such evolution impossible to
implement in an exact way and Eq.~(\ref{eq:ansatz}) represents an
approximated solution.  Thus,
  Eq.~(\ref{eq:ansatz}) is equivalent to assume that the photon
  PDF does not evolve between the lepton mass scale $m_\ell$ and the
  initial scale $Q_0$. However, the DGLAP evolution between $m_\ell$ and  $Q_0$ would
  introduce additional $\mathcal{O}(\alpha)$ terms
  that exceed the LO QED accuracy.

Since the tau mass is typically larger than $Q_0$ ($m_\tau = 1.777$
GeV $\gtrsim Q_0$), we have chosen to determine the $\tau^{\pm}$ PDFs
according to the usual VFN
scheme~\cite{Forte:2010ta,Aivazis:1993pi,Thorne:1997ga},
$i.e.$ by dynamically generating them at the threshold.

\subsection{PDF sets with lepton distributions}\label{sec:results}

\begin{table}
  \centering
  \small
\begin{tabular}{|c|l|c|c|c|c|c|}
  \hline 
  ID & PDF Set & Ref. & QCD & QED & Photon PDF & Lepton PDFs\tabularnewline
  \hline 
  \hline 
  A1 & \texttt{apfel\_nn23nlo0118\_lept0} & \cite{Ball:2012cx} & NLO & LO & $\gamma(x,Q_0) = 0$ & Eq.~(\ref{eq:ansatz0}) \tabularnewline
  \hline 
  A2 & \texttt{apfel\_nn23nnlo0118\_lept0} & \cite{Ball:2012cx} & NNLO  & LO  & $\gamma(x,Q_0) = 0$ &
  Eq.~(\ref{eq:ansatz0}) \tabularnewline
  \hline 
  B1 & \texttt{apfel\_nn23qedlo0118\_lept0} & \cite{Carrazza:2013axa} & LO & LO &
  Internal & Eq.~(\ref{eq:ansatz0}) \tabularnewline
  \hline 
  B2 & \texttt{apfel\_nn23qednlo0118\_lept0} & \cite{Ball:2013hta} & NLO & LO &
  Internal & Eq.~(\ref{eq:ansatz0}) \tabularnewline
  \hline 
  B3 & \texttt{apfel\_nn23qednnlo0118\_lept0} & \cite{Ball:2013hta} & NNLO & LO &
  Internal & Eq.~(\ref{eq:ansatz0}) \tabularnewline
  \hline 
  B4 & \texttt{apfel\_mrst04qed\_lept0} & \cite{Martin:2004dh} & NLO & LO & Internal
  & Eq.~(\ref{eq:ansatz0}) \tabularnewline
  \hline 
  C1 &\texttt{apfel\_nn23qedlo0118\_lept} & \cite{Carrazza:2013axa} & LO & LO &
  Internal & Eq.~(\ref{eq:ansatz}) \tabularnewline
  \hline 
  C2 &\texttt{apfel\_nn23qednlo0118\_lept} & \cite{Ball:2013hta} & NLO & LO &
  Internal & Eq.~(\ref{eq:ansatz}) \tabularnewline
  \hline 
  C3 & \texttt{apfel\_nn23qednnlo0118\_lept} & \cite{Ball:2013hta} & NNLO & LO &
  Internal & Eq.~(\ref{eq:ansatz}) \tabularnewline
  \hline 
  C4 & \texttt{apfel\_mrst04qed\_lept} & \cite{Martin:2004dh} & NLO & LO & Internal & Eq.~(\ref{eq:ansatz}) \tabularnewline
  \hline 
\end{tabular}
\caption{\small Summary of the sets of PDFs including photons and
  leptons generated with {\tt APFEL}.}
\label{tab:sets}
\end{table}

In this section we discuss the results of the implementation of the
lepton PDF evolution in {\tt APFEL}. As discussed in the previous
section, the determination of lepton PDFs from a direct fit to data
seems to be hard to achieve and thus, as an alternative, we adopt a
model based on a theoretical assumption for modelling lepton PDFs at
the initial scale.

The model presented in the previous section is based on the assumption
that leptons are generated in pairs from LL photon splitting at the
respective mass scales. This results in the ansatz in
Eq.~(\ref{eq:ansatz}) for the light lepton PDFs. However, in order to
test how sensitive the results are with respect to the initial scale
distributions, we also consider the ``zero-lepton'' ansatz where the
lepton PDFs at the initial scale $Q_0$ are (artificially) set to zero,
that is:
\begin{equation}\label{eq:ansatz0}
  \ell^-(x,Q_0) =\ell^+(x,Q_0) = 0\,.
\end{equation}

In this context, the construction of PDF sets with leptons requires a
pre-existing PDF set to which we add our models for the lepton
distributions. Of course, in order to apply the ansatz in
Eq.~(\ref{eq:ansatz}), we need PDF sets that already contain a photon
PDF, {\it e.g.} the MRST2004QED set~\cite{Martin:2004dh} or the
NNPDF2.3QED sets~\cite{Ball:2013hta}. We will use both sets to
generate lepton PDFs via the ansatz in Eq.~(\ref{eq:ansatz}). On the
contrary, the ansatz in Eq.~(\ref{eq:ansatz0}) can be applied to any
set so that lepton and photon distributions are generated just by
DGLAP evolution.

In order to assess the effect of considering lepton PDFs in the DGLAP
evolution, we consider three different initial-scale configurations
that are also summarised in Tab.~\ref{tab:sets}:

\begin{itemize}

\item sets where both photon and lepton PDFs are initially absent and
  dynamically generated by DGLAP evolution. For this configuration we
  have constructed the sets A1 and A2 in Table~\ref{tab:sets} based on
  NNPDF2.3 NLO and NNLO, respectively.

\item Sets where the photon distribution is present and the lepton
  PDFs are set to zero at the initial scale $Q_0$ ($i.e.$
  Eq.~(\ref{eq:ansatz0})) and then evolved as discussed in
  Sect.~\ref{sec:DGLAPwithLeptons}. These configurations are based on
  the NNPDF2.3QED and MRST2004QED sets of PDFs and identified by the
  indices B1, B2, B3 and B4 in Table~\ref{tab:sets}.

\item Sets of PDFs generated from NNPDF2.3QED and MRST2004QED using
  the ansatz in Eq.~(\ref{eq:ansatz}) for the light lepton PDFs (sets
  C1, C2, C3 and C4 in Table~\ref{tab:sets}).

\end{itemize}

The evolution of the PDF sets listed in Tab.~\ref{tab:sets} is
performed using {\tt APFEL}, as discussed in
Sect.~\ref{sec:DGLAPwithLeptons}.  All PDF sets are tabulated in the
{\tt LHAPDF6} format, which allows us to include the lepton PDFs in a
straightforward manner. In the following we will present the results
of the implementation of the lepton PDF evolution for the different
ans\"atze discussed above.  In particular, we will quantify momentum
fractions, correlations and the differences among the different sets.
In order to make the discussion more concise, we will only presents
results for the NLO sets. However, the features we observe are common
to all sets and thus all the conclusions we draw apply also to LO and
NNLO sets.

\begin{figure}
  \centering
  \includegraphics[scale=0.5]{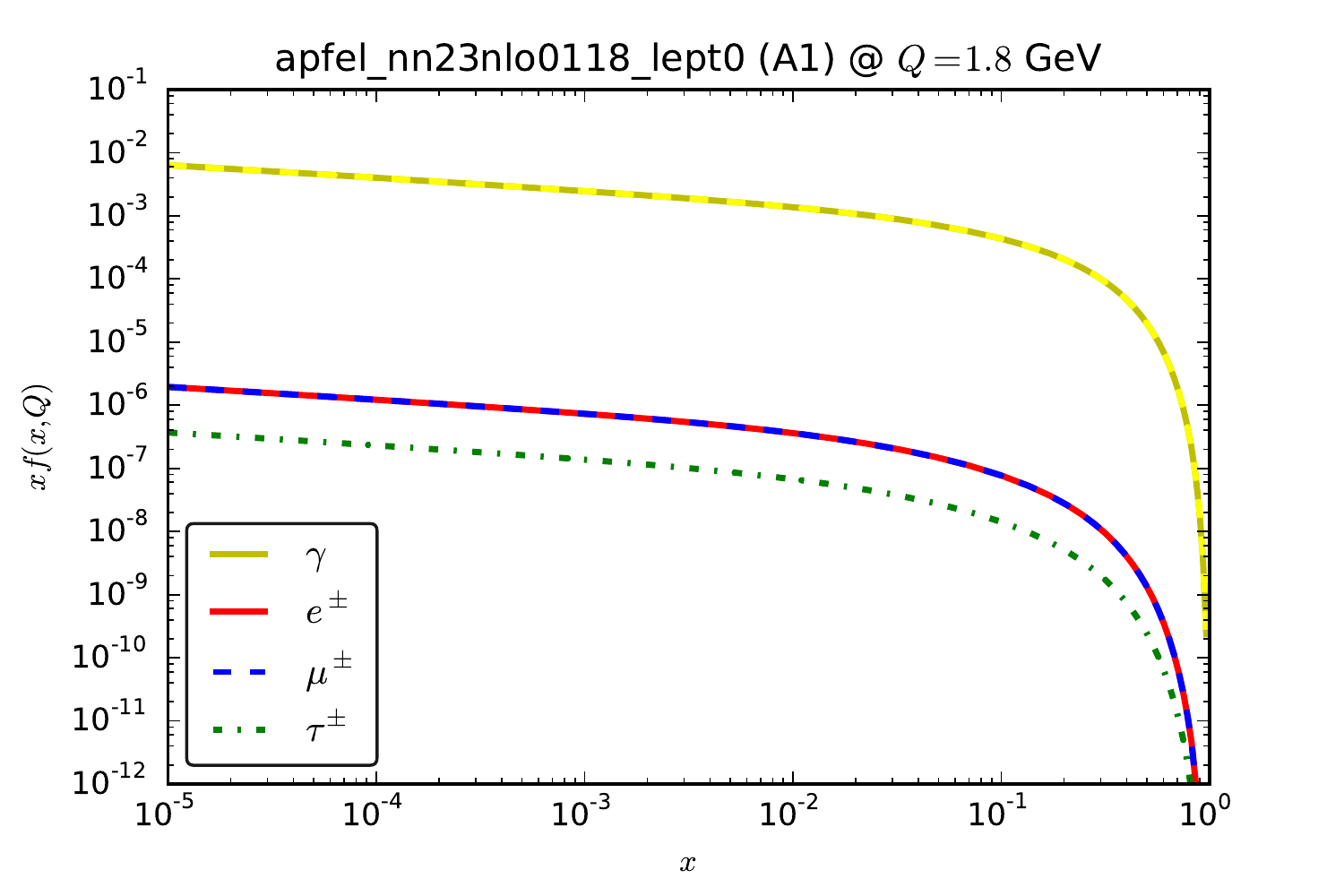}\includegraphics[scale=0.5]{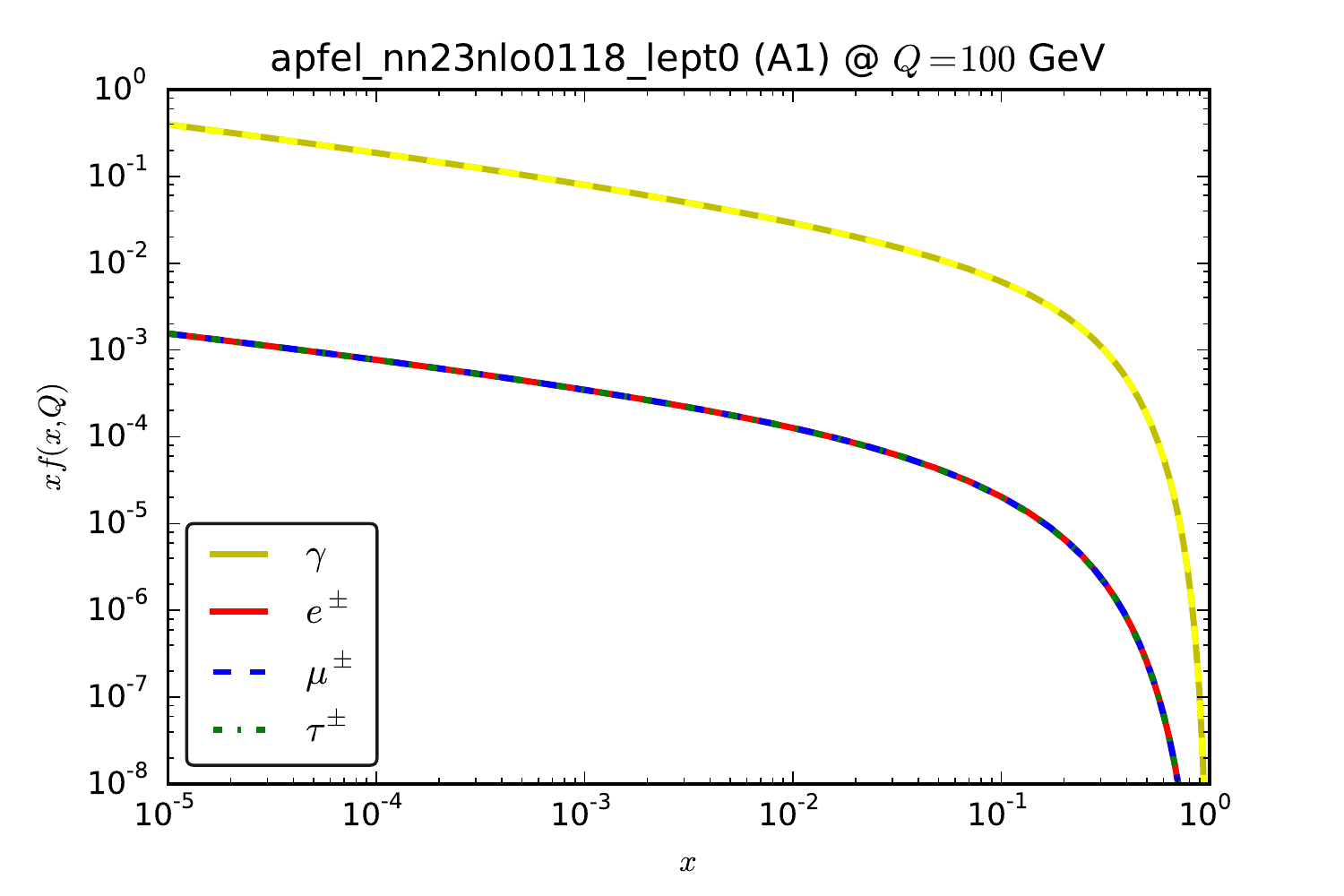}
  \caption{Lepton and photon PDFs for the configuration A1 at $Q=1.8$
    GeV (left) and $Q=100$ GeV (right).\label{fig:dyna}}
  \vspace{5mm}
  \includegraphics[scale=0.5]{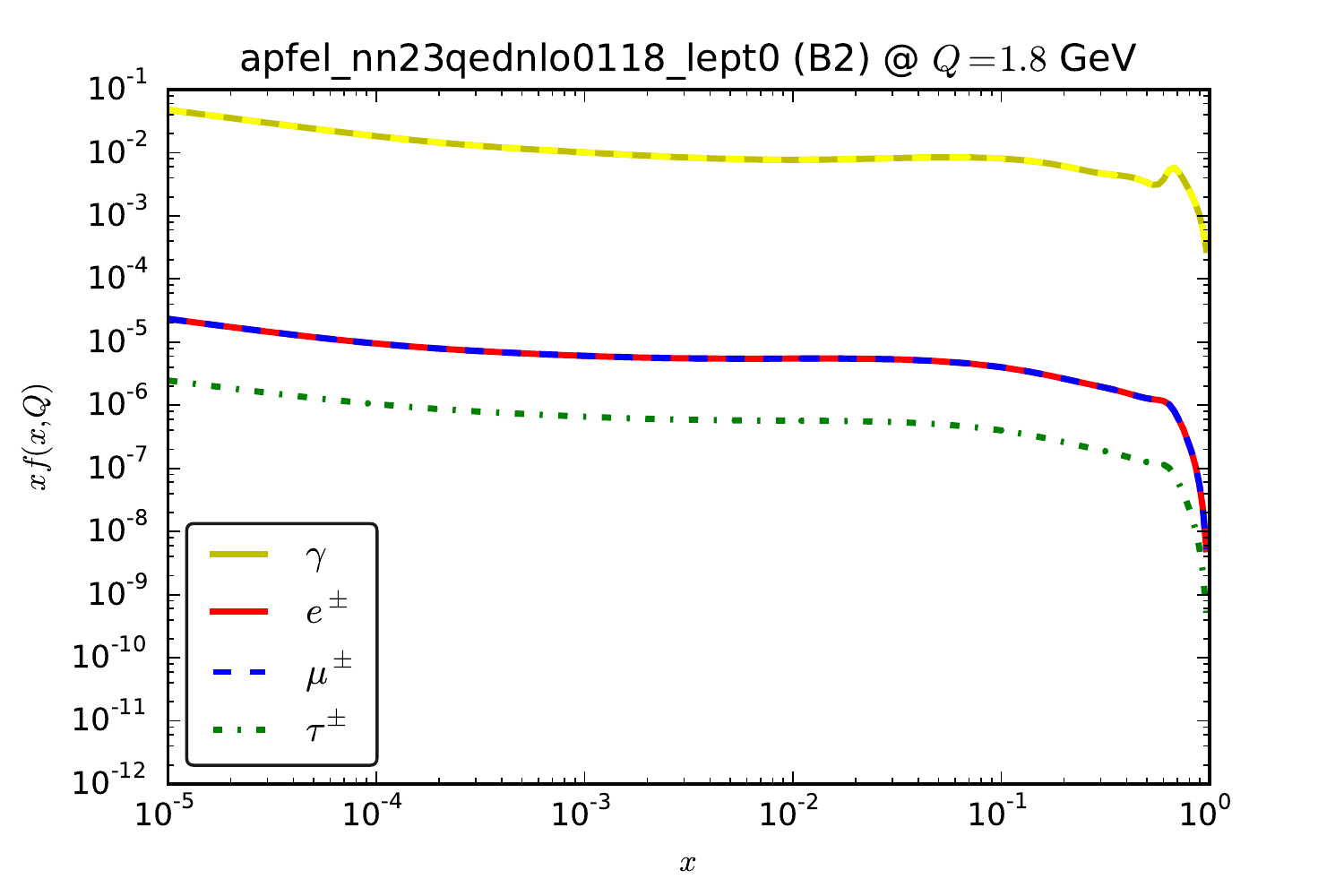}\includegraphics[scale=0.5]{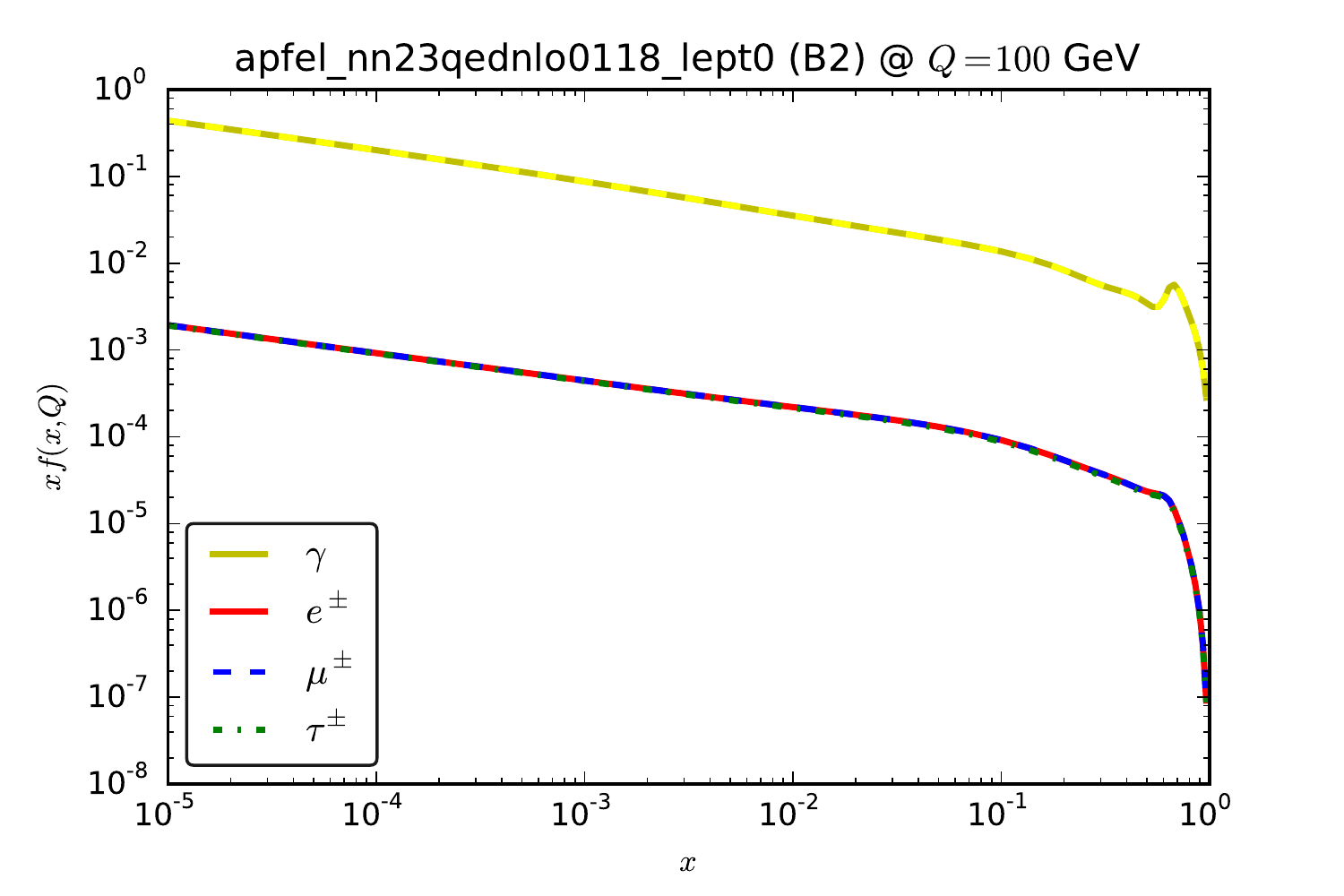}
  \includegraphics[scale=0.5]{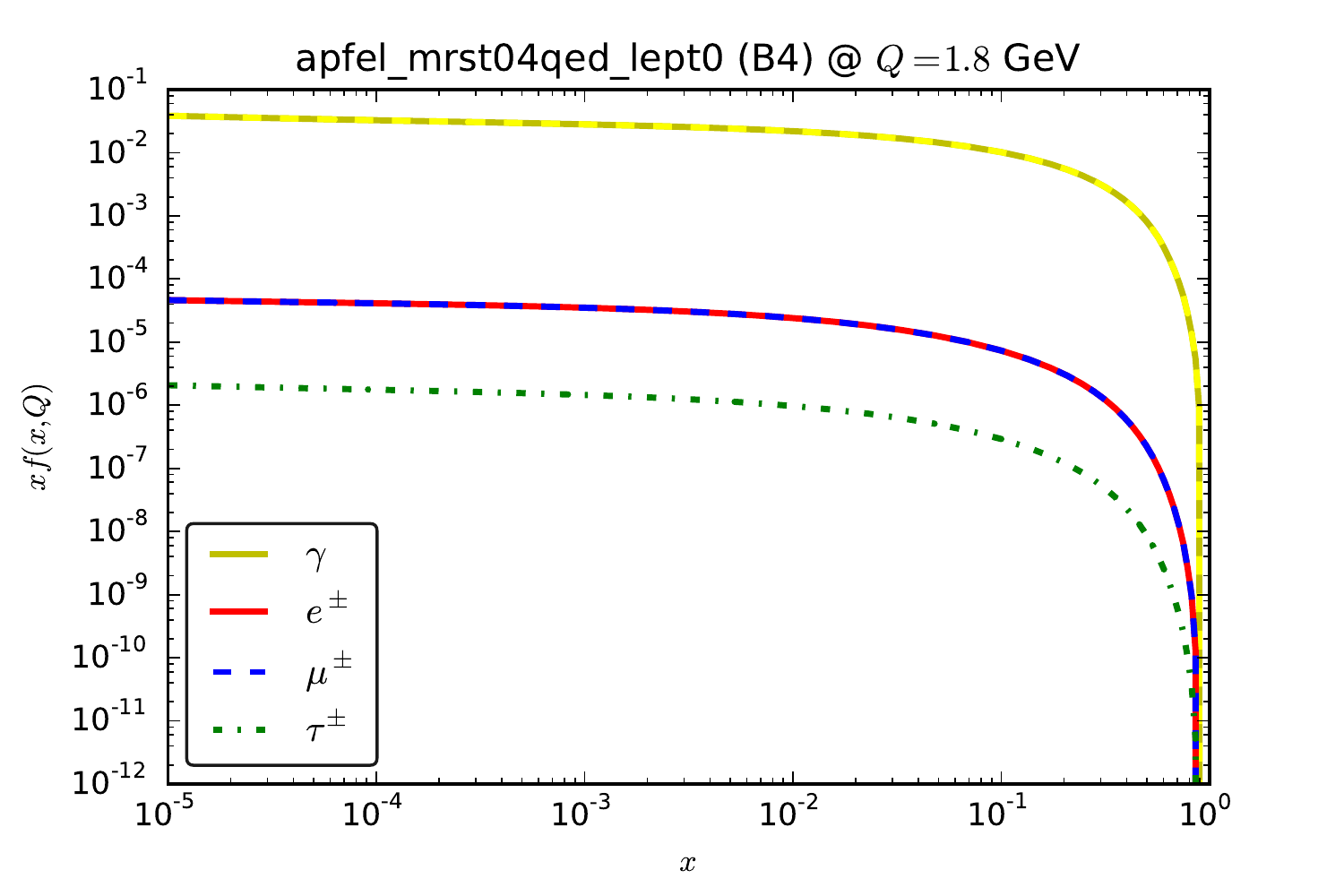}\includegraphics[scale=0.5]{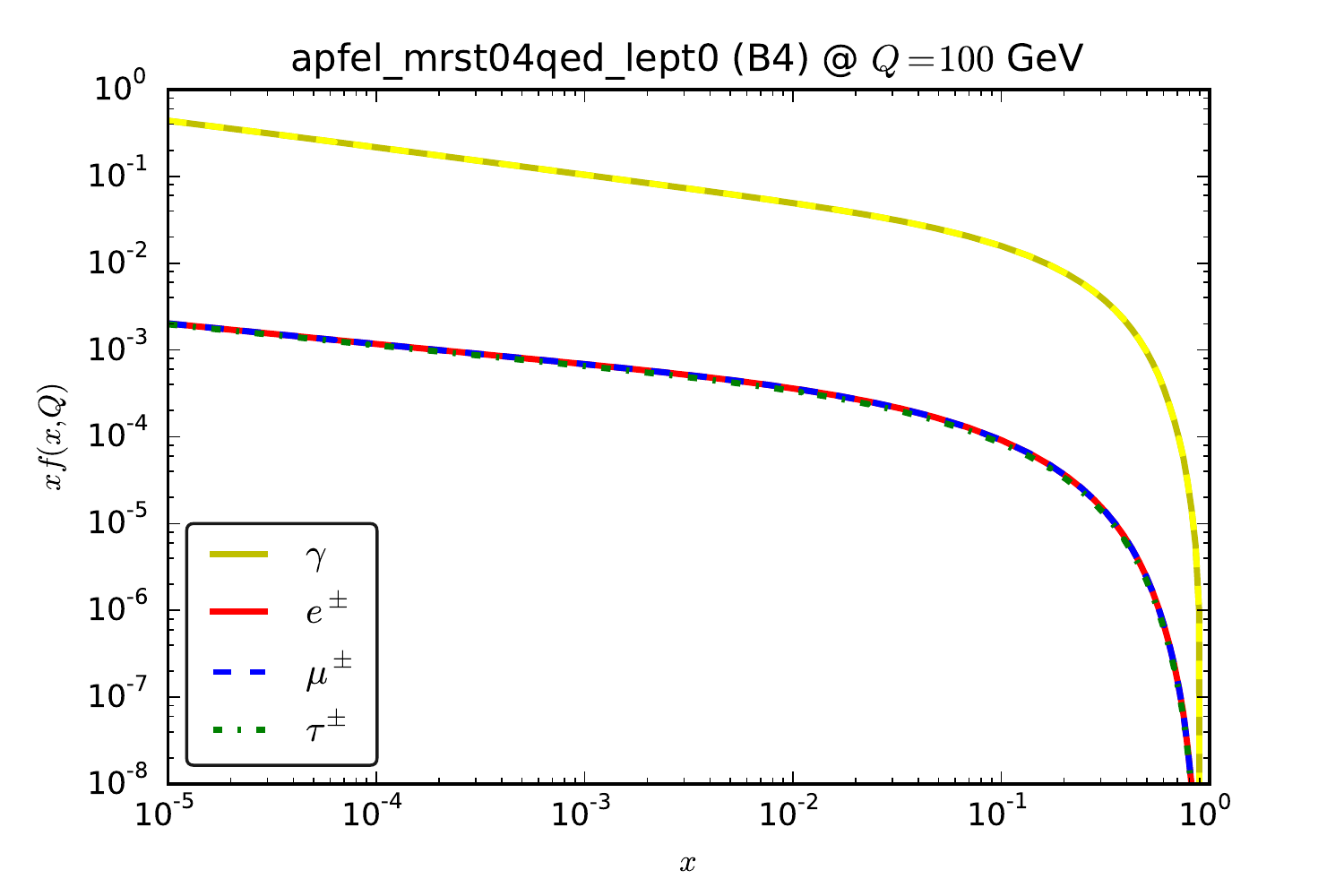}
  \caption{Same as Fig.~\ref{fig:dyna} for the configurations B2 (top)
    and B4 (bottom).\label{fig:photonfit}}
\end{figure}

In Fig.~\ref{fig:dyna} we show the lepton and photon PDF central
values for the set identified by the ID A1 in Table~\ref{tab:sets}. In
this configuration photons and leptons are set to zero at $Q_0=1$ GeV
and then dynamically generated by DGLAP evolution. The left plot shows
PDFs at $Q=1.8$ GeV, in this case electron and muon PDFs are identical
(by definition), and the $\tau$ PDF has just been dynamically
generated ($m_\tau = 1.777$ GeV). In the right plot the same
comparison is displayed at $Q=100$ GeV, showing that all lepton PDFs
are close to each other. Similar results are obtained also with the
NNPDF2.3 NNLO (A2) set.

Configurations B2 and B4 are shown in Fig.~\ref{fig:photonfit}. In
this case, the prior sets of PDFs contain a photon PDF while lepton
PDFs are null at the initial scale $Q_0$ and thus generated
dynamically by DGLAP evolution. Again, similar results are obtained
for the NNPDF2.3QED LO (B1) and NNLO (B3) sets.

\begin{figure}
  \centering
  \includegraphics[scale=0.5]{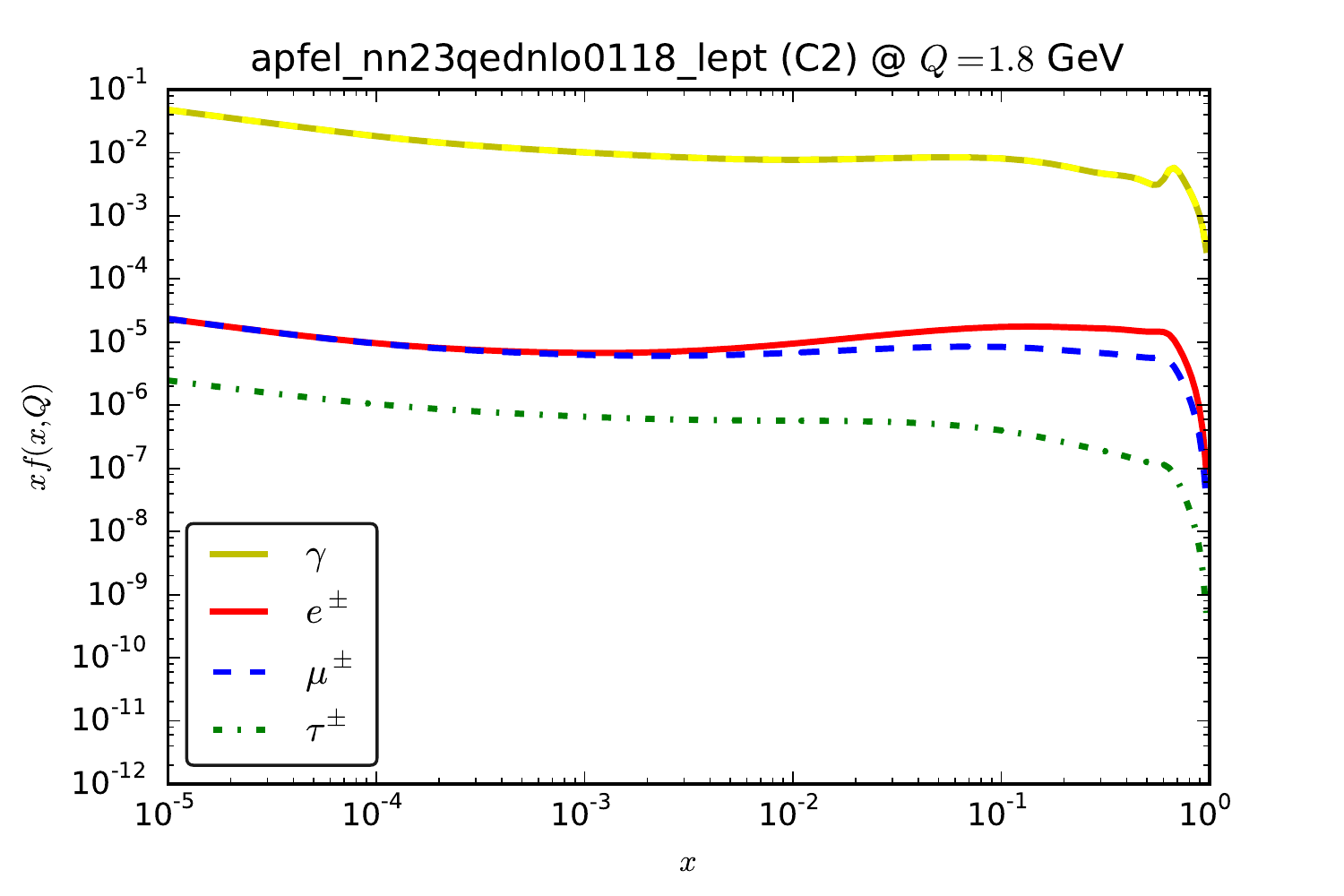}\includegraphics[scale=0.5]{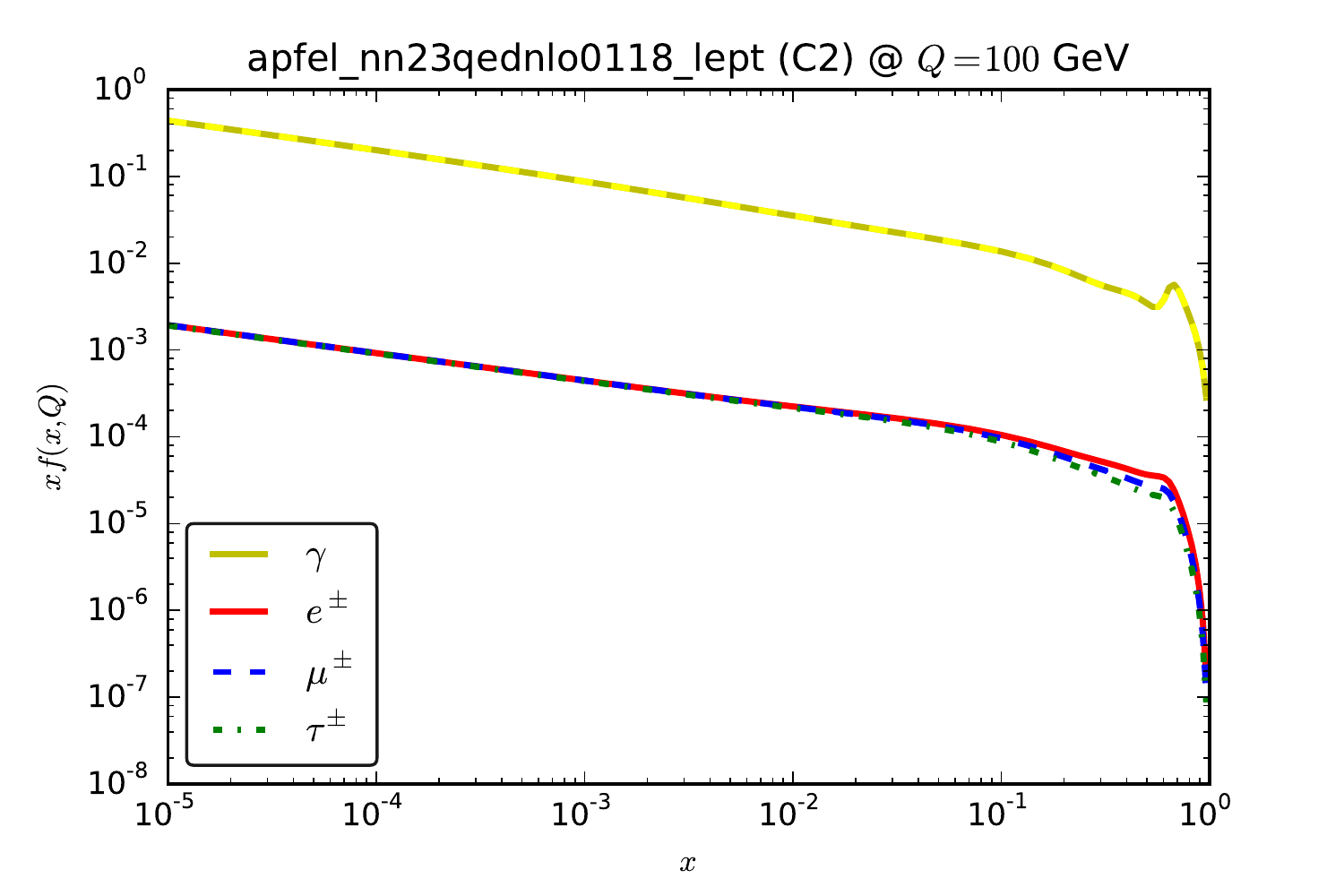}
  \includegraphics[scale=0.5]{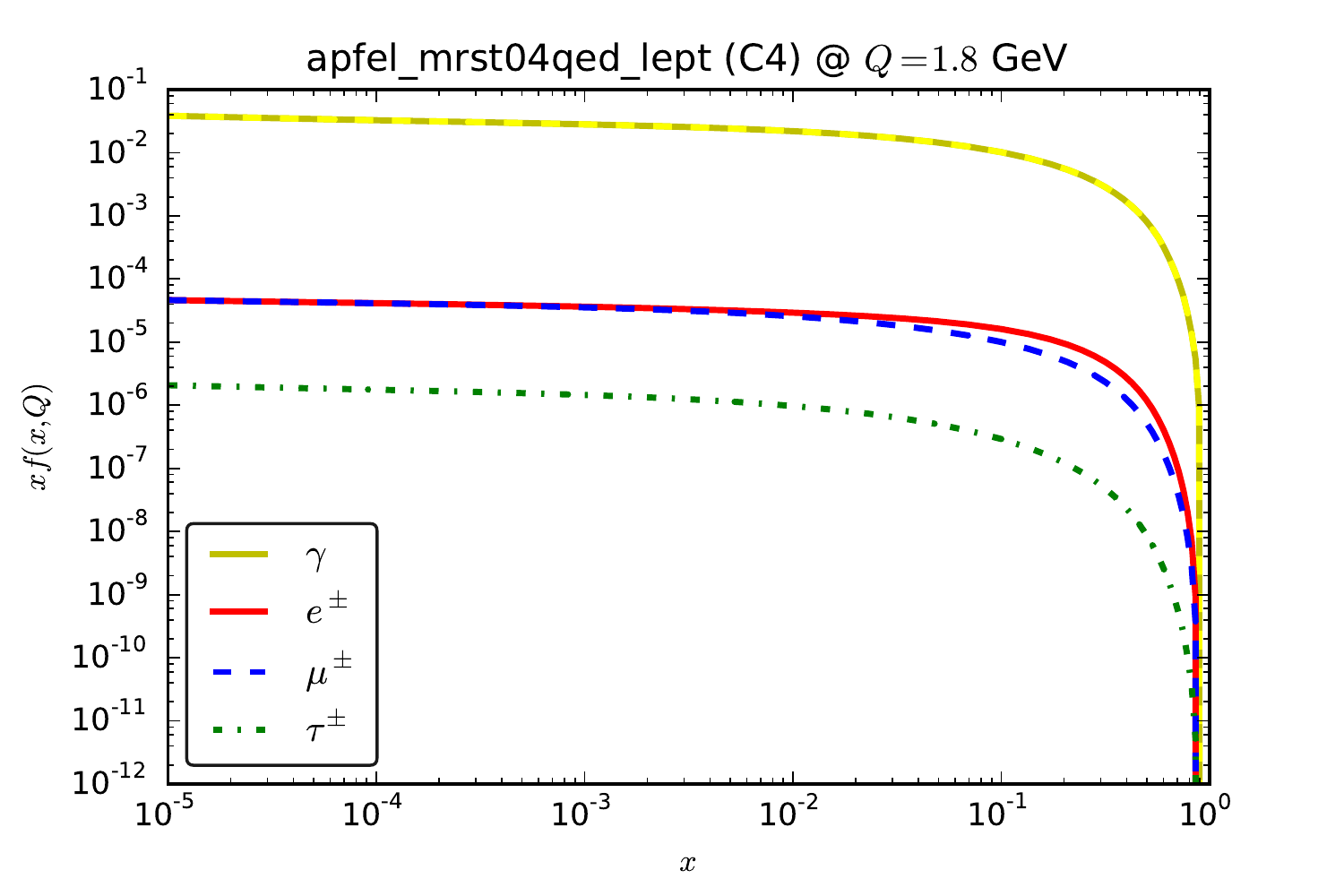}\includegraphics[scale=0.5]{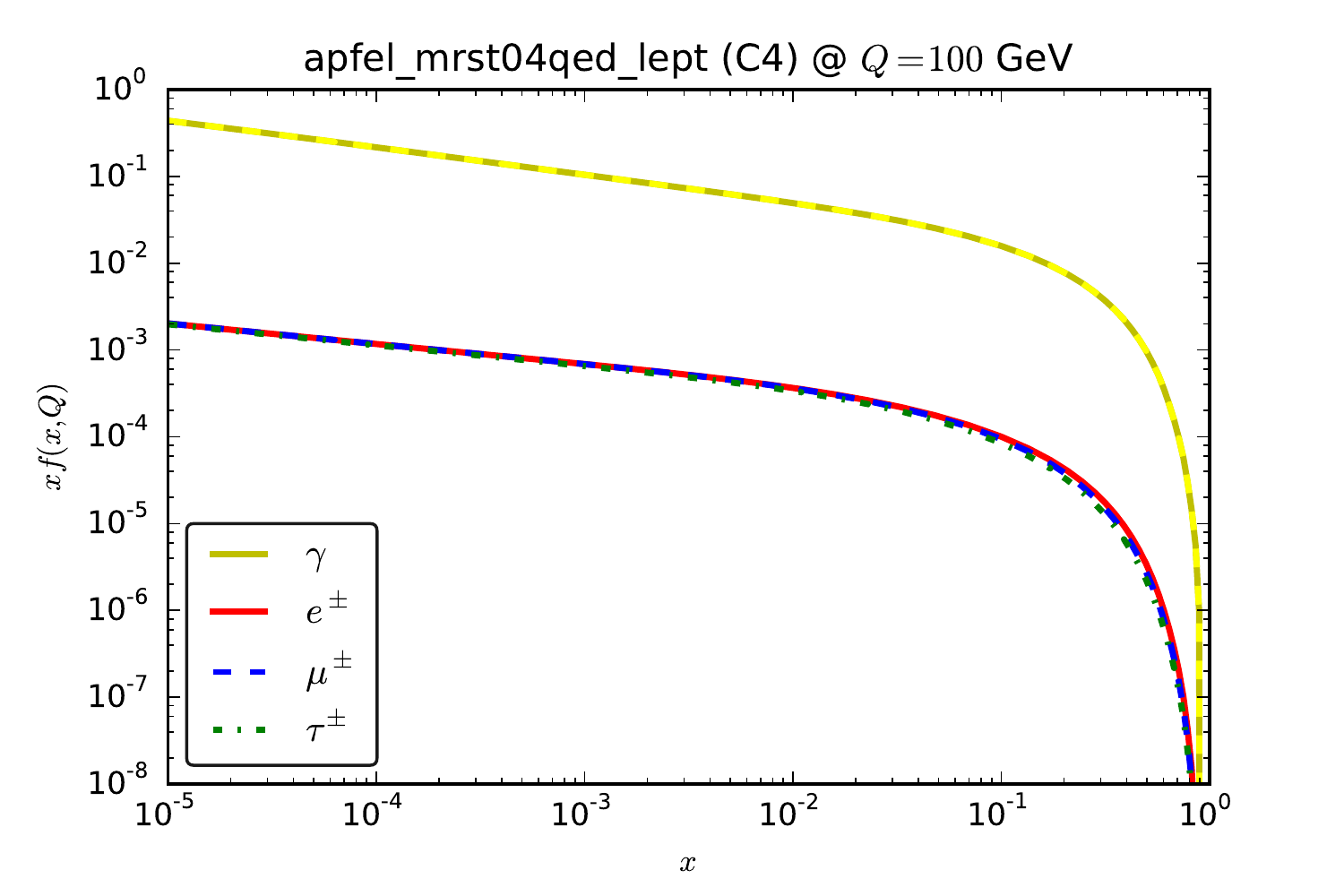}
  \caption{Same as Fig.~\ref{fig:dyna} for the configurations C2 (top)
    and C4 (bottom).\label{fig:leptonevolansatz}}
  \vspace{5mm}
  \includegraphics[scale=0.5]{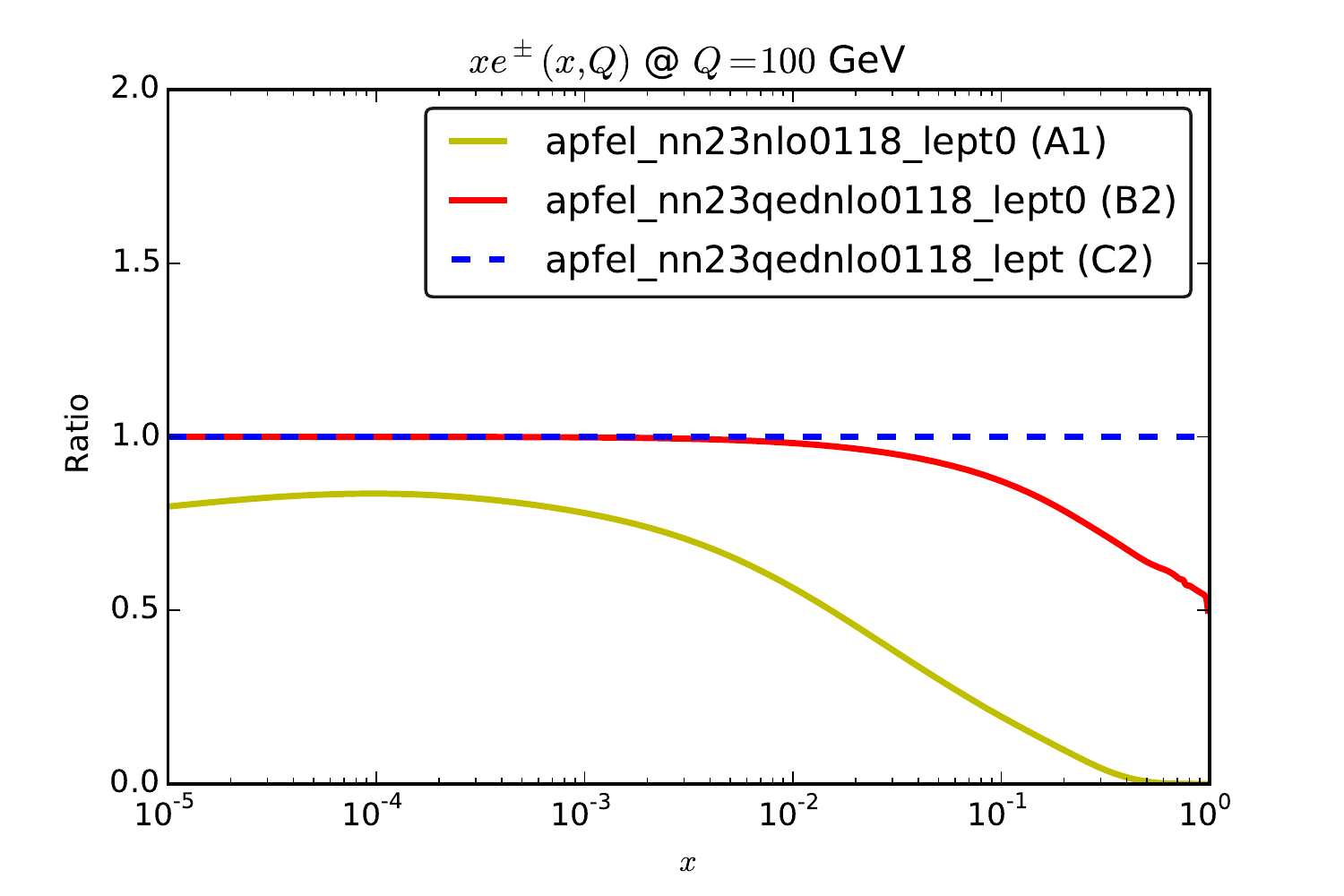}\includegraphics[scale=0.5]{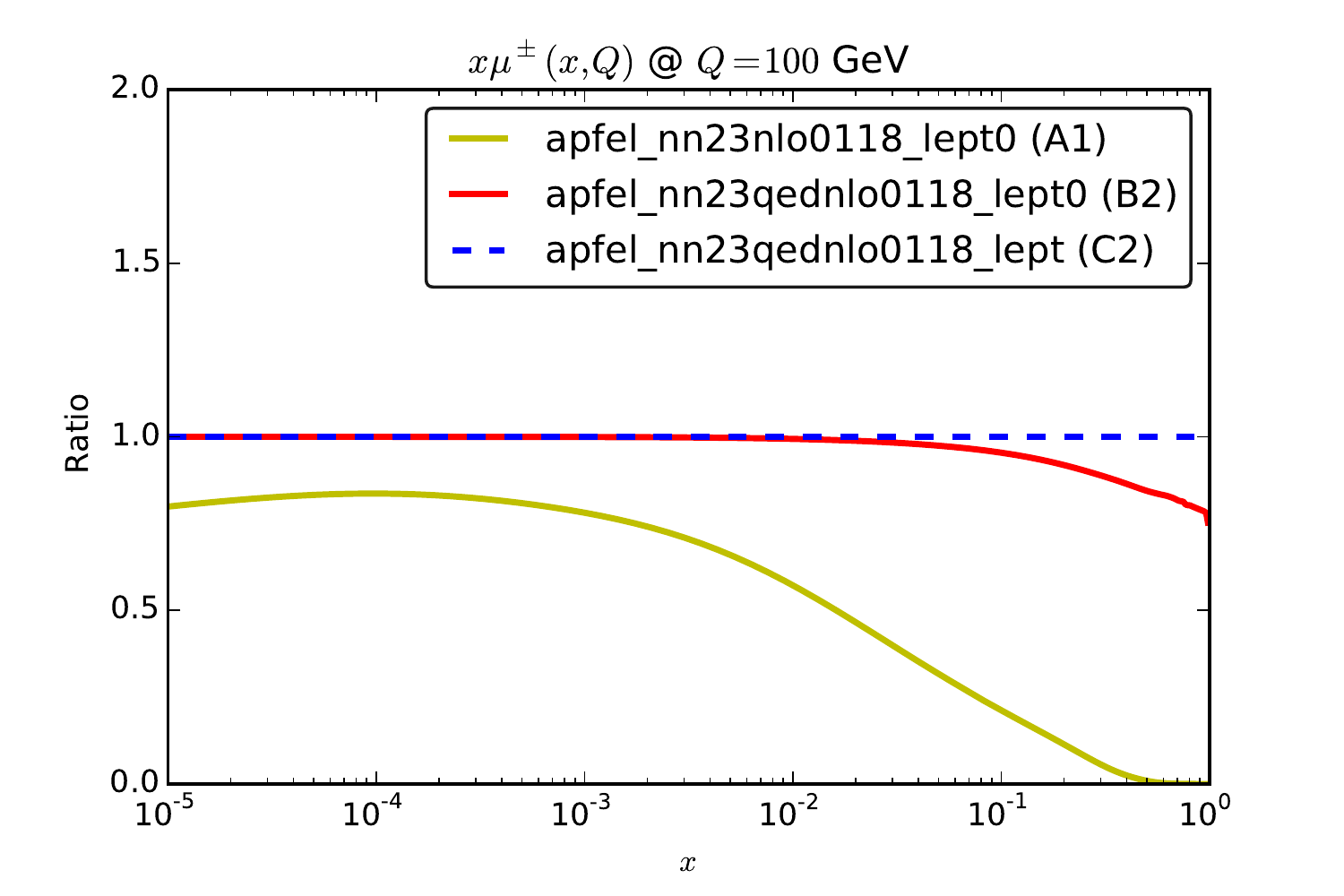}
  \caption{Electron (left) and muon (right) PDFs at $Q=100$ GeV for
    the configurations A1, B2 and C2, displayed as ratios to
    C2.\label{fig:leptratios}}
\end{figure}

Finally, we consider the configurations of type C where, starting from
a prior containing a photon PDF, the initial distributions for the
leptons are determined using the ansatz in Eq.~(\ref{eq:ansatz}).  In
Fig.~\ref{fig:leptonevolansatz} we show the resulting lepton PDFs for
the configurations C2 (top) and C4 (bottom), at $Q=1.8$ GeV (left) and
$Q=100$ GeV (right). Again, the qualitative behaviour for the
configurations C1 and C3 is the same.

Fig.~\ref{fig:leptratios} shows the ratios for the light lepton PDFs
at $Q=100$ GeV for the configurations A1, B2, C2 to C2. In this way we
can quantify the difference generated by the various initial
conditions starting from the same prior set, $i.e.$ NNPDF2.3 at
NLO. For the electron PDFs (left plot), the ans\"atze in
Eqs.~(\ref{eq:ansatz}) and~(\ref{eq:ansatz0}) applied to a set with a
photon PDF leads to similar results in the small-$x$ region, while
differences up to 50\% are observed in the large-$x$ region. The
electron PDFs resulting from the set without a photon PDF are instead
much smaller for all in the entire $x$ range.  The same behaviour is
observed also for the muon PDFs (right plot in
Fig.~\ref{fig:leptratios}), with slightly less enhanced discrepancies
as compared to electrons.

\begin{figure}
  \centering
  \includegraphics[scale=0.5]{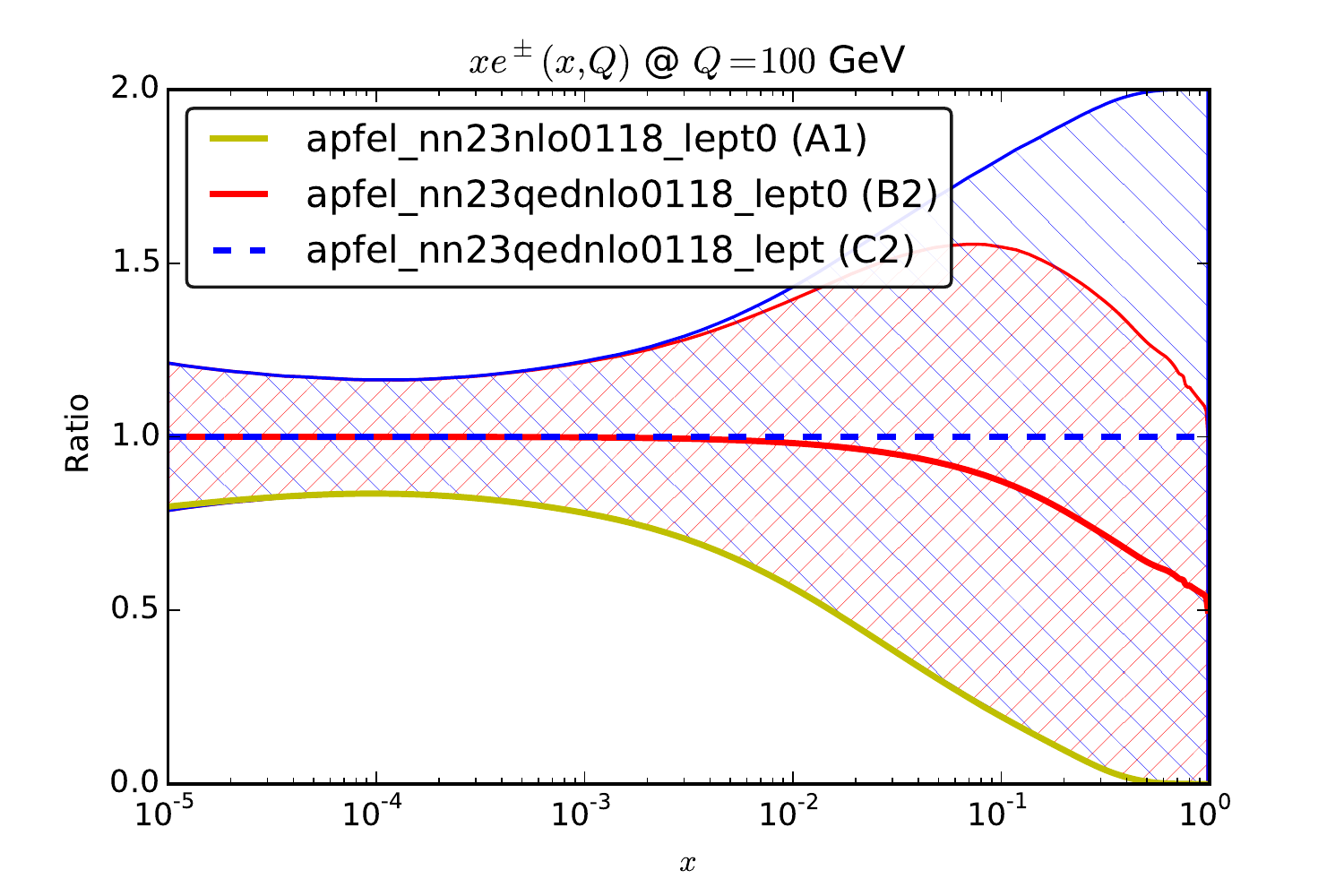}\includegraphics[scale=0.5]{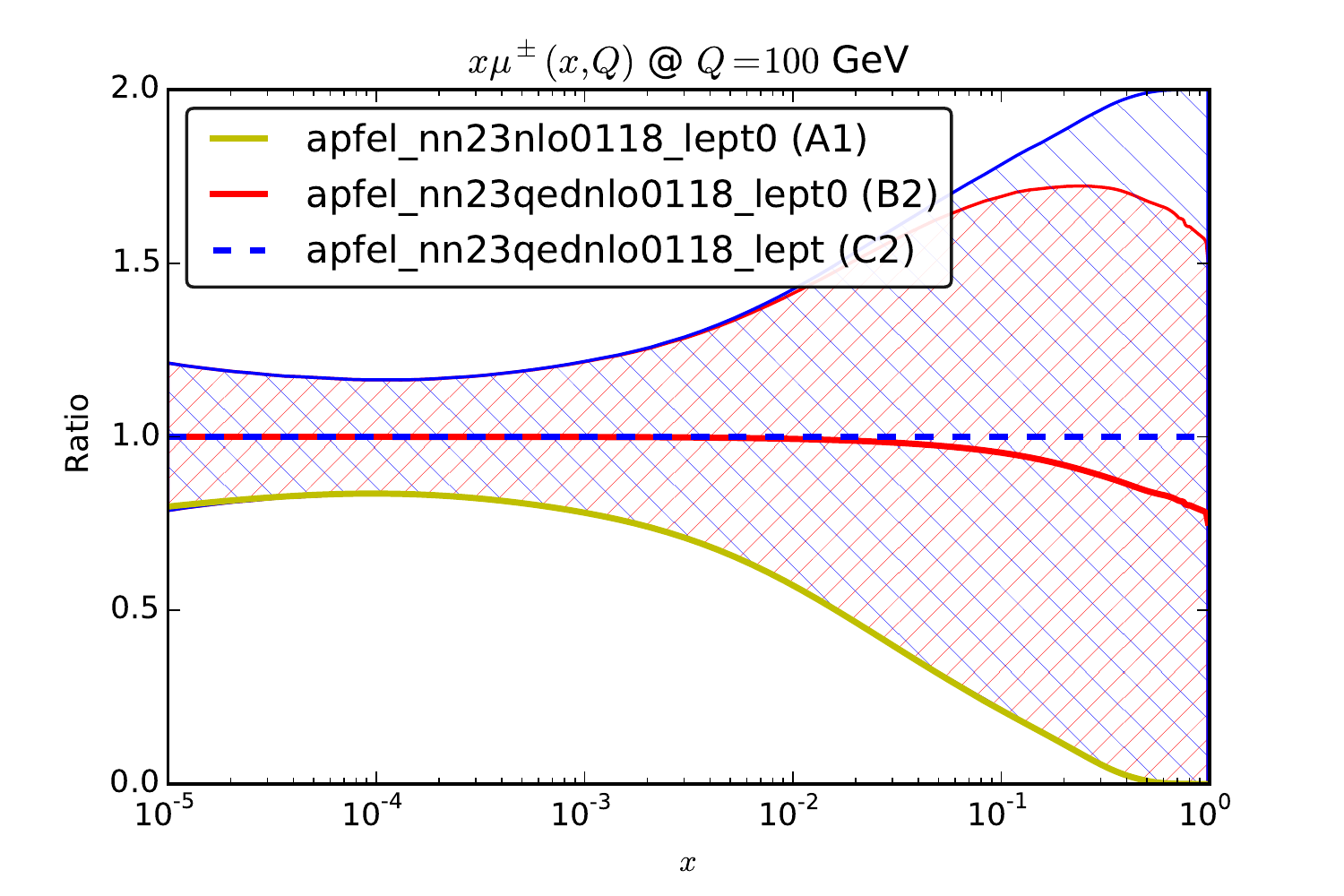}
  \caption{Same as Fig.~\ref{fig:leptratios} but with 68\% CL bands displayed.}\label{fig:leptratioserror}
\end{figure}

As already mentioned, the photon PDF of the NNPDF2.3QED
  sets is provided with an uncertainty. This enables us to assess how
  compatible the lepton PDFs emerging from the three different
  configurations described above are. In
  Fig.~\ref{fig:leptratioserror} we show the same plots presented in
  Fig.~\ref{fig:leptratios} but including the 68\% confidence level
  (CL) bands (when available). We observe that the configurations B2 and
  C2 are perfectly compatible within uncertainties all over the range
  in $x$. We can thus conclude that, even though the ansatz in
  Eq.~(\ref{eq:ansatz}) provides an arguably better estimate of the
  lepton PDFs, the present precision with which the photon PDF is
  determined is such that the difference with respect
  to the ``zero-lepton'' ansatz in Eq.~(\ref{eq:ansatz0}) is still within
  uncertainties. Finally, it is unsurprising that the configuration
  A1, that has no uncertainty band because it assumes that the photon
  PDF is identically zero at the initial scale for all PDF replicas,
  lies right on the lower limit of the uncertainty band of both
  configurations B2 and C2. The reason for this can be traced back to
  the fact the the photon PDF of the NNPDF2.3QED sets at the initial
  scale is compatible with zero and, due to the enforcement of
  positivity, the lower limit of the 68\% CL band essentially concides
  with $\gamma(x,Q_0)=0$.

\begin{figure}
  \centering
  \includegraphics[scale=0.5]{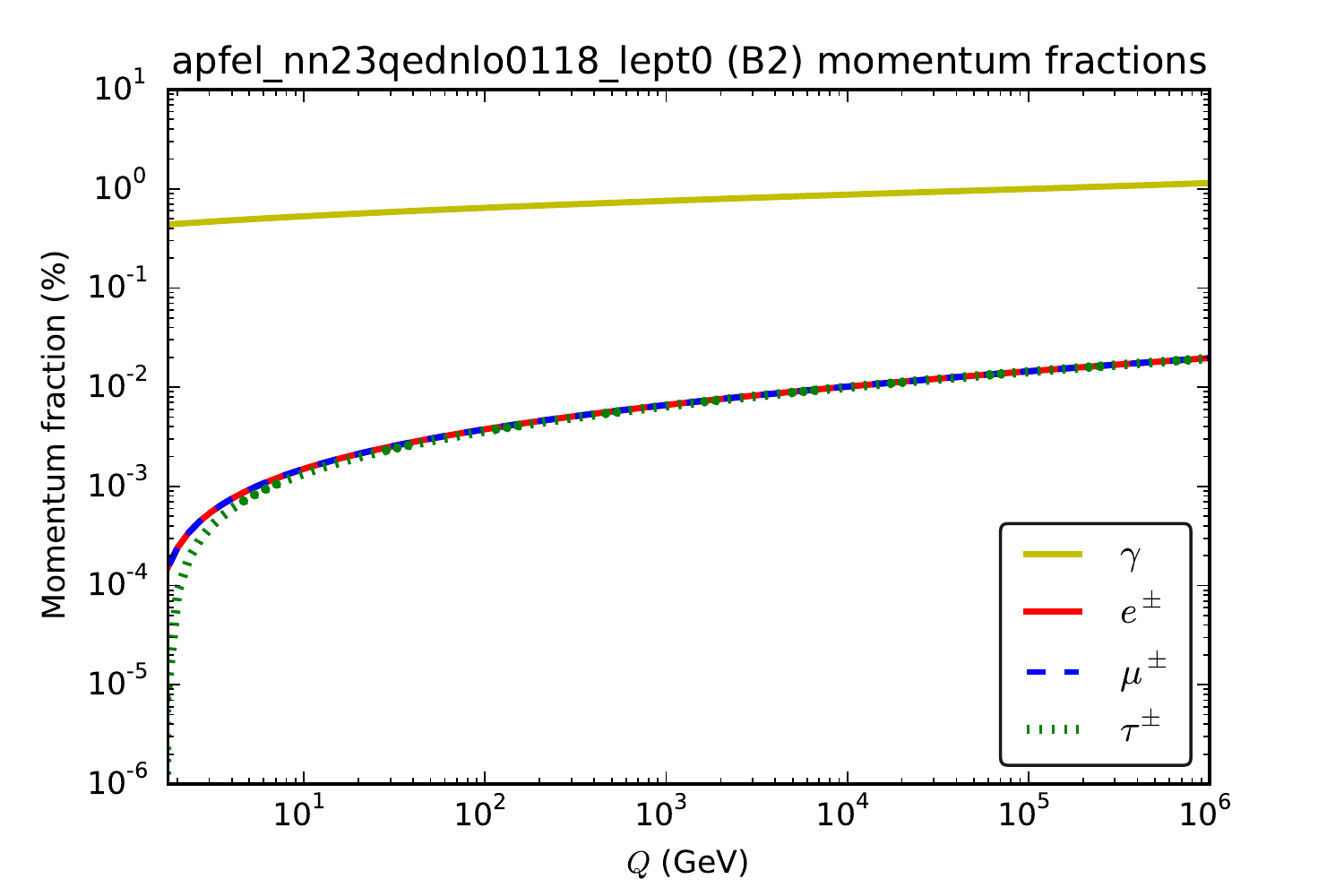}\includegraphics[scale=0.5]{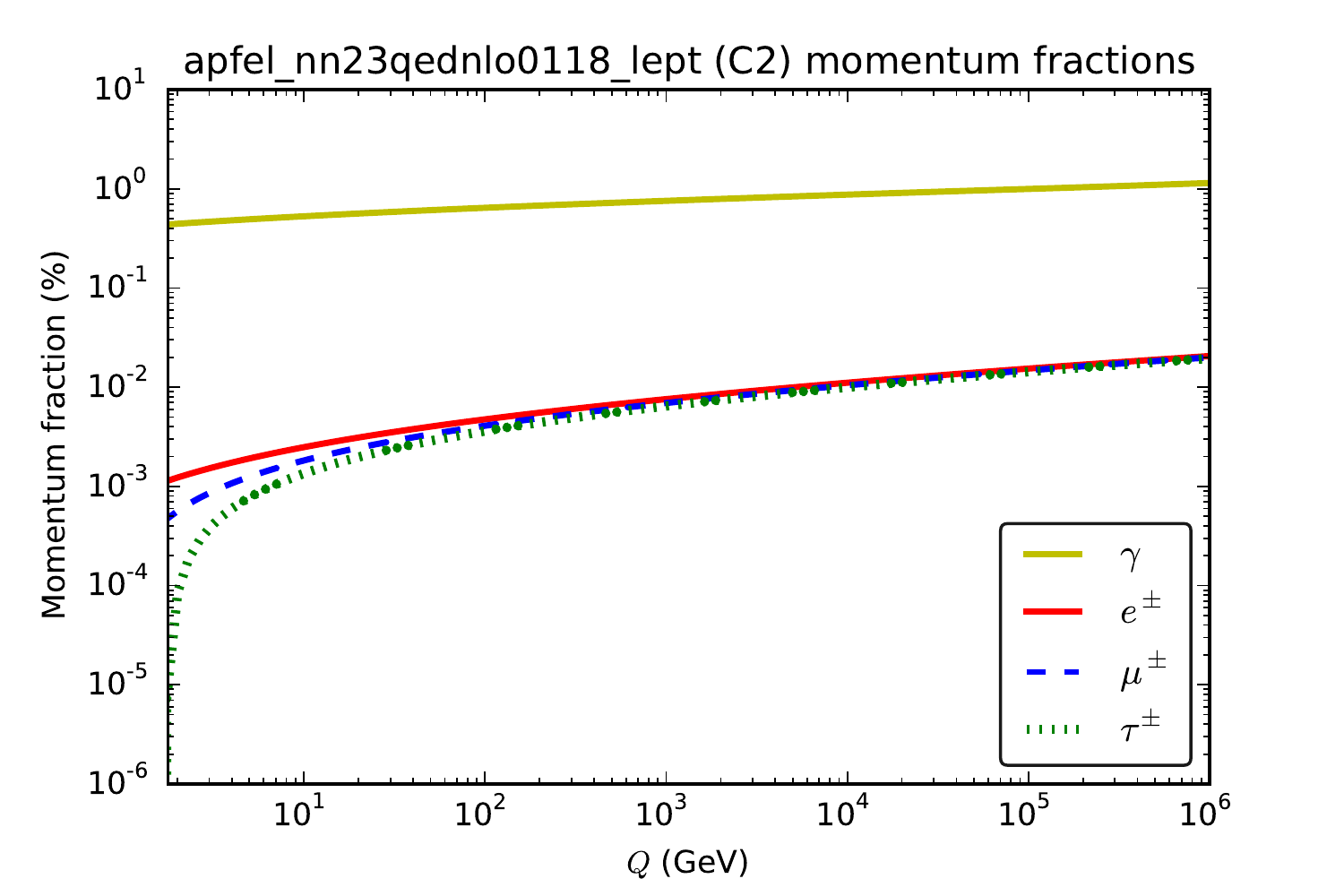}
  \caption{Momentum fractions for the photon and lepton
    PDFs.\label{fig:msrlept}}
  \vspace{5mm}
  \includegraphics[scale=0.5]{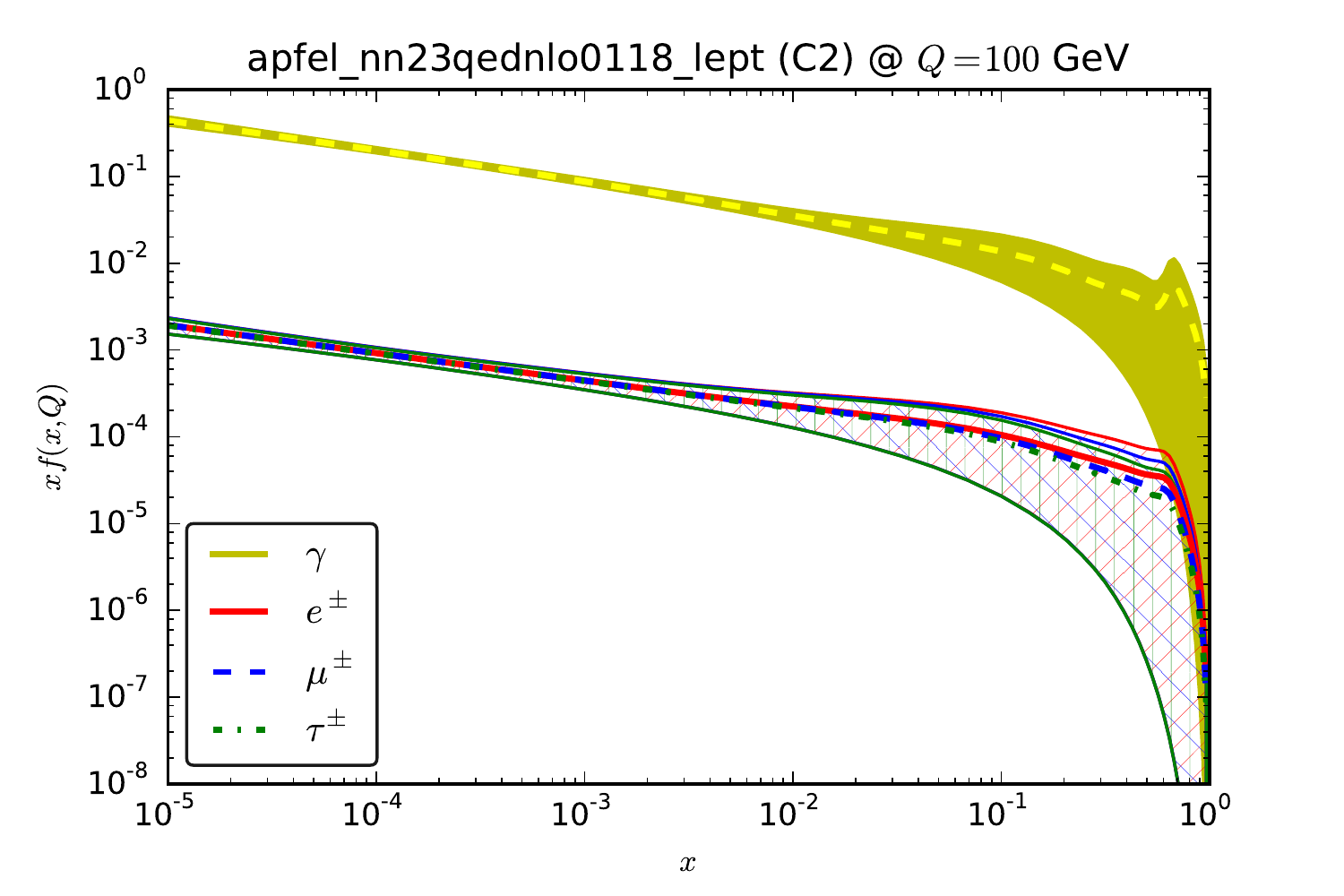}\includegraphics[scale=0.5]{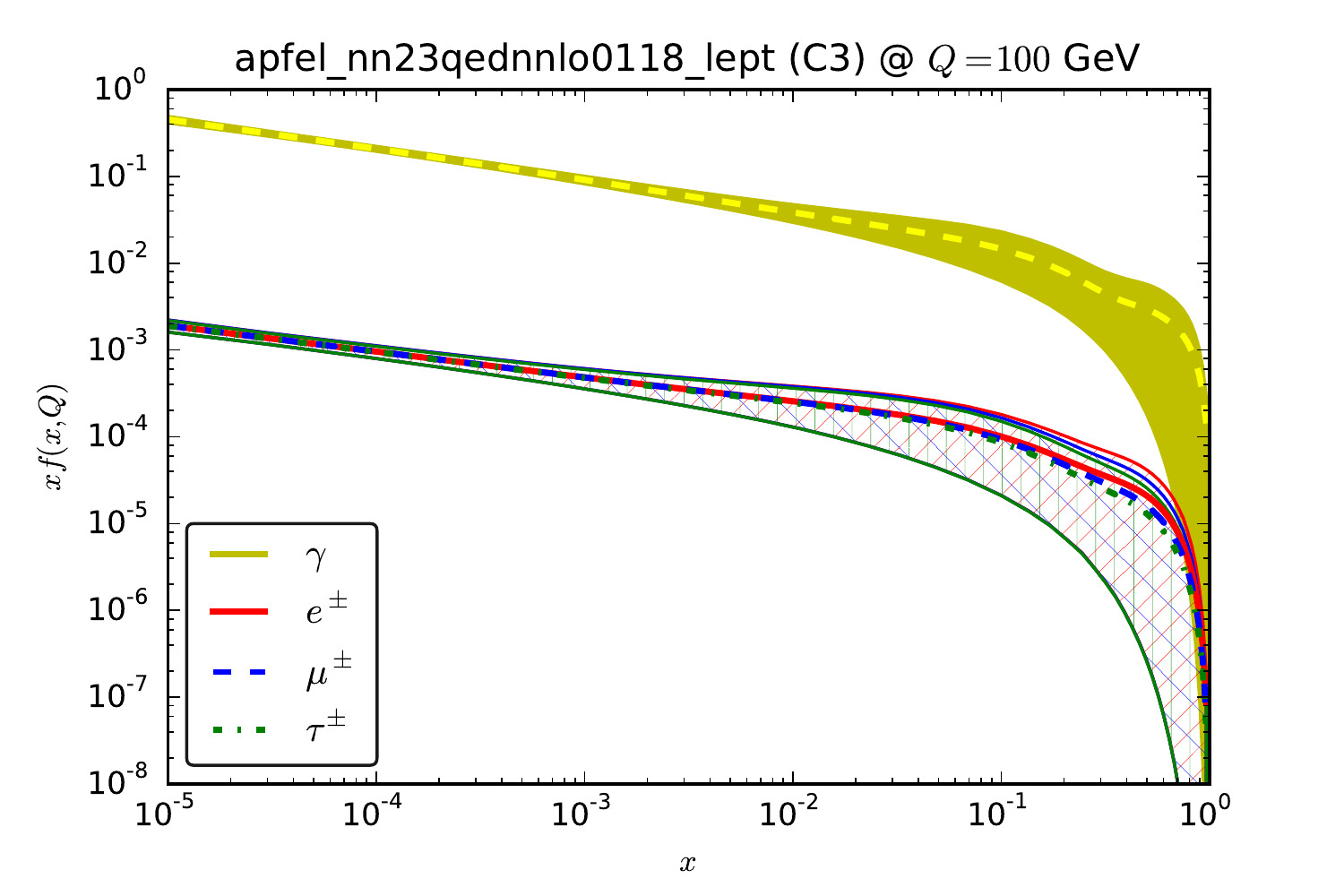}
  \caption{Uncertainty bands for the photon and the lepton PDFs of
    configurations C2 (left) and C3 (right) at $Q=100$
    GeV.\label{fig:leptuncer}}
\end{figure}

Interesting information about the photon and lepton content of the
proton is provided by the respective momentum fractions defined as:
\begin{equation}
  \textrm{MF}_\gamma(Q) = \int_0^1 dx \, x\gamma(x,Q)\,,\quad\mbox{and}\quad
  \textrm{MF}_{\ell^{\pm}}(Q) =  \int_0^1 dx \, x\ell^{\pm}(x,Q)\,.
\end{equation}
In Fig.~\ref{fig:msrlept} we plot the percent momentum fractions as a
function of the scale $Q$ for the configurations B2 (left) and C2
(right). While the photon PDF carries up to around 1\% of the momentum
fraction of the proton, lepton PDFs carry a much smaller fraction (by
about two orders of magnitude), regardless of the initial-scale
parameterisation. This is consistent with the fact that, for both
parameterisations, lepton PDFs are proportional to $\alpha$ times the
photon PDF ($\ell \propto \alpha \times \gamma$). In conclusion,
lepton PDFs carry such a small fraction of the proton momentum that
they do not cause a significant violation of the total momentum sum
rule when attached to a pre-existing PDF set.

Now we turn to consider the uncertainties of the lepton PDFs. A
realistic estimate of the lepton PDFs also requires an estimate of the
respective uncertainties. To this end, using exactly the same
procedure discussed in the previous sections, we have generated lepton
PDFs for all replicas of the NNPDF2.3 family sets. This allows one to
estimate the uncertainty on each lepton PDF in the usual way.  In
Fig.~\ref{fig:leptuncer} we plot the lepton PDFs with the respective
uncertainty for the configurations C2 and C3 at $Q=100$
GeV. Uncertainties are presented as symmetric 68\% CL
bands centred around the mean value of each PDF. As expected, the
lepton PDF uncertainties follow the pattern of the photon PDF,
characterised by a large uncertainty at large $x$.

\begin{figure}
  \centering
  \includegraphics[scale=0.8]{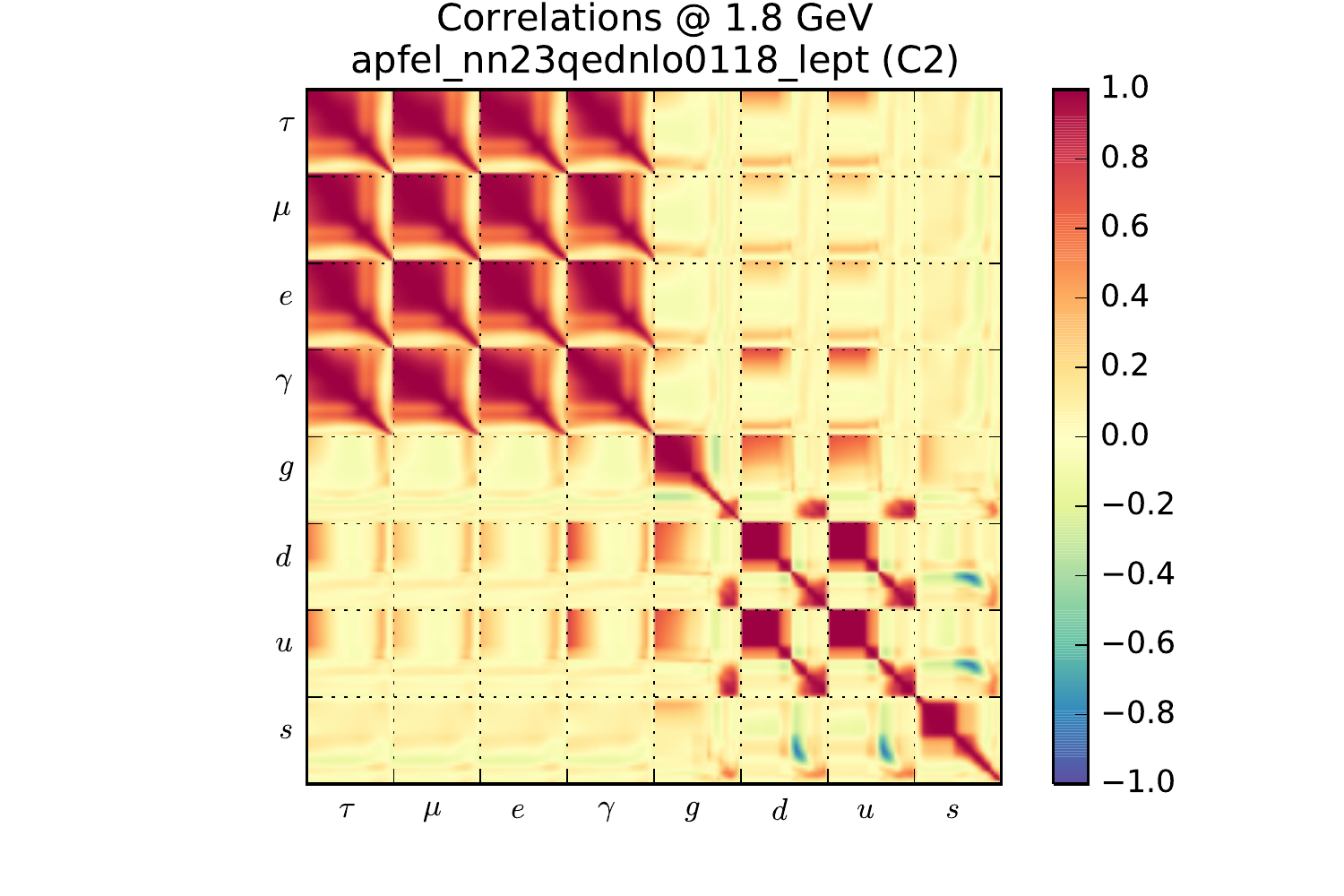}
  \caption{PDF correlation coefficients at $Q=1.8$ GeV for the
    configuration C2.\label{fig:pdfcorrelation}}
\end{figure}

To conclude this section, in Fig.~\ref{fig:pdfcorrelation} we present
the PDF correlation matrix for the configuration C3 at $Q=1.8$ GeV on
a grid of $N_x = 50$ points logarithmically distributed in the
interval $x_1,x_2=[10^{-5},0.9]$, for the partons $\tau$, $\mu$, $e$,
$\gamma$, $g$, $d$, $u$ and $s$. The correlation coefficients are
defined as:
\begin{equation} \rho_{\alpha\beta}(x_1,x_2,Q) = \frac{N_{\rm
      rep}}{N_{\rm rep}-1} \left( \frac{\left\langle
        f_\alpha(x_1,Q)f_\beta(x_2,Q)\right\rangle_{\rm rep}- \left\langle
        f_\alpha(x_1,Q)\right\rangle_{\rm rep}\left\langle
        f_\beta(x_2,Q)\right\rangle_{\rm rep}}{\sigma_\alpha(x_1,Q) \cdot
      \sigma_\beta(x_2,Q) } \right) \, ,
  \label{eq:correlation}
\end{equation}
where averages are taken over the $N_{\rm rep}$ replicas and where
$\sigma_i(x,Q)$ are the corresponding standard deviations.

We note a clear distinction between the QED (top-left square region)
and the QCD sector (bottom-right region). As expected, there are
strong correlations between $\tau$, $\mu$, $e$ and $\gamma$ due to the
fact that leptons are generated by photon splitting. A similar but
milder behaviour is also observed for $g$, $d$, $u$ and $s$. Finally,
the off-diagonal elements show that quark and gluon distributions are
instead very mildly correlated to lepton and photon PDFs.  Similar
results are obtained also for the other configurations.

\section{Phenomenological impact} \label{sec:pheno}

\subsection{Parton luminosities}\label{sec:lumi}

In the computation of the cross sections of hadron-collider processes,
PDF contributions factorise in the form of \textit{parton
  luminosities}.  Thus, before looking at specific processes, it is
useful to study the behaviour of the parton luminosities of the
different initial states, by including also leptons.

Parton luminosities are doubly differential quantities defined as:
\begin{equation}\label{eq:DoubleDiffLumi}
  \frac{d^2\mathcal{L}_{ij}}{dy d\tau} = f_{i}\left(x_1,Q\right) f_{j}\left(x_2,Q\right)\,,\qquad x_1\equiv\sqrt{\tau}e^{y}\, , \qquad  x_2 \equiv \sqrt{\tau}e^{-y}\, ,\qquad \tau \equiv M_X^2/s\, ,
\end{equation}
where $s$ is the squared center-of-mass energy of the hadronic
collision, and $M_X$ and $y$ are the invariant mass and the rapidity
of the partonic initial/final state, respectively.  In
Eq.~\eqref{eq:DoubleDiffLumi}, $f_{i}(x,Q)$ is the PDF of the $i$-th
parton evaluated at the factorisation scale $Q$. Different choices for
$Q$ can be adopted in order to improve predictions of a particular
process and/or distribution. At the level of pure luminosities,
without the convolution with any specific matrix element, the
factorisation scale can be naturally set to $Q=M_X$.
\begin{figure}
  \centering
  \includegraphics[scale=0.5]{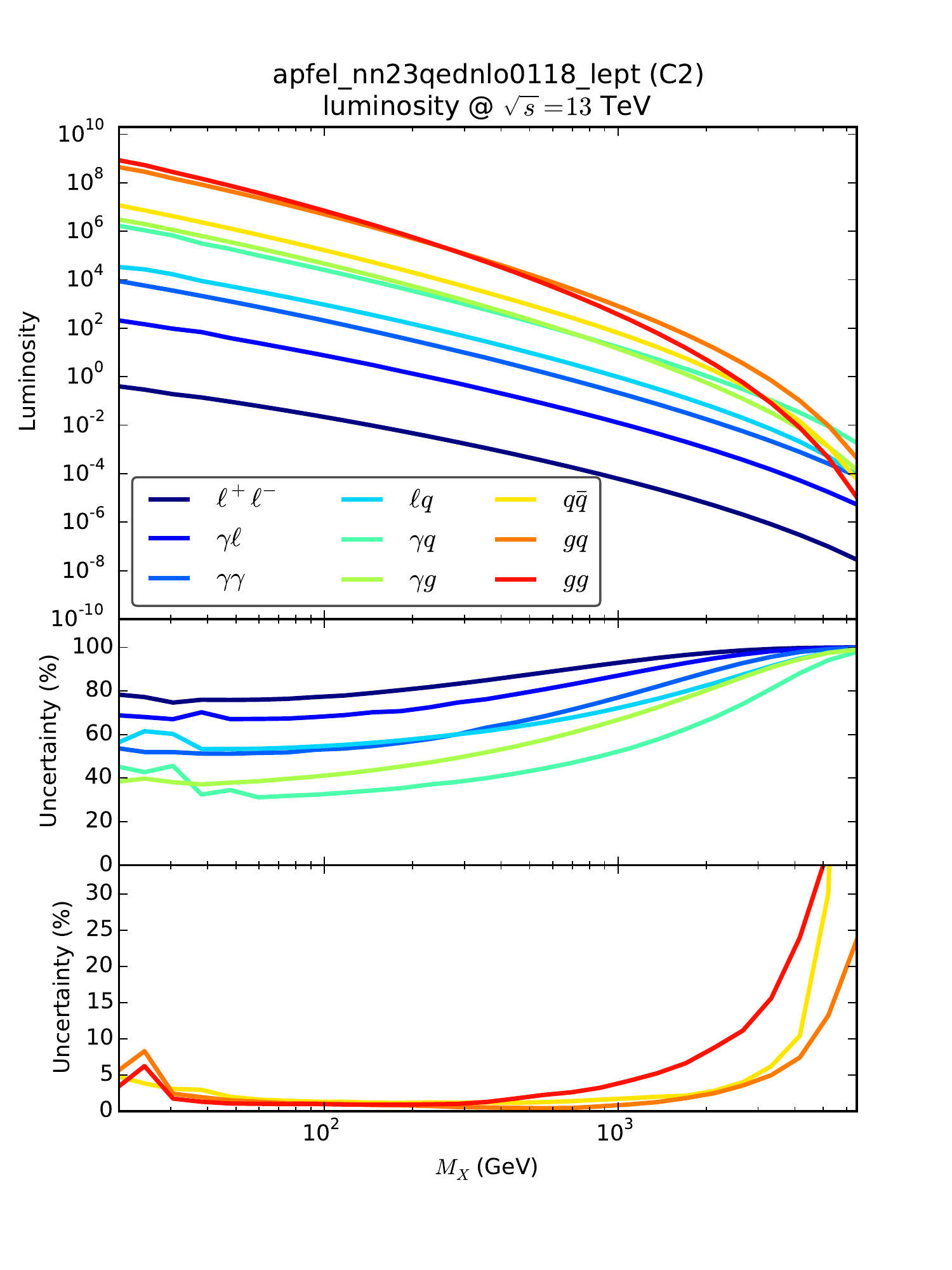}\includegraphics[scale=0.5]{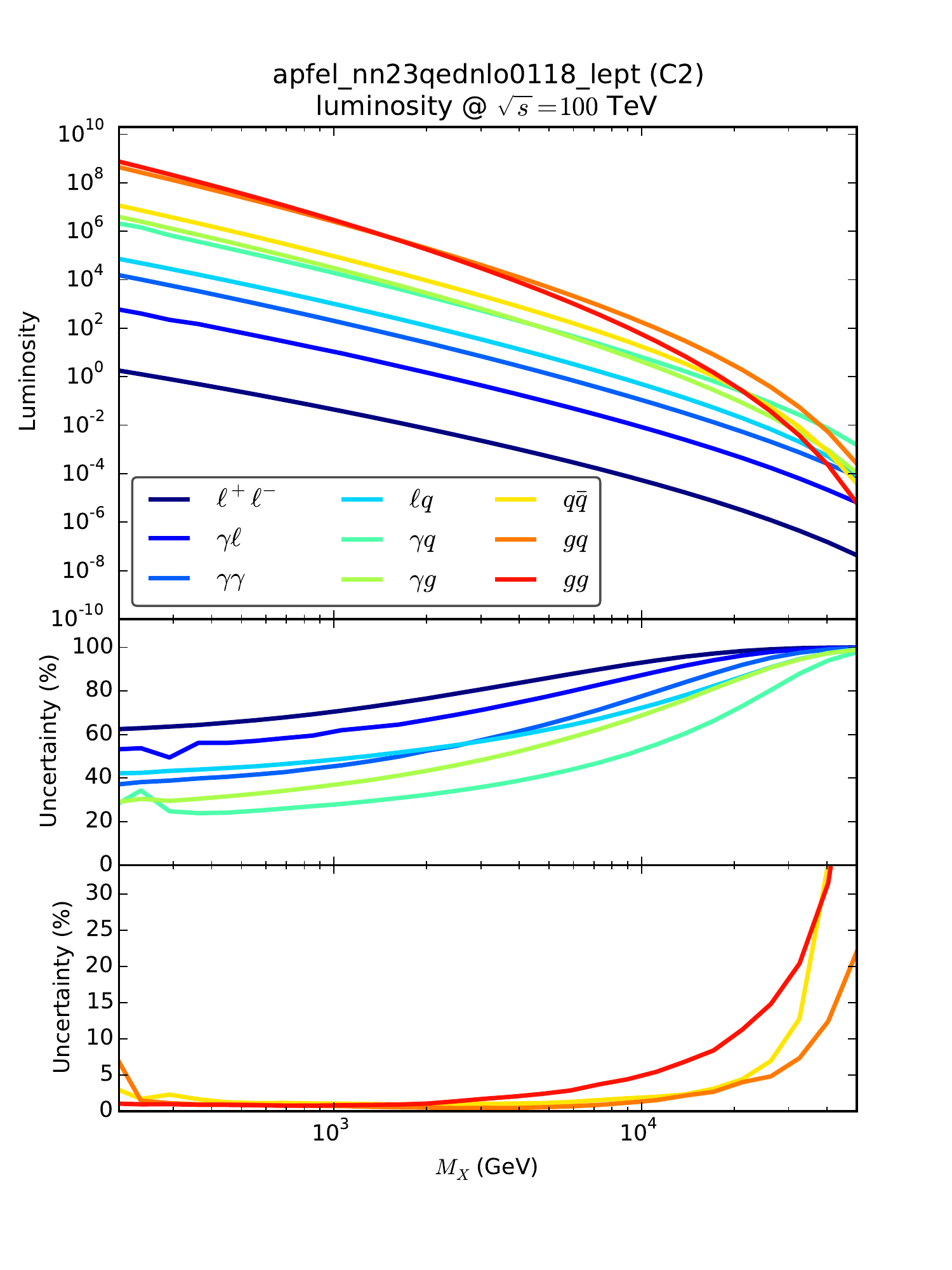}
  \caption{$M_X$-differential parton luminosities involving also
    photon and lepton PDFs.}
  \label{fig:lumileptM}
\end{figure}
\begin{figure}
  \centering
  \includegraphics[scale=0.5]{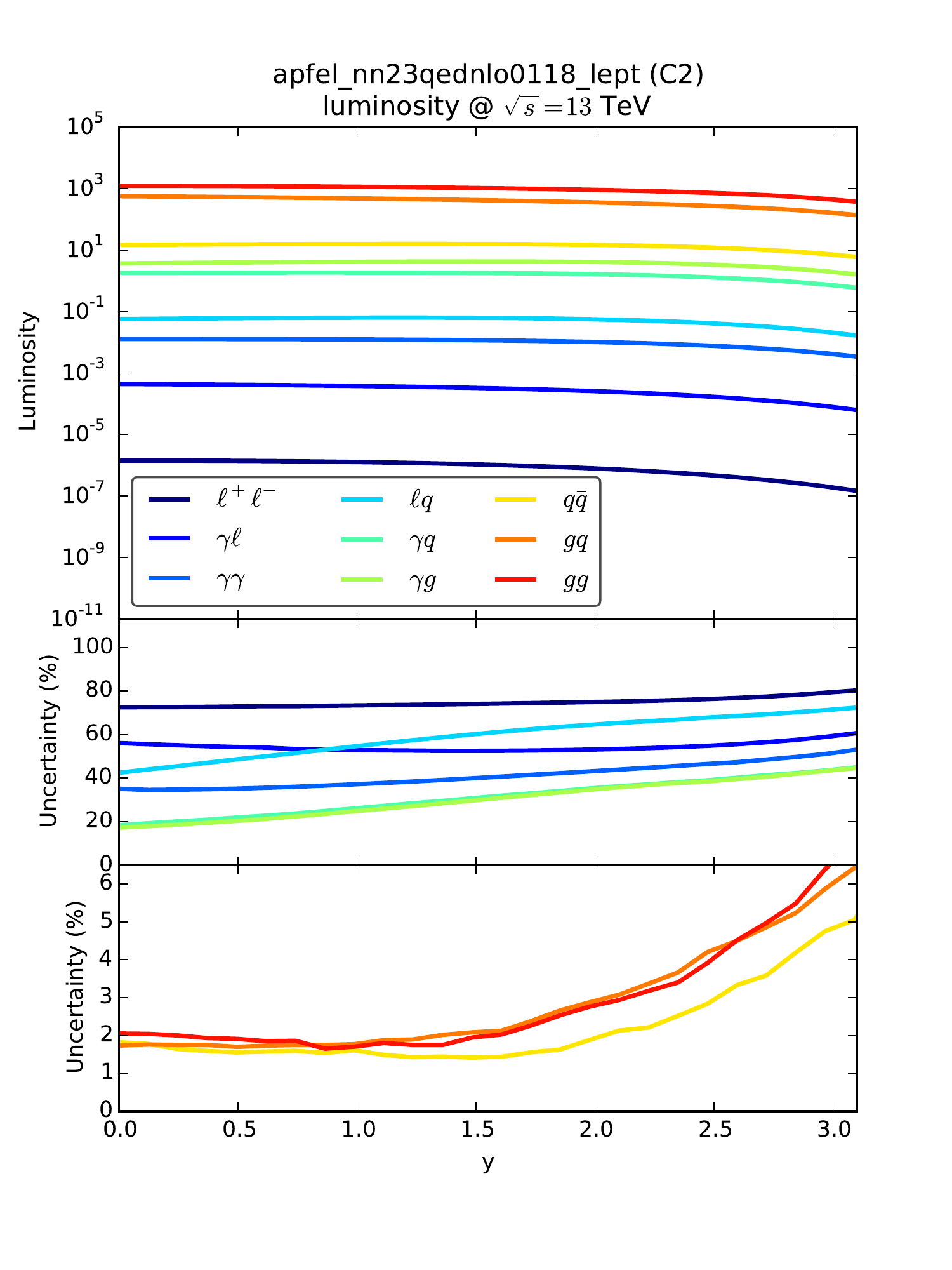}\includegraphics[scale=0.5]{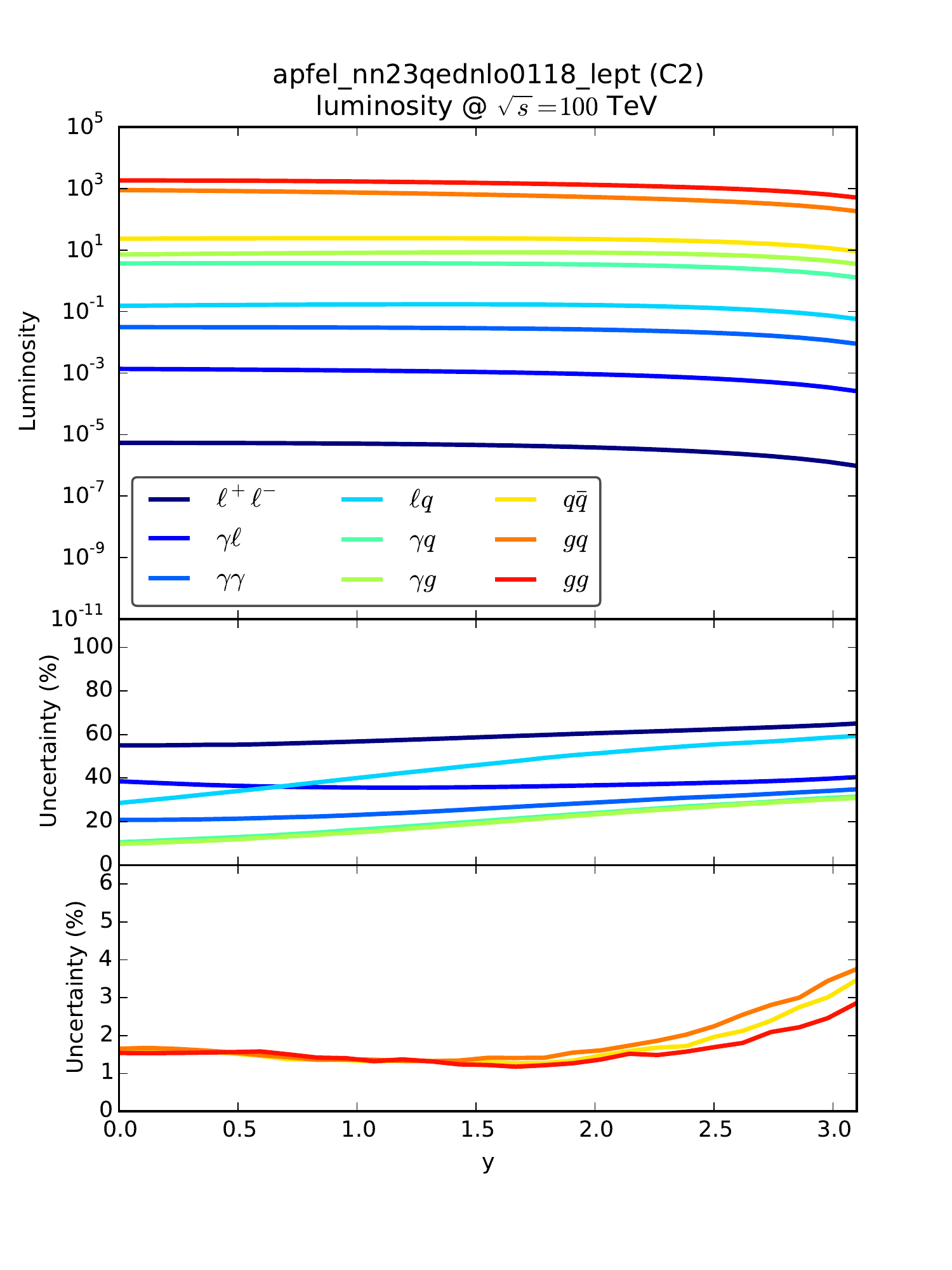}
  \caption{$y$-differential parton luminosities involving also photon
    and lepton PDFs.}
  \label{fig:lumilepty}
\end{figure}

For the numerical evaluation of the different luminosities, it is
practical to integrate out one of the kinematic variables in
Eq.~(\ref{eq:DoubleDiffLumi}), namely, either $\tau$ or $y$.  In
particular, by integrating out $y$, one obtains the $M_X$-differential
luminosities, which read:
\begin{equation}\label{eq:lumdef}
  \Phi_{ij}\left( M_X\right) \equiv
  \frac{d\mathcal{L}_{ij}}{d M_X^2}=\frac{1}{s}\int_{\tau}^1
  \frac{dx}{x} f_i\left(x,M_X\right) f_j\left( \tau/x,M_X\right)\,.
\end{equation}
By integrating out $\tau$, instead, the $y$-differential luminosities
are given by:
\begin{equation}\label{eq:lumdefy}
  \Psi_{ij}(y) \equiv \frac{d\mathcal{L}_{ij}}{dy}  =
  2e^{-2y}\int_{\sqrt{\tau_{\rm cut}}e^{y}}^{e^{-y}} dx\,x
  f_{i}(x,\sqrt{s}xe^{-y}) f_{j}(xe^{-2y},\sqrt{s}xe^{-y})\,,
\end{equation}
with $\tau_{\rm cut} \equiv M_{X,\rm cut}^2/s$. In
Eq.~(\ref{eq:lumdefy}) the lower bound of the integral, proportional
to $\sqrt{\tau_{\rm cut}}$, implies that $M_{X}\geq M_{X,\rm cut}$.
Although artificial, this kind of cut is always present in
phenomenological predictions. Indeed, in the computation of hadronic
cross sections a cut on $M_{X}$ is automatically introduced by the
masses of the particles in the final state or, indirectly, by the cuts
on the massless particles, which are necessary to avoid infrared
and collinear divergencies and ensure the finiteness of the cross sections.

In order to compare the size and the $M_X$-dependence of the different
parton luminosities, in Fig.~\ref{fig:lumileptM} we plot
$\Phi_{\gamma\gamma}$, $\Phi_{\gamma\ell}$, $\Phi_{\ell ^+ \ell ^-}$,
$\Phi_{\ell q}$, $\Phi_{\gamma q}$, $\Phi_{\gamma g}$,
$\Phi_{q\bar{q}}$, $\Phi_{gq}$ and $\Phi_{gg}$ as functions of $M_X$
at $\sqrt{s}=13$ TeV (left) and $\sqrt{s}=100$ TeV (right) for the set
C2.  The quantities $\Phi_{i \ell}$, $\Phi_{\ell ^+ \ell ^-}$,
$\Phi_{i q}$ and $\Phi_{q\bar{q}}$ are defined as:
\begin{eqnarray}
  \Phi_{i \ell}\left(M_X\right) \equiv
  \sum_{j=e^\pm,\mu^\pm,\tau^\pm}\Phi_{i j}\left(
    M_X\right)\, , \qquad  \Phi_{\ell ^+ \ell ^-}\left(M_X\right) \equiv
  \sum_{j=e,\mu,\tau}\Phi_{j^+ j^-}\left( M_X\right)\,, \label{eq:multilum1} \\
    \label{eq:multilum2}
   \Phi_{i q}\left(M_X\right) \equiv
  \sum_{j= u,\bar{u},\ldots,b,\bar{b} }\Phi_{i j}\left(
    M_X\right)\, , \qquad  \Phi_{q\bar{q}}\left(M_X\right) \equiv
  \sum_{j=u,d,c,s,b}\Phi_{j \bar{j}}\left( M_X\right)\,. 
\end{eqnarray}
Similarly, in Fig.~\ref{fig:lumilepty} we plot the
$\Psi_{\gamma\gamma}$, $\Psi_{\gamma\ell}$, $\Psi_{\ell ^+ \ell ^-}$,
$\Psi_{\ell q}$, $\Psi_{\gamma q}$, $\Psi_{\gamma g}$,
$\Psi_{q\bar{q}}$, $\Psi_{gq}$ and $\Psi_{gg}$ parton luminosities as
functions of $y$ at the same energies and for the same PDF set. In the
case of $\sqrt{s}=13$ TeV (left), we set $M_{X,\rm cut} = 10$ GeV and
for $\sqrt{s}=100$ TeV (right) we set $M_{X,\rm cut} = 100$ GeV. The
$\Psi_{i \ell}$, $\Psi_{\ell ^+ \ell ^-}$, $\Psi_{i q}$ and
$\Psi_{q\bar{q}}$ luminosities are defined in full analogy with
Eqs.~\eqref{eq:multilum1} and \eqref{eq:multilum2}. In
Figs.~\ref{fig:lumileptM} and \ref{fig:lumilepty} we also plot with
consistent colours the 68\% uncertainty of the luminosities involving
photon or lepton PDFs (lower insets) and of the luminosities involving
only (anti)quarks or gluons (central insets).

The relative size of the plotted luminosities follows the expected
pattern. In general, the photon PDF suppresses the luminosity by a
factor of $\alpha$ w.r.t. the (anti)quark PDFs and, analogously, the
lepton PDFs suppress the luminosity by an additional factor of
$\alpha$ w.r.t. the photon PDF.  This can be easily seen in
Fig.~\ref{fig:lumileptM}(\ref{fig:lumilepty}), {\it e.g.} by comparing
$\Phi_{\gamma\ell}$($\Psi_{\gamma\ell}$) with
$\Phi_{\gamma\gamma}$($\Psi_{\gamma\gamma}$) and
$\Phi_{\ell ^+ \ell ^-}$($\Psi_{\ell ^+ \ell ^-}$), the three lowest
curves. However, from Fig.~\ref{fig:lumileptM} we also notice that
this hierarchy is not clearly respected at large invariant masses. In
this kinematic region the QCD luminosity combinations,
$\Phi_{q\bar{q}}$, $\Phi_{gq}$ and $\Phi_{gg}$, get closer to the
luminosities involving photons and leptons, suggesting possibly
non-negligible phenomenological implications due to lepton and photon
channels.  The major part of this effect is caused by the
  relative behaviour of the strong coupling $\alpha_s$ with respect to
  the QED coupling $\alpha$ as functions of the scale $M_X$. As is
  well known, DGLAP evolution leads to an increase of PDFs in the
  small-$x$ region and to a decrease in the large-$x$ region as the
  evolution scale increases. The magnitude of
  such a behavior is driven by the rate of change of the coupling, no
  matter whether it is negative, as for $\alpha_s$, or positive, as
  for $\alpha$. Given that
  the absolute value of the QCD $\beta$-function is larger than the
  QED one up to very large scales, $\alpha_s$-driven PDFs, like quark
  and gluon PDFs, are relatively more suppressed at larger values of
  $x$ as compared to $\alpha$-driven PDFs, like lepton and photon
  PDFs. It is clear from Eq.~(\ref{eq:lumdef}) that the behavior of
  the $M_X$-differential luminosities at large values of $M_X$
  reflects the behavior of PDFs at large values of $x$ and thus,
  following the argument given above, QCD luminosities are more
  suppressed than QED luminosities in this region and eventually they
  get overcome. On the other hand, it should be noticed that in this
  region the  PDF uncertainty for the
  QED-induced channels is almost as large as the luminosity central value, as shown in the second inset of the plots in Fig.~\ref{fig:lumileptM}. Thus, by taking into account PDF uncertainties, QED luminosities turn out to be compatible at high $M_X$ with QCD luminosities. This feature again points to the fact that meaningful comparisons can be pursued only including PDF errors, especially when those are large, which is in general true for photon and lepton PDFs.
Finally, contrary to the
$\Phi_{ij}$ luminosities, the $\Psi_{ij}$ luminosities maintain the
same hierarchy all over the range in $y$. Indeed, the value of $y$ is
not directly related to the value of $M_X$, which also in this case is
used as factorization scale. Thus, the entire previous argument on the
suppression of the QCD luminosities with respect to the QED ones does not apply for the case of $\Psi_{ij}$.

In Fig.~\ref{fig:lumileptM} the value of $M_X$ on the $x$-axis have
been chosen in such a way that similar values of $\tau$ are spanned in
the plots for $\sqrt{s}=13$ TeV and $\sqrt{s}=100$ TeV. For this
reason the curves look very similar in the two plots. The only
differences are the size of the 68\% uncertainties, which are larger
in the 13 TeV case. This effect is due to the different values of the
factorisation scale ($M_X$) for the same values of $\tau$, leading to
smaller uncertainties at 100 TeV, where $M_X$ is larger.  As expected,
PDF uncertainties are much smaller for the QCD initial states (lower
inset) than for those involving photons and leptons (central
inset).

Similar arguments hold for the comparison of the plot at $\sqrt{s}=13$
TeV and $\sqrt{s}=100$ in Fig.~\ref{fig:lumilepty}. Also in this case the
values of $M_{X,{\rm cut}}$, and consequently of $\tau_{\rm cut}$,
have been chosen in such a way that similar values of $\tau$ are
integrated out in the two cases. With higher values of
$M_{X,{\rm cut}}$ the difference among the luminosities and even their
hierarchy can change, consistently with what is shown in plots of
Fig.~\ref{fig:lumileptM}. However, even for large values of
$M_{X,{\rm cut}}$, partonic luminosities depend very mildly on $y$.

\subsection{Production processes} \label{sec:processes}
\begin{figure}
  \centering
  \includegraphics[scale=0.4]{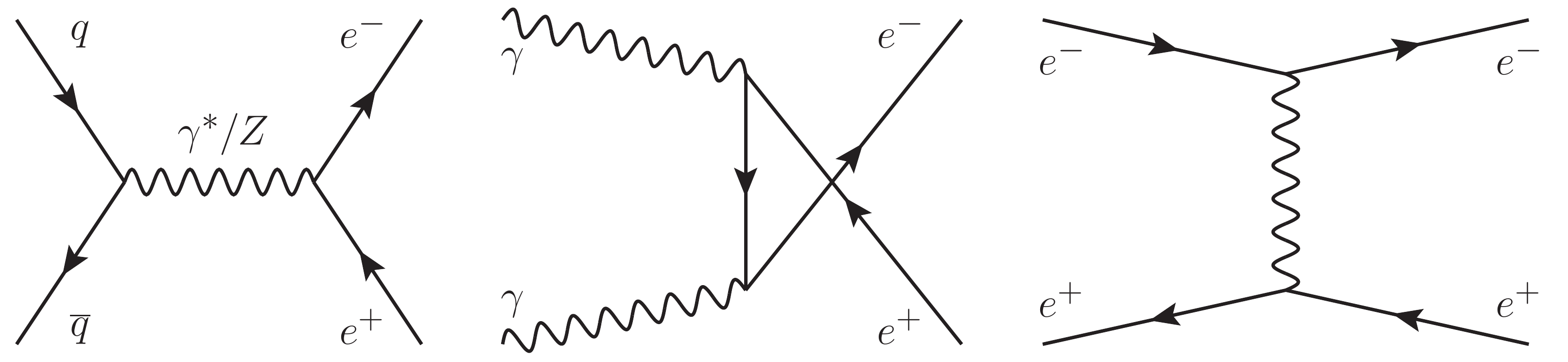}
  \caption{Representative diagrams for partonic processes in $e^+ e^-$
    production.}
  \label{fig:diagrams}
\end{figure}
We now turn to study the effect of lepton PDFs on some key processes
at the 13 TeV LHC and at the 100 TeV FCC-hh. The analysis of the
parton luminosities presented in Section~\ref{sec:lumi} points to a
dependence of the relative size of different parton luminosities on
the invariant mass of the final-state particles. For this reason, we
concentrate on differential distributions w.r.t. this kinematic
variable.

All the results have been produced with the help of {\aNLO}
\cite{Alwall:2014hca} using the PDF set C2 in Tab.~\ref{tab:sets},
{\it i.e.} {\tt apfel\_nn23qednlo0118\_lept}. The relevant SM input
parameters have been set to the following values:
\begin{eqnarray}
&\alpha_s(m_{Z})=0.118\, ,\quad &G_{F}=1.16639\times 10^{-5}\,, \quad m_{Z}=91.1876~\gev\, , \quad m_{W}=80.385~\gev\, , \nonumber\\
& m_{H}=125~\gev\,, \quad &\Gamma_{Z}=2.4952~\gev\, , \quad \Gamma_{W}=2.085~\gev\, . \label{eq:inputs}
\end{eqnarray} 
Since all leptons and quarks, besides the top quark, are considered as
partons, they are all treated as massless.  As a general approach, for
all processes we set renormalisation and factorisation scales as
$\mu_{F}=\mu_{R}=H_{T}/2$, where $H_{T}$ is the scalar sum of the
transverse masses $m_T(i)$ of the final-state particles, defined as:
\begin{equation}
  m_T(i) = \sqrt{m(i)^2 + p_T(i)^2}\,.
\end{equation}
For the sake of simplicity, all simulations are done at LO and parton
level. However, we include all the contributions induced by tree-level
diagrams, {\it i.e.} not only those featuring the largest power of
$\alpha_s$ but also the subleading contributions. Examples will be
given for the specific processes.

We remark that here we focus on the impact of the lepton PDFs at the
LHC and at the FCC-hh.  In order to perform a trustworthy similar
study for the case of the photon PDF, NLO EW corrections cannot be
neglected. Notable examples can be found in the context of the SM,
{\it e.g.} in Ref.~\cite{Boughezal:2013cwa} for the neutral current
Drell-Yan or in Ref.~\cite{Baglio:2013toa} for $WW$ production, but
also in the context of BSM in Ref.~\cite{Hollik:2015lha} for
squark-antisquark production. Nevertheless, in the following, in order
to assess the impact of lepton and photon PDFs, we will present
results also for photon-induced processes. In general, we will keep
separated the contributions from initial states with only (anti)quarks
and gluons, initial states with at least one photon and no leptons,
and initial states with at least one lepton. In all the plots of this
section these three categories are easily identifiable by their
distinctive colours, which are red, green and blue respectively.

\subsubsection*{Neutral Drell-Yan}

We start by considering the case of the production of an
electron-positron pair at the LHC and at the FCC-hh. Besides the
quark-antiquark and photon-photon initial states, this process
receives, for instance, a new contribution from the electron-positron
initial state, as can be seen in Fig.~\ref{fig:diagrams}. Similarly,
also $\mu^+ \mu^-$ and $\tau^+ \tau^-$ initial states can
contribute. Fig.~\ref{fig:diagrams} suggests that different
kinematical distributions can arise from different partonic channels
such as $q\bar{q}$, with only an $s$-channel diagram, $\gamma\gamma$,
with $t$-and $u$-channel diagrams, and $e^+e^-$ with $s$- and
$t$-channel diagrams. It is important to note that all these partonic
processes yield LO cross sections of $\ord(\alpha^2)$, thus they all
contribute to the same perturbative order.
\begin{figure}
  \centering
  \includegraphics[clip=true, trim=0.cm 3.5cm 0.7cm 1cm, width=0.49\textwidth]{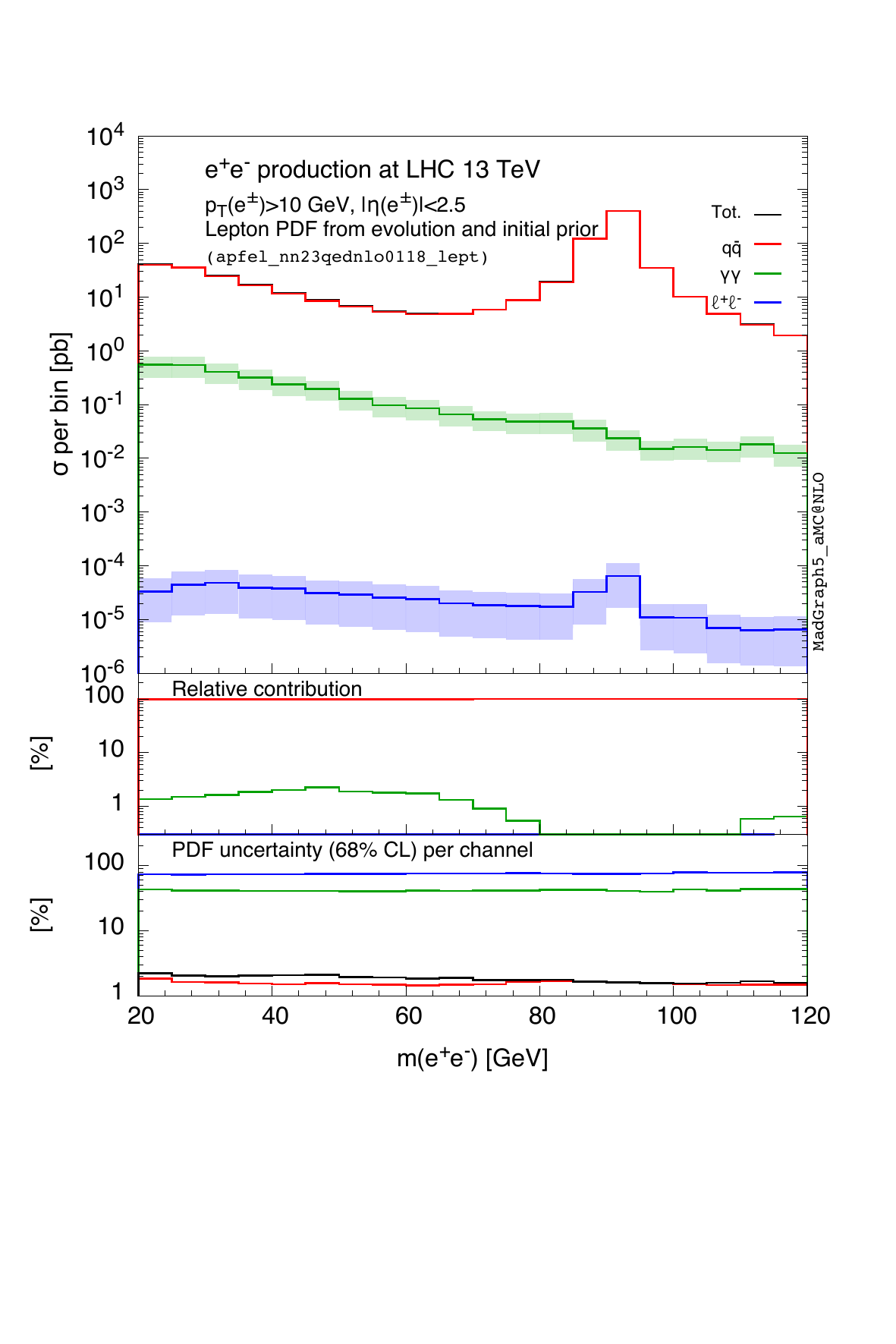}
  \includegraphics[clip=true, trim=0.cm 3.5cm 0.7cm 1cm, width=0.49\textwidth]{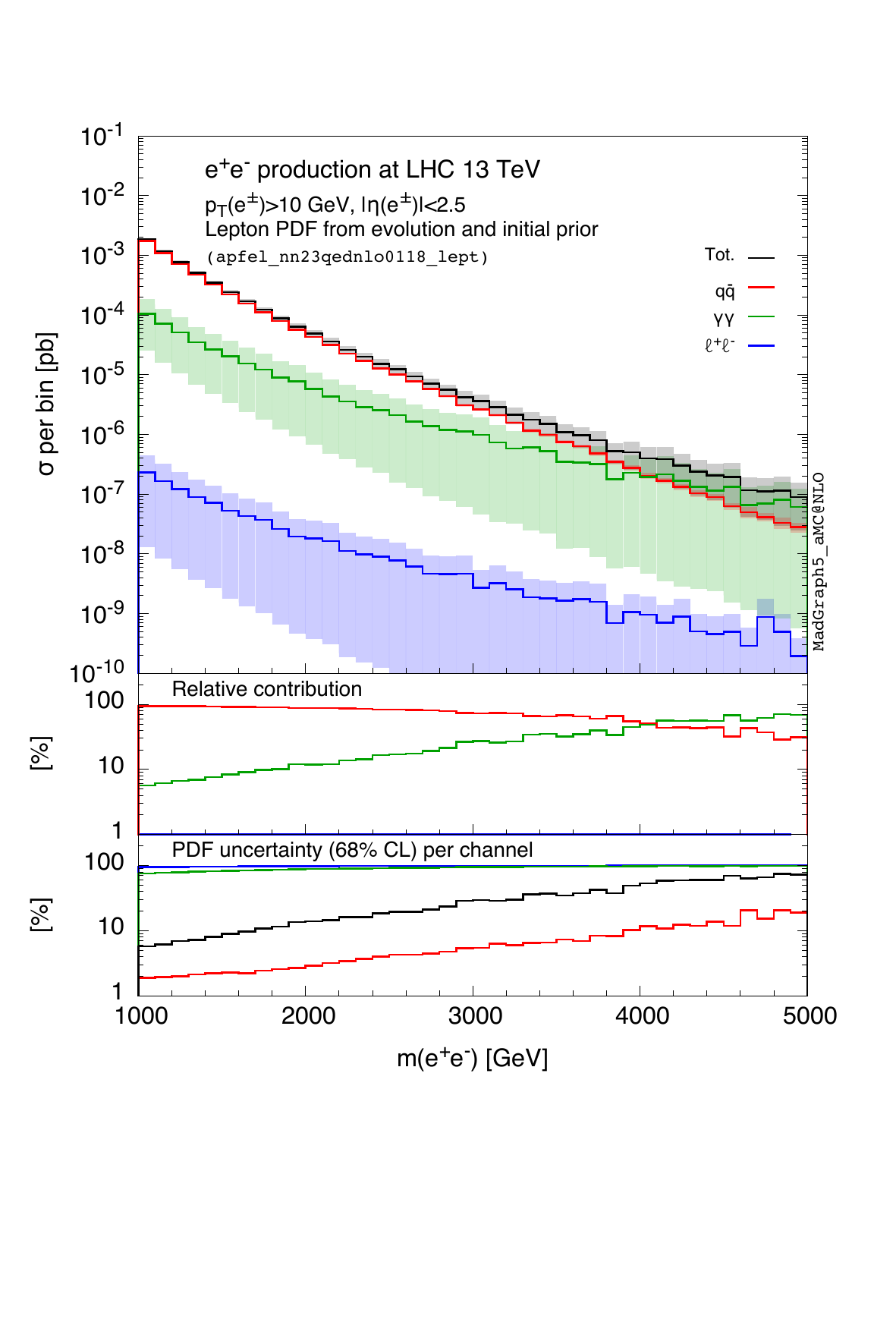}\\
  \caption{\label{fig:epem-lhc} $e^+e^-$ production at the LHC. Low
    invariant mass (left) and high invariant mass (right) of the
    lepton pair.}
\end{figure}
\begin{figure}
  \centering
  \includegraphics[clip=true, trim=0.cm 3.5cm 0.7cm 1cm, width=0.49\textwidth]{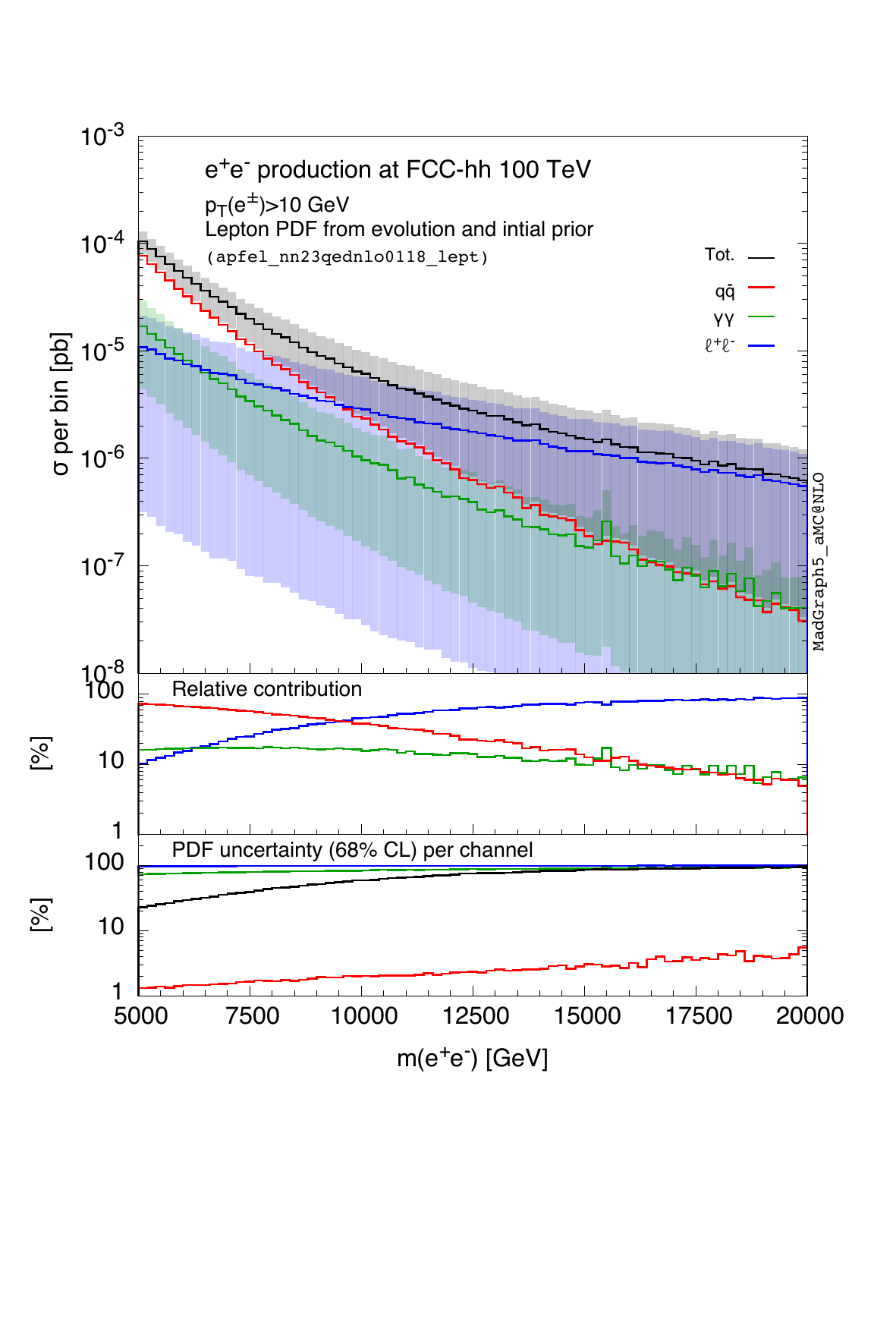}
  \includegraphics[clip=true, trim=0.cm 3.5cm 0.7cm 1cm, width=0.49\textwidth]{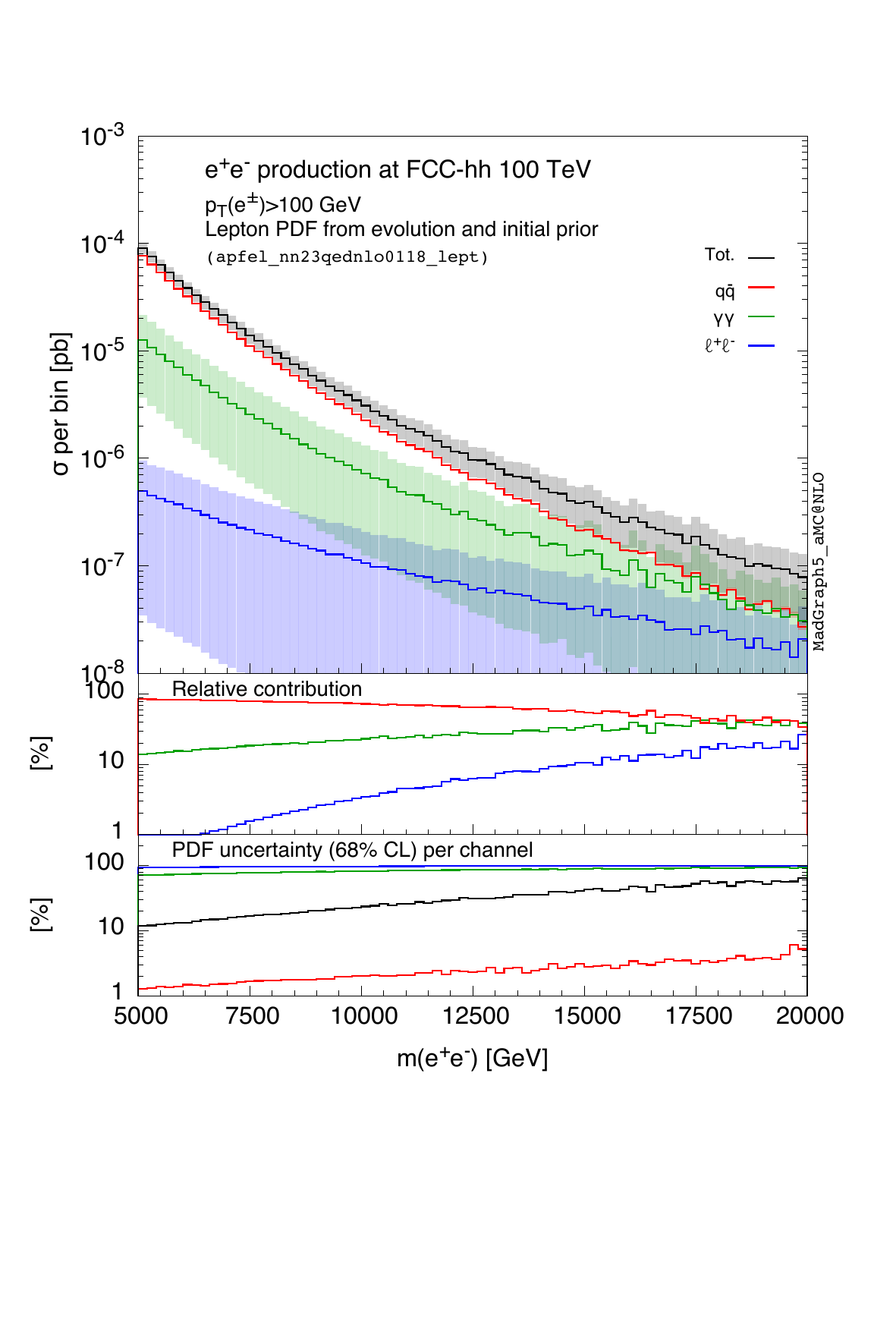}\\
  \includegraphics[clip=true, trim=0.cm 3.5cm 0.7cm 1cm, width=0.49\textwidth]{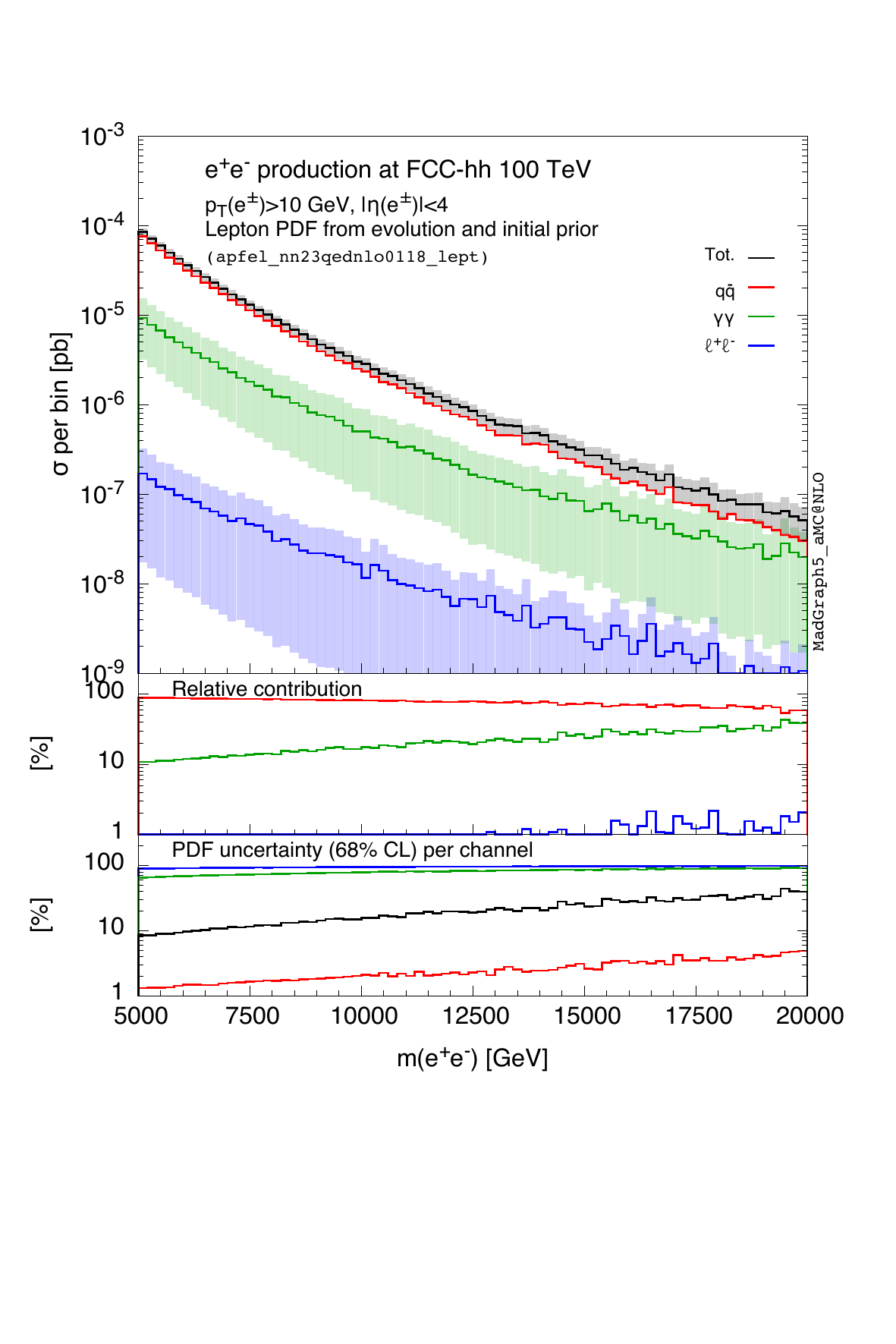}
  \includegraphics[clip=true, trim=0.cm 3.5cm 0.7cm 1cm, width=0.49\textwidth]{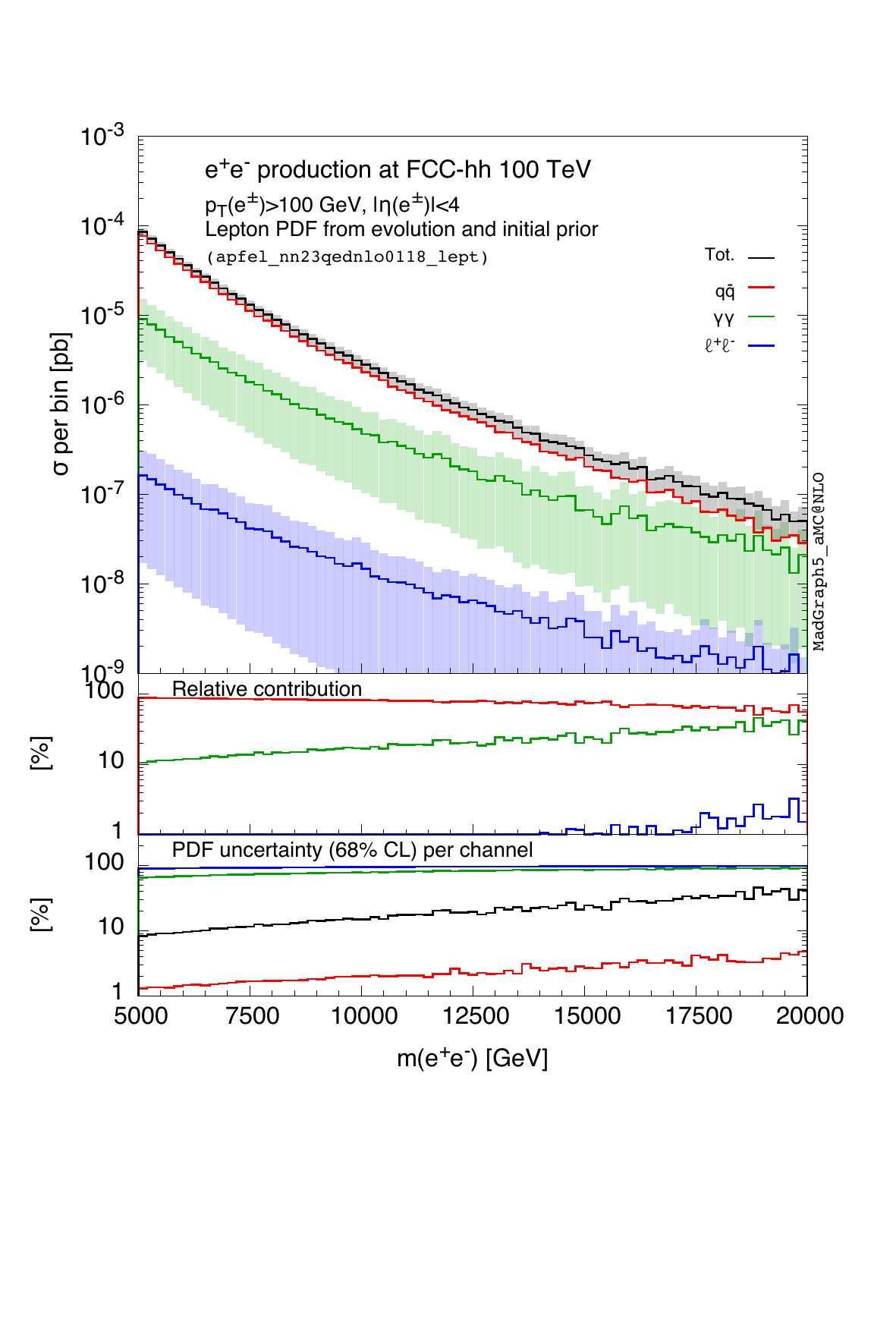}
  \caption{\label{fig:epem-fcc1} $e^+e^-$ production at the FCC-hh for
    very high invariant masses with a $p_T>10~\gev$ cut (left column),
    a $p_T>100~\gev$ cut (right column), without (top row) and with
    (bottom row) the $|\eta(e^\pm)|< 4$ cut.}
\end{figure}

In Fig.~\ref{fig:epem-lhc} we show the invariant mass distribution of
the lepton pair in neutral Drell-Yan production at the 13 TeV LHC,
with the following cuts:
\begin{equation}
  p_T(e^\pm) > 10\gev, \quad |\eta(e^\pm)|< 2.5\, .
\end{equation}
In particular, we consider two invariant mass regions: the low
invariant mass region, $m(e^+e^-) \in [20\gev,120\gev]$ and the large
invariant mass region $m(e^+e^-) > 1\tev$, shown in the left and right
plot, respectively.  In the main panel the $q\bar q$, $\gamma\gamma$
and $\ell^+\ell^-$ contributions are separately displayed with red,
green and blue lines, respectively. Similarly, the black line is the
hadronic cross section, {\it i.e.} the sum of the blue, green and red
lines. The bands displayed around each contribution correspond to the
PDF uncertainty at 68\% CL. In both the high and low invariant-mass
regions we find the contribution of initial-state leptons to be fully
negligible with respect to the main $q\bar q$ one, with a suppression
of more than six and more than two orders of magnitude,
respectively. As expected, photon-induced processes are not negligible
for large values of $m(e^+e^-)$.  This can be seen also in the first
inset of the two plots, which display the relative contribution to
the hadronic cross section for each of the three channels, with
$\ell^+\ell^-$ too small to be visible in the $[1\%,100\%]$ range.  In
the third inset we display the relative size of the 68\% PDF
uncertainty band for the three separate channels and for the hadronic
cross section.  The $\gamma\gamma$ and $\ell^+\ell^-$ channels entail
much larger PDF uncertainties than $q\bar q$, however the PDF
uncertainty of the hadronic cross section degrades only at large
values of $m(\ell^+\ell^-)$ due to the $\gamma\gamma$ contribution.

In Fig.~\ref{fig:epem-fcc1} we show similar plots at the 100 TeV
FCC-hh for very large invariant masses, $m(e^+e^-) > 5\tev$. In this
case, we also investigate how the results would change by loosening the
selection cuts down to values that are presumably not achievable at the
experimental level. In the first row of plots in
Fig.~\ref{fig:epem-fcc1}, we apply only the cuts $p_T(e^\pm) > 10\gev$
(left plot) and $p_T(e^\pm) > 100\gev$ (right plot).  In both plots
the contribution of the $\ell^+\ell^-$ channel is not negligible and
in the left plot, for $m(e^+e^-) > 10\tev$, it is even dominant.  This
behaviour is due to the $e^+e^- \rightarrow e^+e^-$ partonic cross
section with massless electrons, which diverges for electrons
collinear to the beam pipe. However, once a reasonable cut
$|\eta(e^\pm)|< 4$ on the lepton pseudorapidity is set, the
contribution of the $\ell^+\ell^-$ initial state is strongly
suppressed. The corresponding distributions with $|\eta(e^\pm)|< 4$
are shown in the lower plots of Fig.~\ref{fig:epem-fcc1}, again with
$p_T(e^\pm) > 10\gev$ (left plot) and $p_T(e^\pm) > 100\gev$ (right
plot). In conclusion, although there are kinematic configurations
where $\ell^+\ell^-$ initial states are relevant, they are presumably
excluded by standard experimental cuts. We observed a similar feature
in the $p_T(\ell)$ distributions with $m(e^+e^-) > 2\tev$ at the LHC;
at small $p_T$'s the $\ell^+\ell^-$ contribution is dominant, but it
is completely suppressed if a cut on $\eta(e^\pm)$ is imposed. For the
$\gamma\gamma$ initial state, a detailed analysis of the impact of the
PDF uncertainties at differential level has been performed in
Ref.~\cite{Boughezal:2013cwa}.

\subsubsection*{Dijet}

The next process we consider is dijet production, the dominant process
at hadron colliders. We look at dijet invariant mass ($m(jj)$)
distributions, for jets with $p_T(j) > 10(100) \gev$ at the LHC
(FCC-hh) and with or without a cut on the jet pseudo-rapidity
($|\eta(j)|< 4$) for the FCC-hh. At the LHC, instead, we always apply
the pseudorapidity cut $|\eta(j)| < 2.5$. We consider two cases: in
the first we assume zero probability for leptons and photons to
fake final state jets, while in second we assume such
a probability to be 100\%. In other words, in the latter case we
consider leptons and photons in the final state on the same footing of
quarks and gluons. The realistic value of the fake rate depends 
on the identity of the particle and on the detector characteristics. At 
LHC experiments it is typically negligible for photons, electrons and muons, 
while for taus it is about 40\%, see e.g. Ref.~\cite{Khachatryan:2015dfa}. Moreover, the value
of the fake rate is also analysis dependent. Thus, by presenting 
results for the two extreme cases, we provide a conservative estimate 
of possible effects coming from leptons in the initial state.\\
Since we consider $2\to2$ processes at LO and
parton level, the cut on the transverse momentum of the final state
particles is sufficient to obtain a finite cross section and each
parton leads to one jet.

\begin{figure}
  \centering
  \includegraphics[clip=true, trim=0.cm 3.5cm 0.7cm 1cm, width=0.49\textwidth]{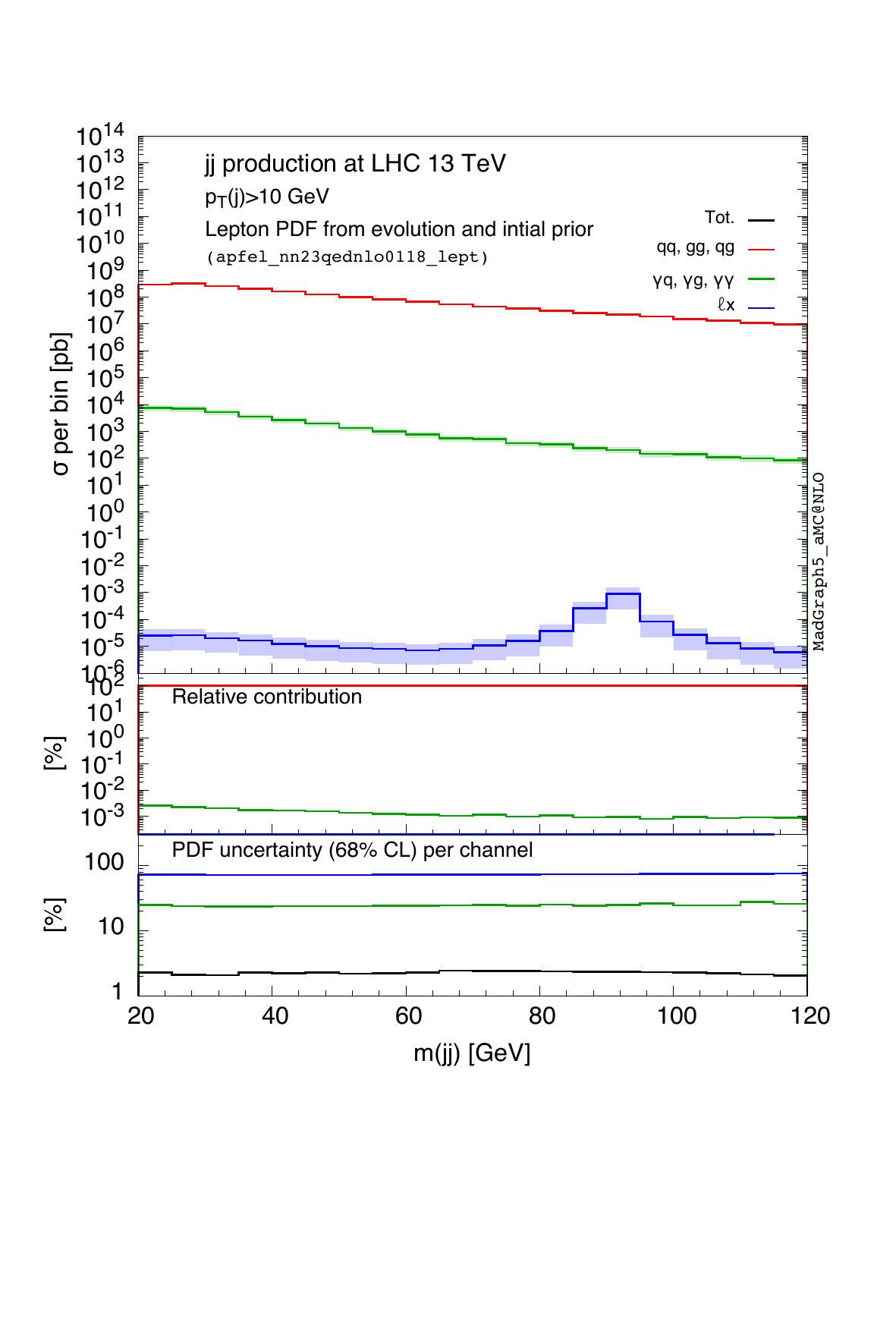}
  \includegraphics[clip=true, trim=0.cm 3.5cm 0.7cm 1cm, width=0.49\textwidth]{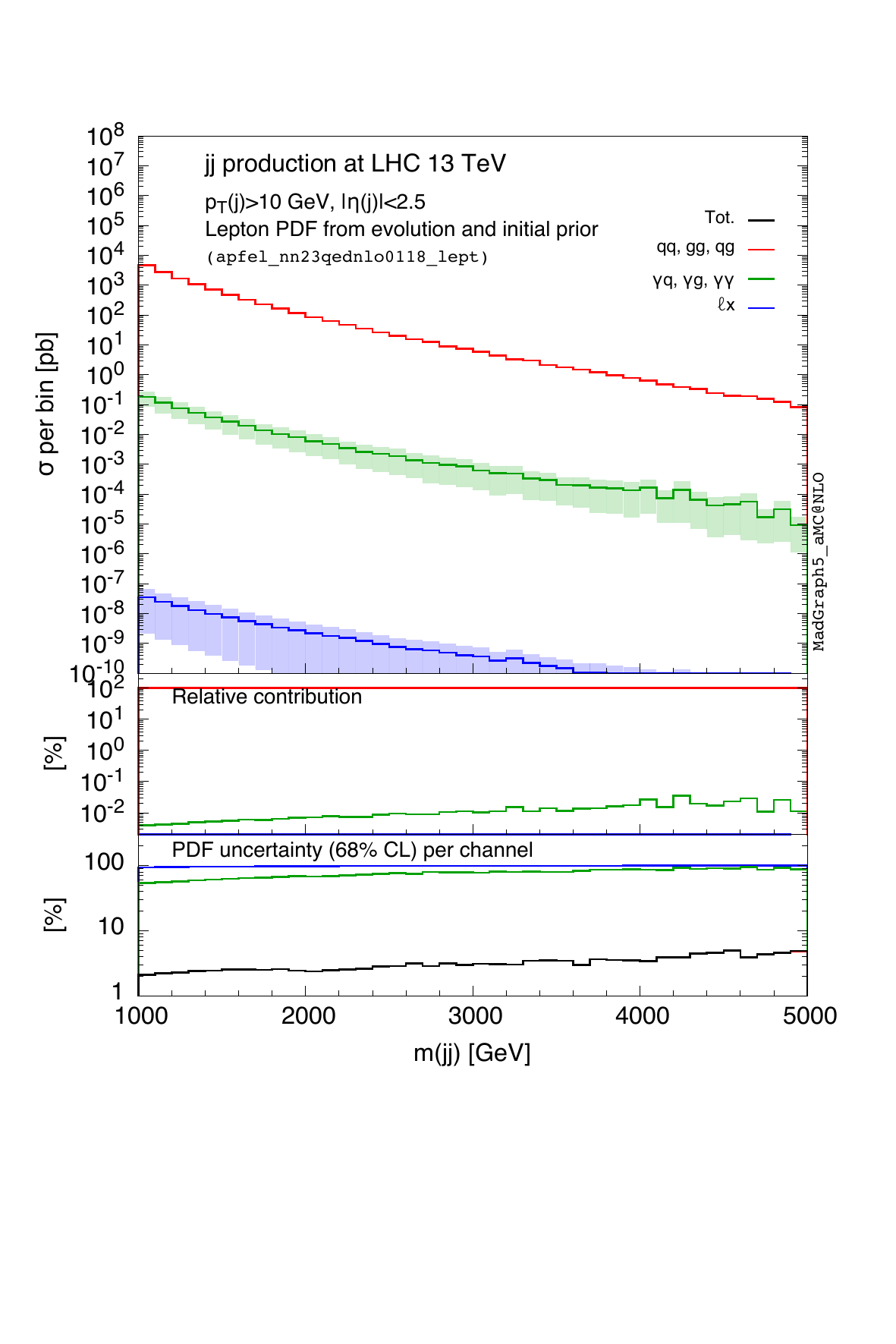}
  \caption{\label{fig:dijet-lhc} Dijet production at the LHC. Low
    (left) and high (right) invariant mass of the dijet pair. Zero
    probability of photons/leptons faking jets is assumed.}
\end{figure}

\begin{figure}
  \centering
  \includegraphics[clip=true, trim=0.cm 3.5cm 0.7cm 1cm, width=0.49\textwidth]{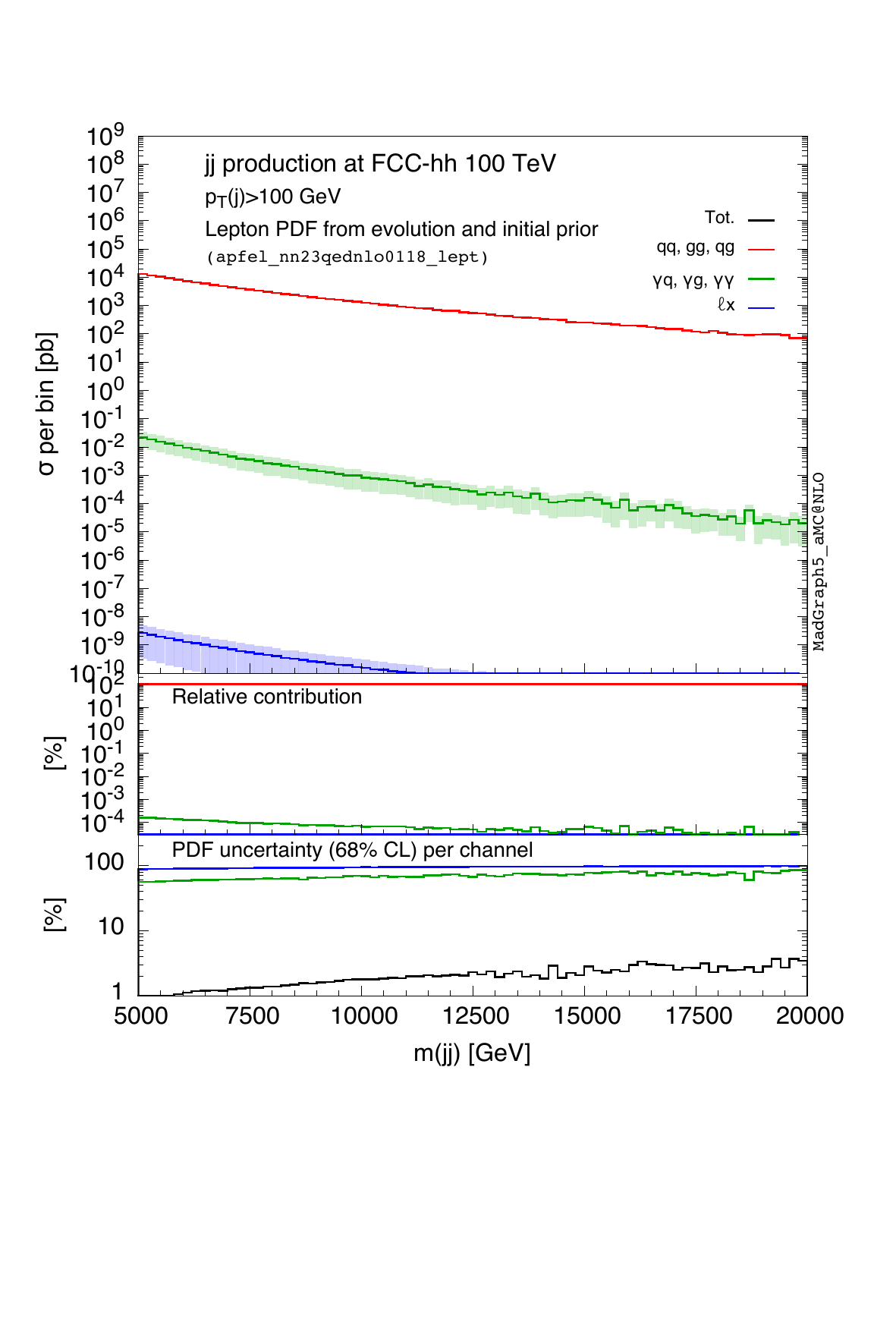}
  \includegraphics[clip=true, trim=0.cm 3.5cm 0.7cm 1cm, width=0.49\textwidth]{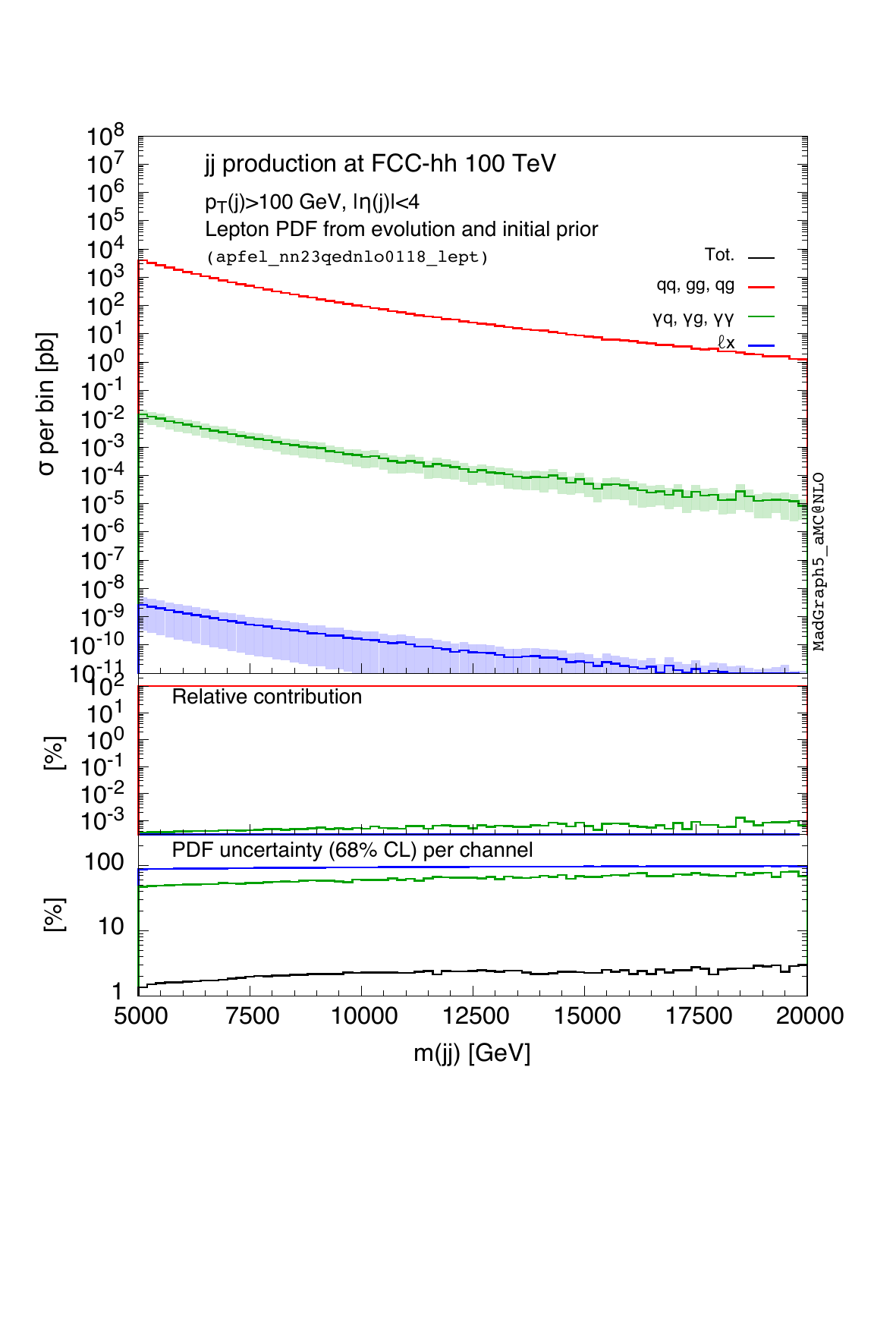}\\
  \caption{\label{fig:dijet-fcc} Dijet production at the FCC-hh in the
    very-high invariant mass region, without (left) or with (right)
    the $|\eta(j)|<4$ cut.  Zero probability of photons/leptons faking
    jets is assumed.}
\end{figure}

We start by showing plots for the first case (no photons/leptons
faking jets), in the low and high dijet invariant-mass regions at the
LHC and in the very high invariant-mass region at the FCC-hh with and
without the $|\eta(j)|< 4$ cut. The LHC and FCC-hh plots are shown in
Fig.~\ref{fig:dijet-lhc} and Fig.~\ref{fig:dijet-fcc}. In both cases
the lepton-initiated contribution is suppressed by several orders of
magnitude.  However, part of the suppression does not originate from
the PDFs, but from the different perturbative orders of the various partonic
channels.  The gluon--gluon initial state yields $\ord(\alpha_s^2)$
terms, which are the dominant ones at LO, while for instance the photon--gluon and
lepton-lepton initial states yield only $\ord(\alpha_s \alpha)$ and
$\ord(\alpha^2)$ contributions, respectively. Therefore, the total
rates of these subprocesses are suppressed also by the subleading
order in the perturbative expansion. It is worth mentioning that
quark--antiquark initial states lead to contributions of
$\ord(\alpha_s^2)$, $\ord(\alpha_s \alpha)$ and $\ord(\alpha^2)$,
which we consistently take into account in our analysis.\\

\begin{figure}
  \centering
  \includegraphics[clip=true, trim=0.cm 3.5cm 0.7cm 1cm, width=0.49\textwidth]{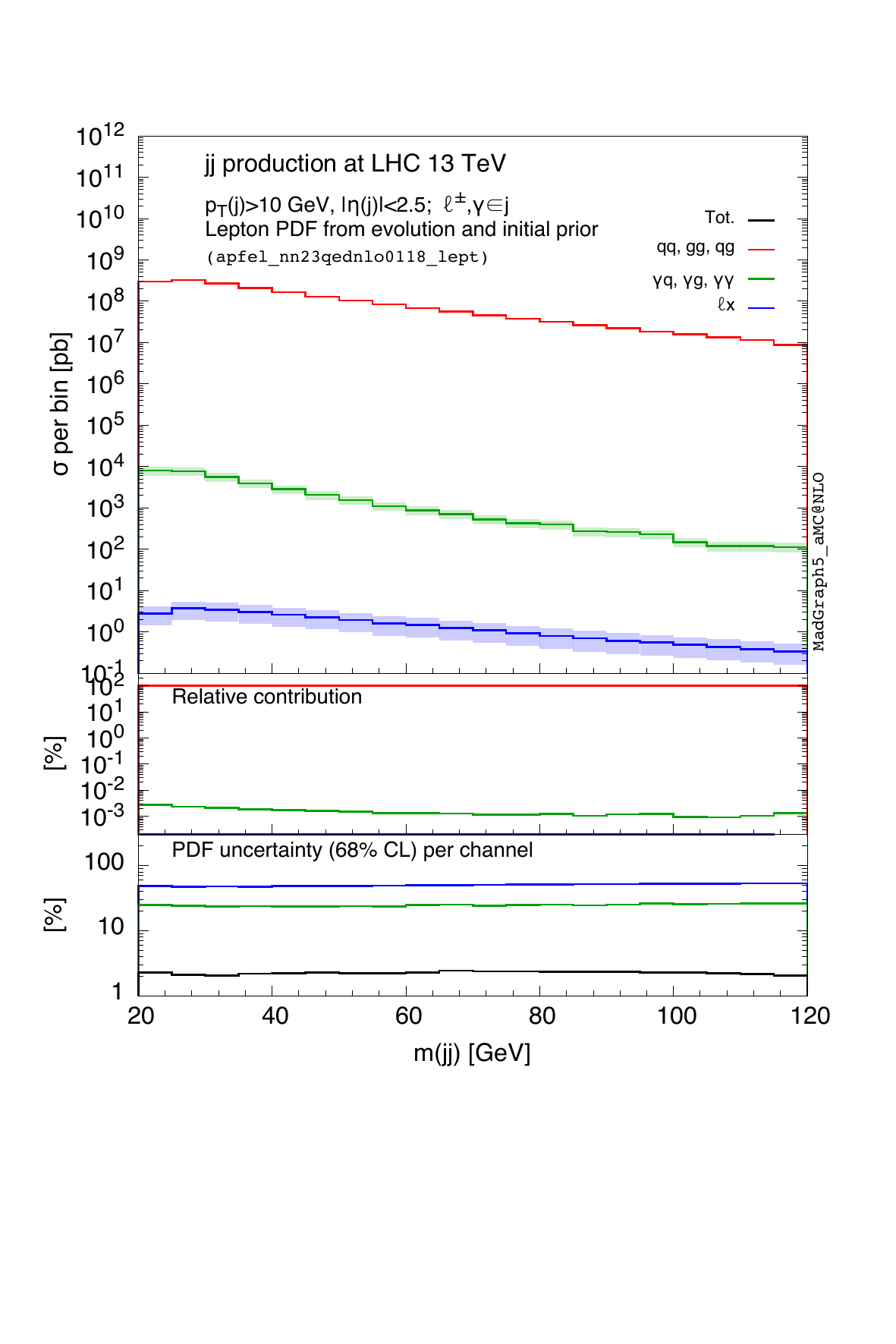}
  \includegraphics[clip=true, trim=0.cm 3.5cm 0.7cm 1cm, width=0.49\textwidth]{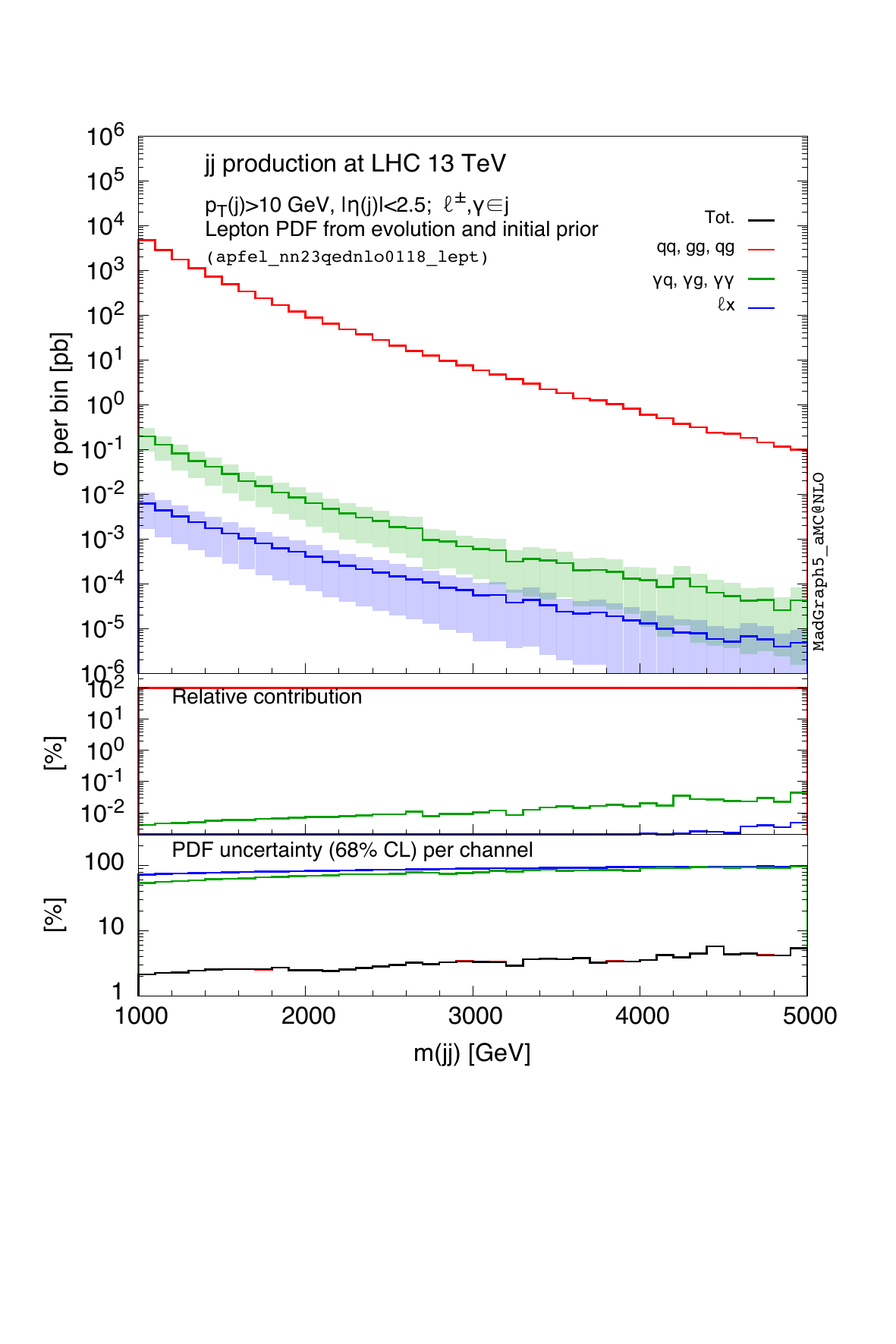}
  \caption{\label{fig:dijetall-lhc} Dijet production at the LHC in the
    low (left) and high(right) invariant mass region.  100\%
    probability of photons/leptons faking jets is assumed.}
\end{figure}

\begin{figure}
  \centering
  \includegraphics[clip=true, trim=0.cm 3.5cm 0.7cm 1cm, width=0.49\textwidth]{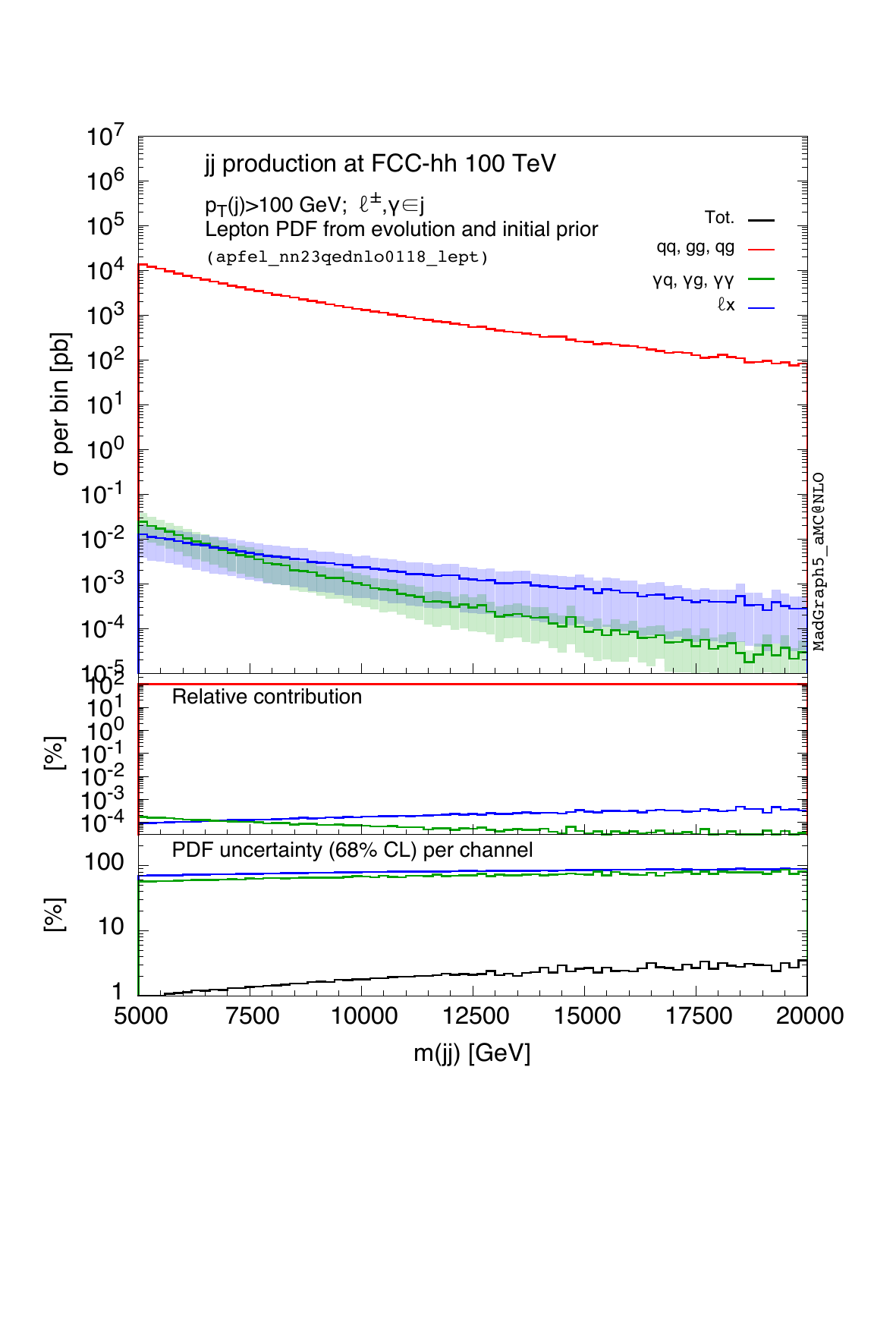}
  \includegraphics[clip=true, trim=0.cm 3.5cm 0.7cm 1cm, width=0.49\textwidth]{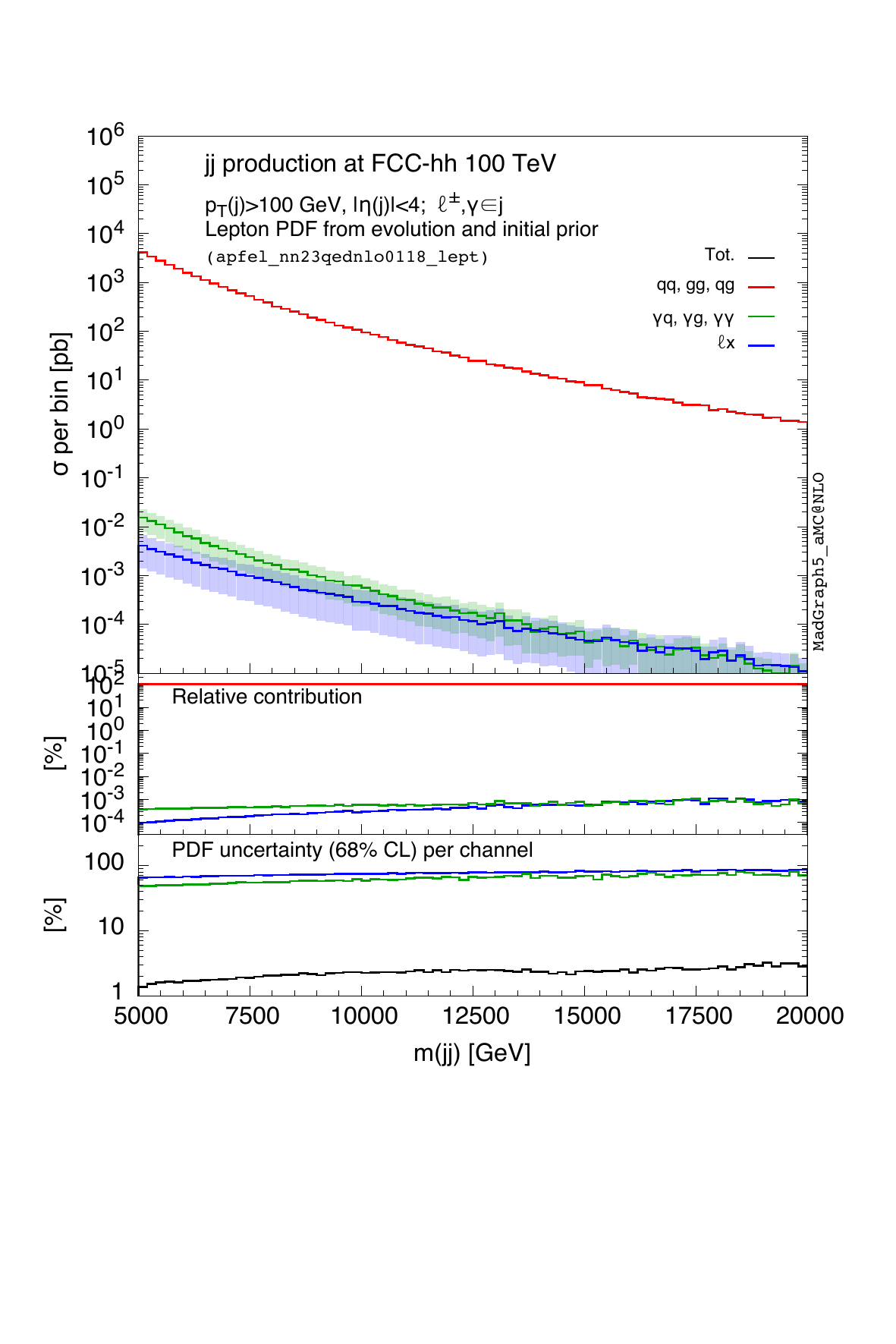}\\
  \caption{\label{fig:dijetall-fcc} Dijet production at the FCC-hh in
    the very-high invariant mass region, without (left) or with
    (right) the $|\eta(j)|<4$ cut.  100\% probability of
    photons/leptons faking jets is assumed.}
\end{figure}

The second case, in which we assume that photons and leptons in the
final state are always reconstructed as jets, is particularly
interesting because of new $t$-channel diagrams appearing, {\it e.g.},
in the $\ell^+\ell^-$ initial state. In general, all the $\ell q$,
$\ell\gamma$ or $\ell^{\pm}_1\ell^{\pm}_2$ initial states give a
contribution. The relevant plots for the LHC and the FCC-hh are
displayed in Fig.~\ref{fig:dijetall-lhc} and
Fig.~\ref{fig:dijetall-fcc}, respectively. Assuming 100\% probability for
leptons and photons to fake final-state jets, the contributions from initial states with at least one
lepton ($\ell x$) are enhanced in such a way that they become as important as those
involving photons and no leptons. However, both contributions are
suppressed by at least four orders of magnitude w.r.t. those initiated
by QCD partons.

\medskip 

\subsubsection*{$\mathbf{W^+W^-}$ and $\mathbf{ZZ}$ into 4 leptons}

Next we turn to diboson production at hadron colliders.  Firstly, we
look at $W^+W^-$ production with undecayed $W$ bosons. Then we analyse
four-lepton production which is the signature emerging from $ZZ$
production with both $Z$ bosons decaying into two leptons.

In the case of $W^+W^-$ production, we keep the $W$ boson stable so
that we can gauge the effects due only to the $\ell^{+}\ell^{-}$
luminosity and not to matrix-element enhancements. The latter effects
will be instead analysed in the case of $ZZ$ production. Moreover,
with stable $W$ bosons, it is possible to check if any effect from
lepton PDFs could be related to the anomaly that has been recently
found by the ATLAS Collaboration in the production of dibosons with 2
TeV invariant mass in the Run-I of the LHC \cite{Aad:2015owa}.

\begin{figure}
  \centering
  \includegraphics[clip=true, trim=0.cm 3.5cm 0.7cm 1cm, width=0.49\textwidth]{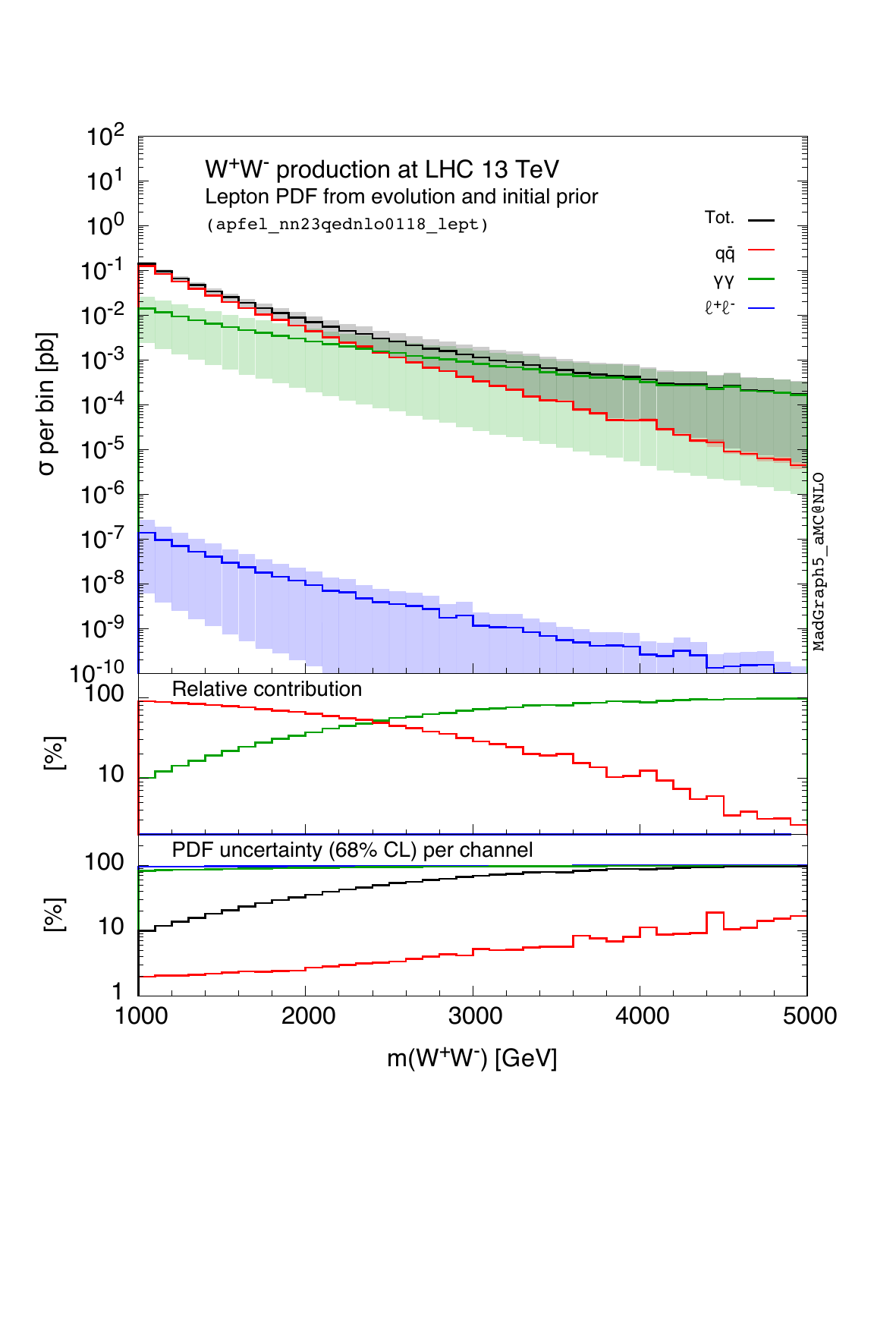}
  \includegraphics[clip=true, trim=0.cm 3.5cm 0.7cm 1cm, width=0.49\textwidth]{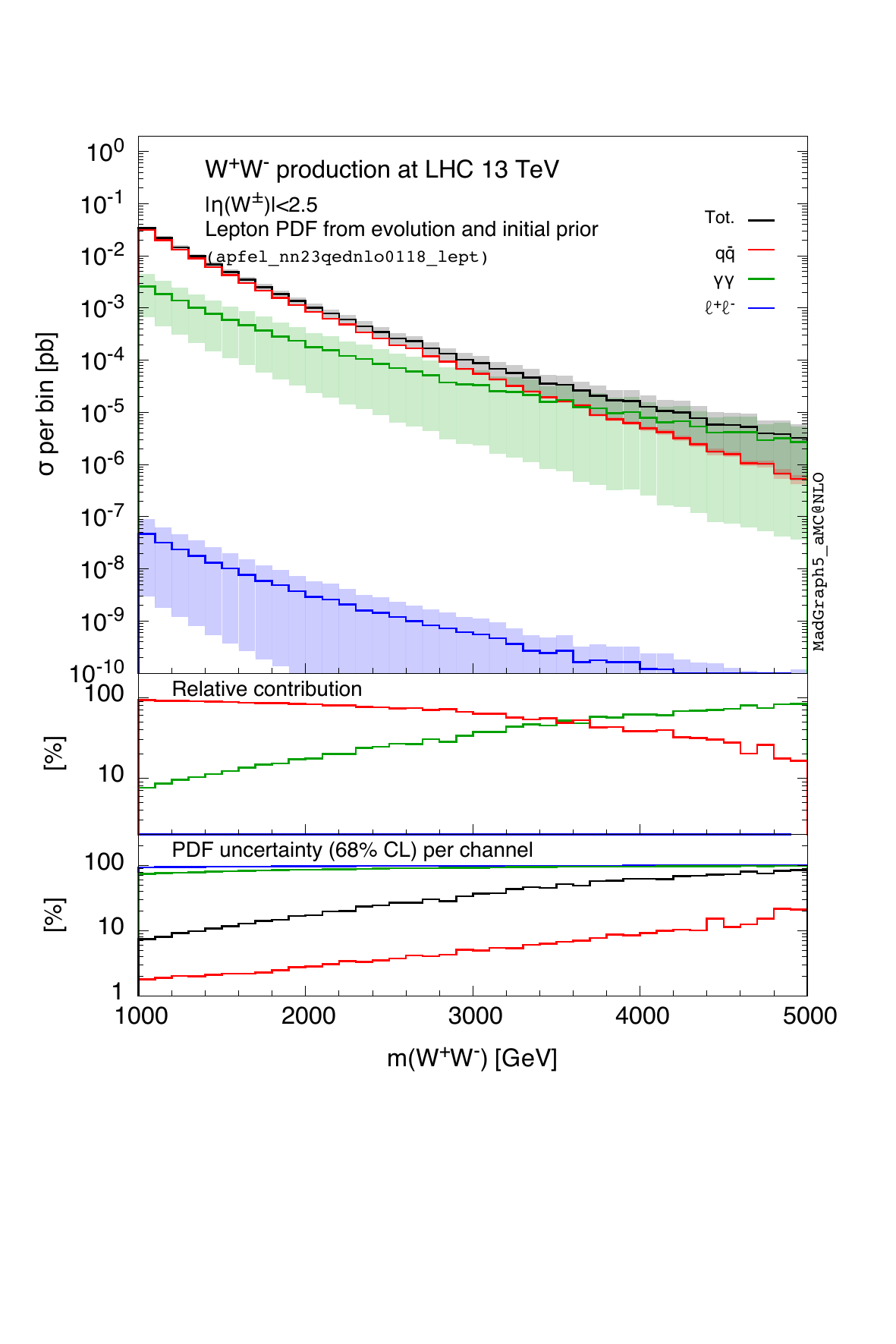}\\
  \includegraphics[clip=true, trim=0.cm 3.5cm 0.7cm 1cm, width=0.49\textwidth]{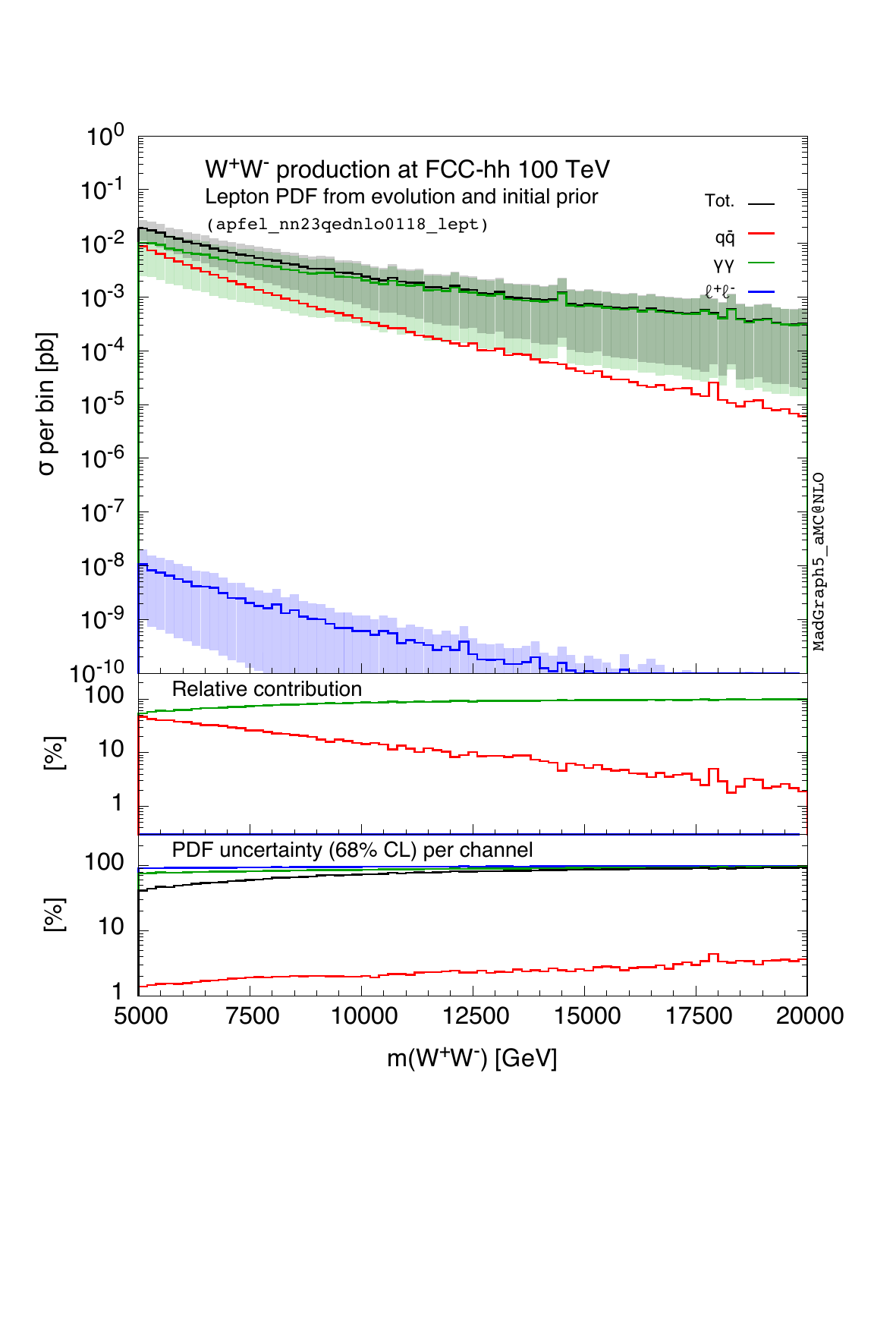}
  \includegraphics[clip=true, trim=0.cm 3.5cm 0.7cm 1cm, width=0.49\textwidth]{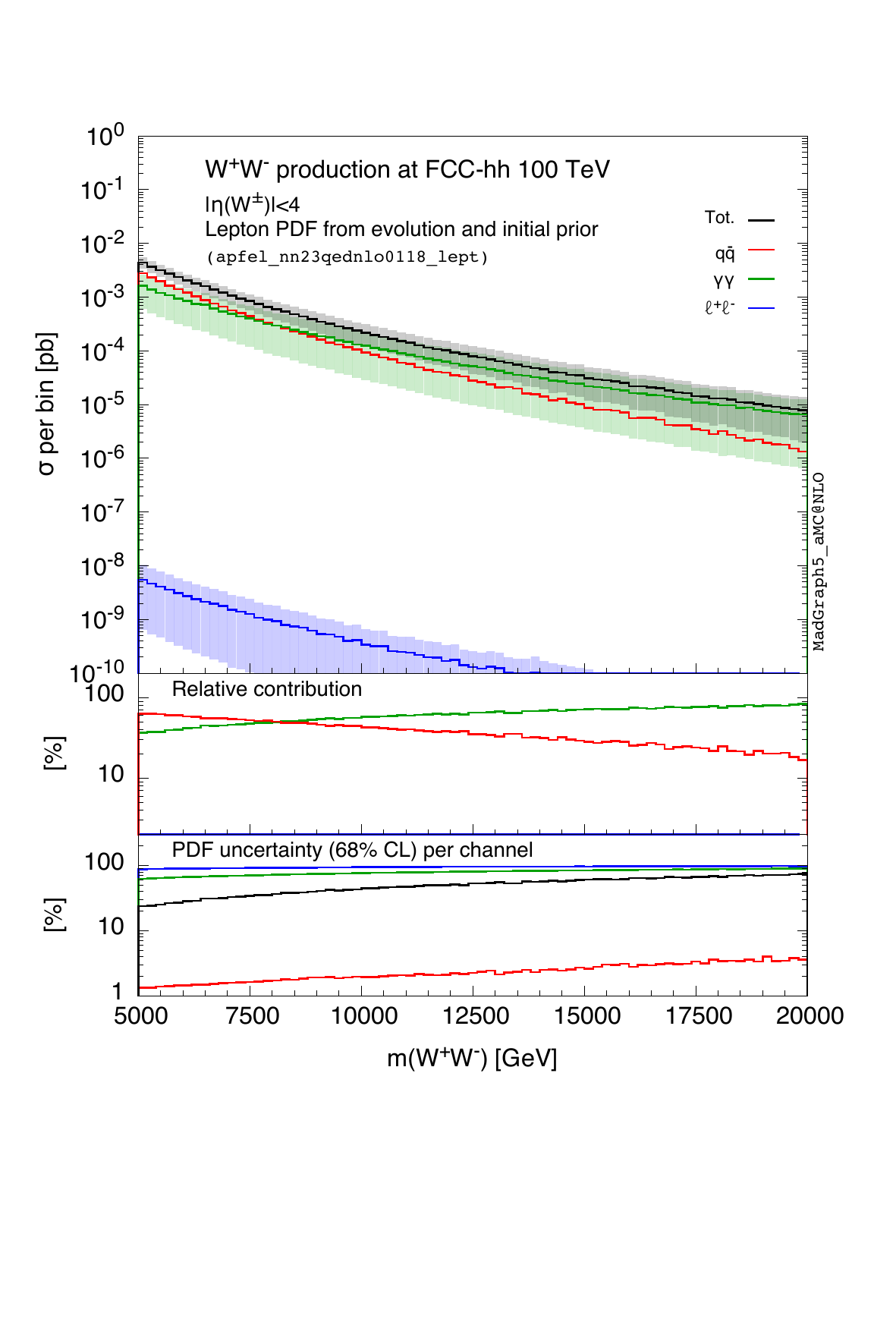}
  \caption{\label{fig:ww} $W$ pair production at the LHC (top row) and
    at the FCC-hh (bottom row), both without cuts (left column) or
    with a rapidity cut (right column). The rapidity cut is
    $|\eta(W^\pm)|<2.5(4)$ at the LHC (FCC-hh).}
\end{figure}

Differential distributions for the invariant mass $m(W^+ W^-)$ of the
$W^+W^-$ pair are shown in Fig.~\ref{fig:ww}, for the LHC in the upper
plots and for the FCC-hh in the lower plots. For both colliders we
consider the case without cuts (plots on the left) and with
$|\eta(W^\pm)|<2.5(4)$ at the LHC (FCC-hh) (plots on the right). Also
for this process the lepton PDF contribution can be in general safely
neglected. On the contrary, the photon-induced processes have to be
taken into account for a reliable estimate of the central value of the
cross section and the associated uncertainty due to PDFs in the large
invariant-mass region. A similar study at the LHC, including NLO QCD
and EW corrections but with no PDF uncertainty, have been pursued also
in Ref.~\cite{Baglio:2013toa}.  The increase of the relative size of the $\gamma\gamma$-channel for high $m(W^+ W^-)$ is consistent with the behaviour of the $\Phi_{\gamma \gamma}$ and
$\Phi_{q\bar{q}}$ luminosities shown in Fig.~\ref{fig:lumileptM}. In addition, no suppression from $s$-channel diagrams is present in $\gamma \gamma \rightarrow W^+ W^-$ production, leading to a further relative enhancement with respect to the $q\bar q$-channel at high $m(W^+ W^-)$. On the other hand, in the $\gamma\gamma$-channel the $W$ bosons are produced more peripherally than in the $q\bar q$-channel. Thus, the cut in pseudorapidity reduces the
relative impact of the $\gamma\gamma$ channel, but it does not
dramatically change the qualitative picture.

\medskip In the case of $ZZ$ production, we analyse the four-lepton
final state emerging from the decay of the $Z$ bosons. Specifically,
we consider either four leptons of the same flavour ($e^+e^-e^+e^-$),
or two lepton pairs of different flavours ($e^+e^-\mu^+\mu^-$). All
diagrams, with and without intermediate $Z$ bosons, are taken into
account. Thus, all the resonant and non-resonant contributions have
been properly included. It is worth noting that, if we were requiring
stable $Z$ bosons, no $\gamma\gamma$-initiated contribution would
appear, since the process $\gamma\gamma \rightarrow Z Z$ has no
tree-level diagrams. Moreover, as in the case of neutral Drell-Yan,
final-state leptons collinear to the beam axis lead to divergent cross
sections for the $\ell^{+}\ell^{-}$ channel.

\begin{figure}
  \centering
  \includegraphics[clip=true, trim=0.cm 3.5cm 0.7cm 1cm, width=0.49\textwidth]{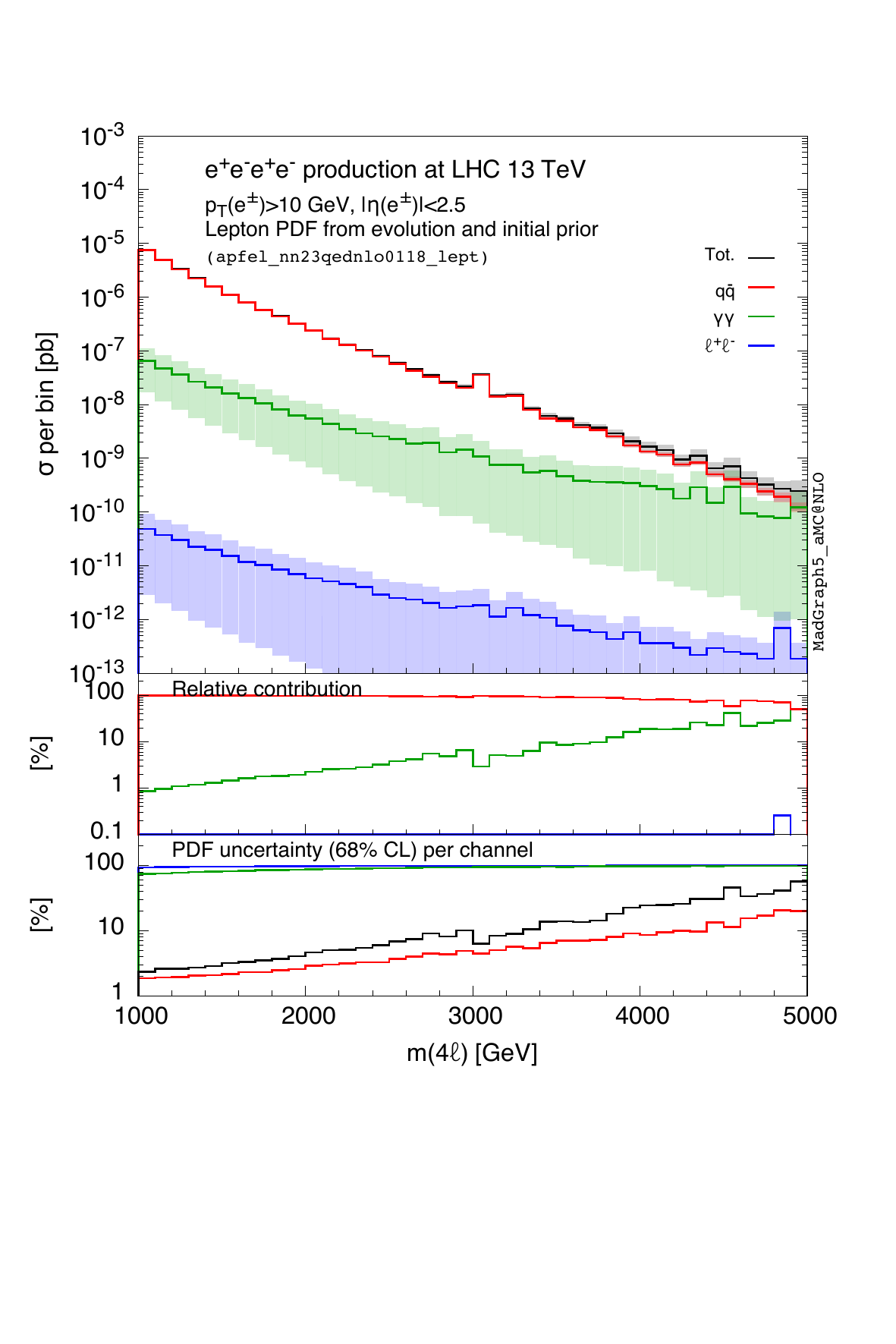}
  \includegraphics[clip=true, trim=0.cm 3.5cm 0.7cm 1cm, width=0.49\textwidth]{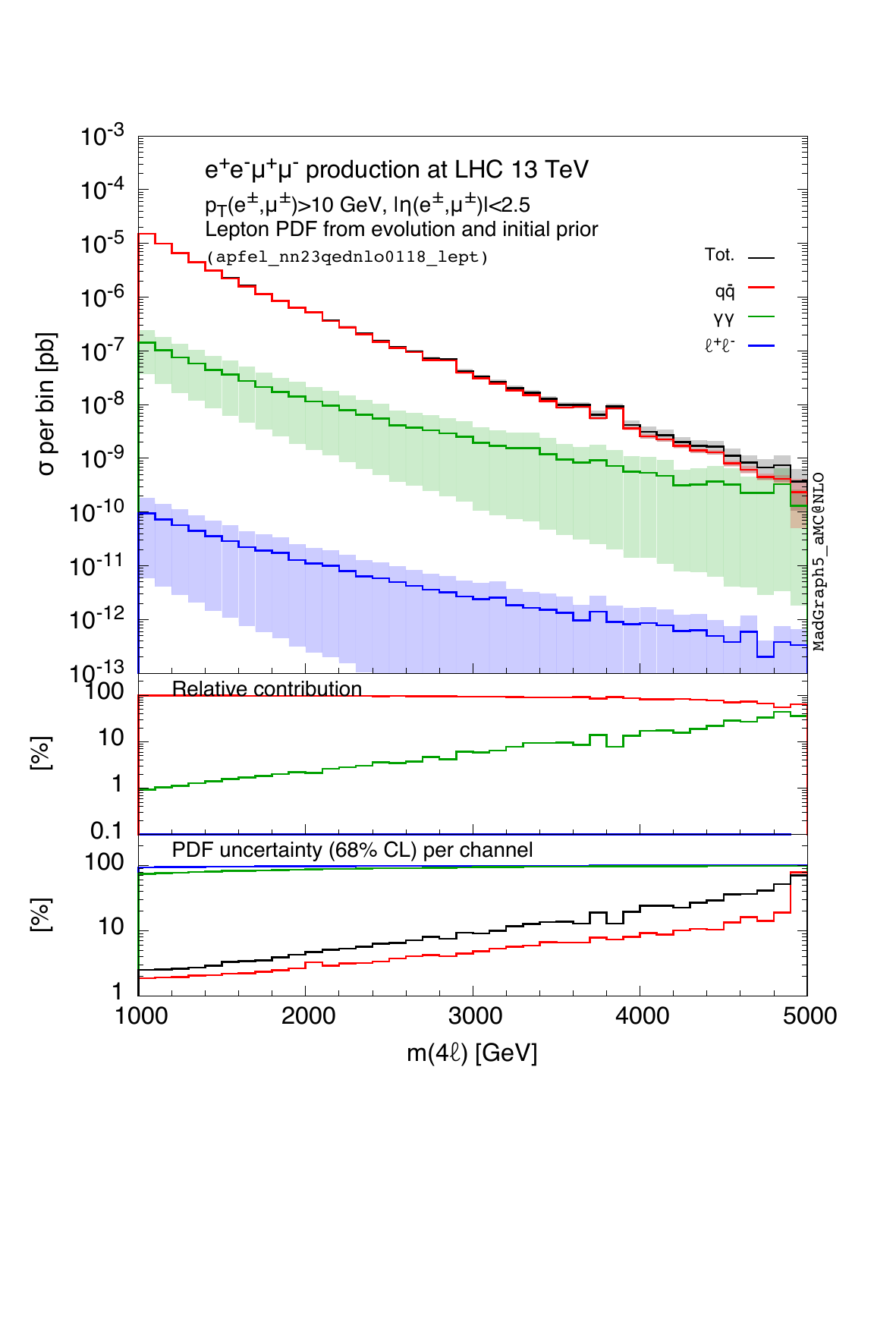}\\
  \caption{\label{fig:4lep-lhc} Four lepton production at the LHC: the
    case of all four leptons of the same flavour (left) and of two
    different flavour pairs (right) is considered.}
\end{figure}

\begin{figure}
  \centering
  \includegraphics[clip=true, trim=0.cm 3.5cm 0.7cm 1cm, width=0.49\textwidth]{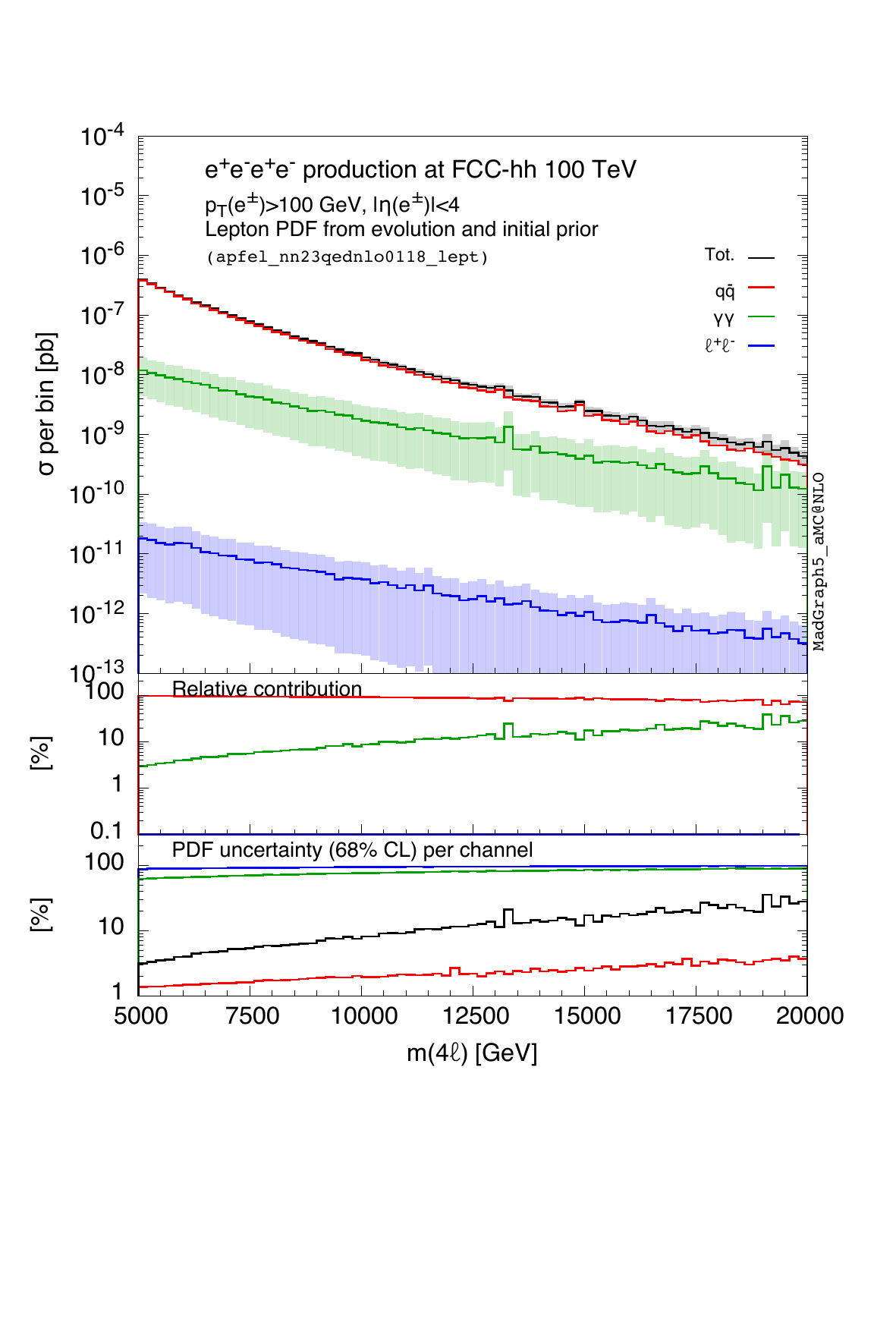}
  \includegraphics[clip=true, trim=0.cm 3.5cm 0.7cm 1cm, width=0.49\textwidth]{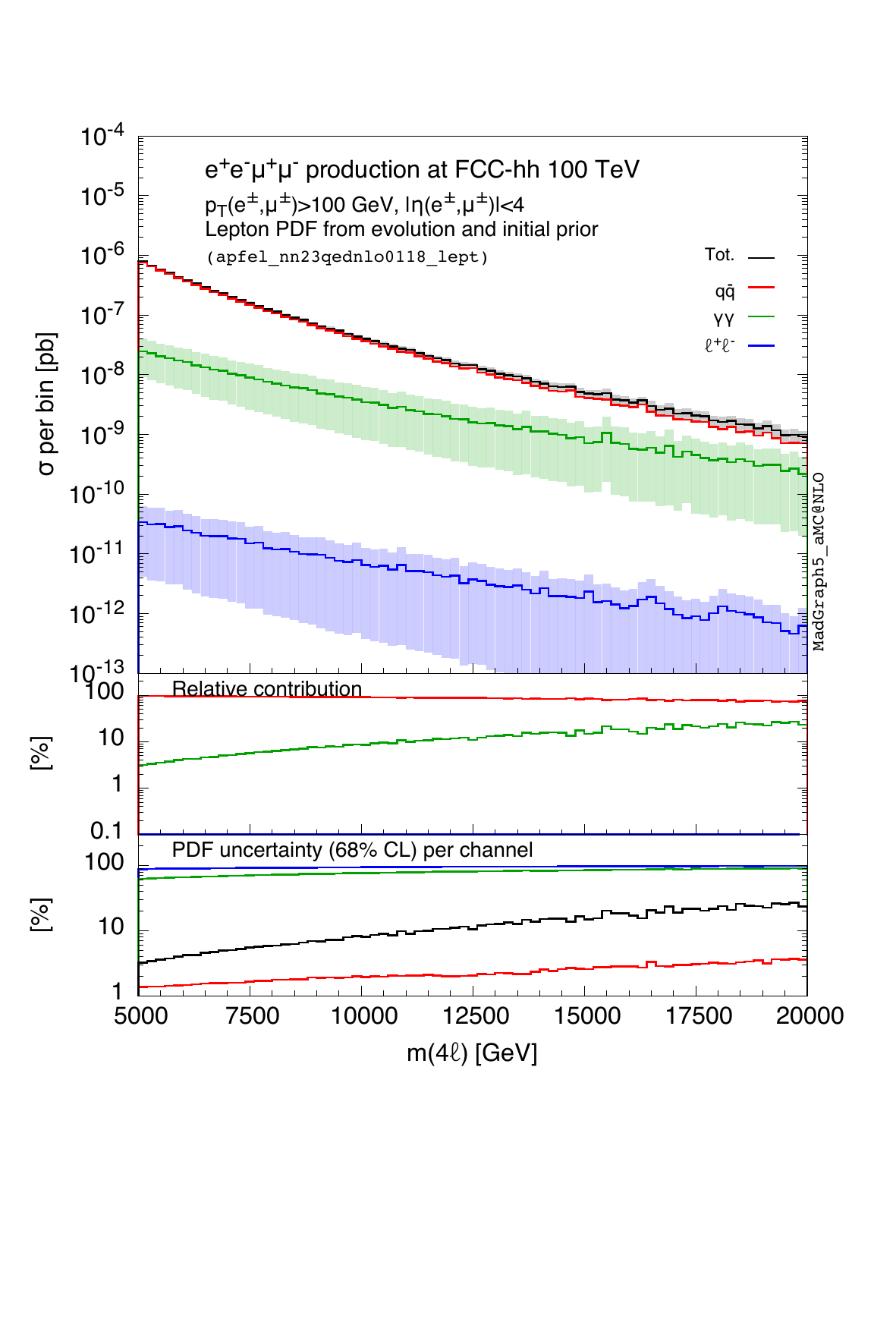}\\
  \caption{\label{fig:4lep-fcc} Four lepton production at the FCC-hh:
    the case of all four leptons of the same flavour (left) and of two
    different flavour pairs (right) is considered.}
\end{figure}

In Figs.~\ref{fig:4lep-lhc} and~\ref{fig:4lep-fcc} we show
differential distributions for the invariant mass of the four-lepton
system at the LHC and FCC-hh, respectively.  In our simulations at the
LHC we required four leptons with:
\begin{equation}
  p_T(\ell) > 10 \gev, \quad |\eta(\ell)|<2.5\,,\label{eq:lepcutslhc}
\end{equation}
while at the FCC-hh they must satisfy:
\begin{equation}
  p_T(\ell) > 100 \gev, \quad |\eta(\ell)|<4\,.\label{eq:lepcutsfcc}
\end{equation}
Furthermore, we require that all opposite-sign lepton pairs have an
invariant mass $m(\ell^\pm,\ell'^\mp) > 20 \gev$ to avoid collinear
singularities due to photon splittings.  Again, we clearly observe
that the $\ell^{+}\ell^{-}$-initiated contribution is completely
negligible. The largest relative contribution to the cross section is
at most at the per-mil level at very large values of the four-lepton
invariant mass and tends to become even smaller in the lower
invariant-mass region.


\subsubsection*{Same-Sign and/or Different-Flavour lepton pair}

We conclude our review of SM processes discussing two cases which are
relevant for both SM physics and BSM searches: the production of a
same-sign lepton pair and of a different-flavour lepton pair. For such
searches in the BSM context, the SM background processes that are
typically considered are those featuring two vector bosons in the
final state ($ZZ$, $WZ$, $W^\pm W^\pm j j$, $W^\pm W^\mp$,
$t \bar t \to W^\pm W^\mp b \bar b$, \ldots), where the
same-sign/different-flavour pair of leptons comes together with jets,
other (undetected) leptons and/or missing transverse energy. Therefore
it is worth estimating the lepton-initiated contribution to the SM
background for these processes.

Before presenting our results, we want to stress one important fact to
be kept in mind: the introduction of lepton PDFs does not open new
production mechanisms. Rather, it has to be considered as a different
scheme to tackle a given computation, on the same footing as the four-
and five-flavour schemes in processes involving bottom quarks. For
example, the process $e^+ e^+ \to e^+ e^+$, computed with lepton PDFs,
corresponds to the process $\gamma\gamma \to e^+ e^- e^+ e^-$ where
leptons are considered as {\it massive} and the $e^-$ degrees of
freedom are integrated out. This latter scheme has several drawbacks
if compared to the one with leptons in the initial state: first, the
process has a larger final-state multiplicity and is therefore more
challenging from the computational point of view. On top of this,
lepton masses are very small as compared to the typical scales probed
at the LHC and FCC-hh. As a consequence, the numerical evaluation and
integration of matrix elements can be hampered by instabilities. In
particular, large logarithms of the form $\alpha \log(Q/m_\ell)$,
where $Q$ is a typical hard scale of the hadronic process, could spoil
the convergence of the numerical integration. On the contrary, using
lepton PDFs, such logarithms are analytically taken into account from
$m_\ell$ up to $Q_0$ by the initial condition in Eq.~\eqref{eq:ansatz}
and consistently resummed from $Q_0$ to $Q$ by DGLAP evolution.
Therefore, the usage of a scheme with lepton PDFs simplifies the
computation, and consequently the phenomenological studies, for the
classes of processes considered here.

After this clarification, we start by considering the same-sign
lepton-pair production. We compare the $\ell^{+}\ell^{+}$-initiated
case to the process $qq\to W^\pm W^\pm qq$, where both the $W$ bosons
decay leptonically. For this process, which has also been proposed as
a way to search for and calibrate double-parton scattering at the
LHC~\cite{hep-ph/9912232}, we do not need to consider additional
untagged leptons and the cross section is finite even without cuts on
jets.

We quote total rates for
$qq\to W^+ W^+ qq\to \ell^+\ell^+\nu_\ell\nu_\ell q q$ and for
$\ell^+\ell^+ \to \ell^+\ell^+$ in Tab.~\ref{tab:sslep_rates}. For the
former, we include all coupling combinations entering at LO, {\it
  i.e.} we evaluate this process at the orders
$|\alpha_s \alpha^2 + \alpha^3|^2$.~\footnote{ In
  Ref.~\cite{hep-ph/9912232} it is shown that the electroweak
  contribution to the cross section is of the same order of magnitude
  of the QCD one. Thus, it cannot be ignored.}  As usual, we consider
the case of the LHC and of the FCC-hh, requiring the two charged
leptons to satisfy the cuts of Eqs.~(\ref{eq:lepcutslhc})
and~(\ref{eq:lepcutsfcc}), respectively.  We can see that in this case
the lepton-initiated contribution is not entirely negligible as
compared to the same-sign $W$ pair production rate, at least at the
LHC.  In this case, the former contribution is $7\%$ of the latter,
while this ratio reduces to 1\% at the FCC-hh. In both cases PDF
uncertainties affecting the lepton-initiated contribution are quite
large.  The quoted results are shown for the case of positively
charged leptons. In the case of negatively charged leptons, the
lepton-initiated cross section does not change, while the cross
section for $qq\to W^- W^- qq$ is reduced by approximately a factor 2
(1.4) with respect to the positively charged case at the LHC (FCC-hh).
The quoted numbers include the contributions from all the three
leptonic flavours. For $qq\to W^\pm W^\pm qq $, different-flavour and
same-flavour leptons yield two thirds and one third of the total cross
section, respectively, due to lepton universality. Instead, for
$\ell^\pm\ell^\pm \to \ell^\pm\ell^\pm$ the relative contributions are
60\% and 40\% because of non-negligible interference effects that
appear only for same-flavour and same-sign lepton pair production.
\begin{table}
    \begin{centering}
    \begin{tabular}{l|cc}
        $\ell = e, \mu, \tau $                 & LHC, $\sigma$[pb]         & FCC-hh, $\sigma$[pb]        \\\hline
        $qq \to \ell^+\ell^+ \nu \nu j j $    & $ 3.54 \cdot 10^{-2} \pm 0.4\% $    & $ 6.33 \cdot 10^{-2} \pm 0.3\% $   \\
        $\ell^+\ell^+ \to \ell^+\ell^+             $    & $ 2.64 \cdot 10^{-3} \pm 37\%  $    & $ 5.86 \cdot 10^{-4} \pm 22\% $ 
    \end{tabular}
    \caption{\label{tab:sslep_rates} Rates for same-sign lepton pair
      production at the LHC and FCC-hh. Leptons are required to
      satisfy the cuts $p_T>10(100) \gev$, $|\eta|<2.5(4)$ at the LHC
      (FCC-hh). Quoted uncertainties are from PDFs, at 68\% CL.}
    \end{centering}
\end{table}

As far as the case of opposite-sign and different-flavour lepton-pair
production is concerned, the total rates for the various flavour
combinations are shown in Tab.~\ref{tab:diffflavlep_rates}. In the table we show
lepton-initiated processes and compare them with a non-lepton-initiated reaction with the same signature,
$p p > \ell^+ \ell'^{-} \nu \nu$. Leptons
are required to satisfy the same cuts as in the same-sign case, {\it
  i.e.} Eqs.~(\ref{eq:lepcutslhc}) and~(\ref{eq:lepcutsfcc}). 
For what concerns $p p > \ell^+ \ell'^{-} \nu \nu$, besides the cuts on leptons we also 
study the effect of a cut on the missing transverse energy $\slashed E_T< 20 \gev$, 
in order to enhance configuratons with the
two leptons back-to-back. This cut can be considered realistic for the LHC\footnote{
The typical missing transverse energy resolution in events with a low-$p_T$ lepton pair at 
LHC experiments is 10\gev or 
less~\cite{Chatrchyan:2011tn,Khachatryan:2014gga,Aad:2012re,ATL-PHYS-PUB-2015-023,ATL-PHYS-PUB-2015-027}}, 
while it may be optimistic at the FCC, for which the precise experimental setup is presently unkown. In all cases, lepton-initiated 
contributions are suppressed with 
respect to $p p > \ell^+ \ell'^{-} \nu \nu$, and remain below the $1\%$ level even with the $\slashed E_T$ cut.

\begin{table}
    \begin{centering}
      \begin{tabular}{l|cc}
        & LHC, $\sigma$[pb]         & FCC-hh, $\sigma$[pb]        \\\hline
        $e^+    \mu^-  \to e^+    \mu^-   $ & $ 5.35 \cdot 10^{-4} \pm 38\% $    & $ 9.93 \cdot 10^{-5} \pm 22\% $   \\
        $e^+    \tau^- \to e^+    \tau^-  $ & $ 5.06 \cdot 10^{-4} \pm 37\% $    & $ 9.65 \cdot 10^{-5} \pm 22\% $   \\
        $\tau^+ \mu^-  \to \tau^+ \mu^-   $ & $ 4.95 \cdot 10^{-4} \pm 38\% $    & $ 9.46 \cdot 10^{-5} \pm 23\% $   \\
        \hline
        $p p  \to e^+ \mu^- \nu_e \tilde\nu_\mu                           $ & $ 4.90 \cdot 10^{-1} \pm 0.6\% $    & $2.01 \cdot 10^{-1} \pm 0.4\% $   \\
        $p p  \to e^+ \mu^- \nu_e \tilde\nu_\mu,  \slashed E_T>20 \gev  $ & $ 9.96 \cdot 10^{-2} \pm 0.6\% $    & $4.27 \cdot 10^{-2} \pm 0.5\% $   \\
      \end{tabular}
      \caption{\label{tab:diffflavlep_rates} Rates for
        different-flavour lepton-pair production at the LHC and
        FCC-hh. Final-state leptons are required to satisfy the cuts
        $p_T>10(100) \gev$, $|\eta|<2.5(4)$ at the LHC
        (FCC-hh). Quoted uncertainties are from PDFs, at 68\% CL.}
    \end{centering}
\end{table}

\subsubsection*{Slepton pair $\mathbf{\tilde{\ell}\tilde{\ell}^*}$}

The last case that we analyse is relevant for BSM; we consider the
hadroproduction of a slepton pair ($\tilde{\ell}\tilde{\ell}^*$)
within the Minimal Supersymmetric Standard Model (MSSM).  Again, we
are mainly interested in assessing whether contributions initiated by
leptons can affect the discovery searches of these particles. Anyway,
as done in the previous cases, we consider also photon-initiated
processes. A similar and more detailed study for $\gamma\gamma$
initial state has been performed in Ref.~\cite{Hollik:2015lha} for the
case of squark-antisquark production, where the $\gamma\gamma$
contribution has been found to be non-negligible. Here, for
slepton-antislepton production, we just provide results for total
rates. However, at variance with the squark-antisquark case, in
slepton-antislepton production $\gamma\gamma$- as well as
$\ell^+ \ell^-$-initiated processes are of the same perturbative order
of $q \bar q$, {\it i.e.} $\ord(\alpha^2)$. Therefore, we expect a
larger relative impact from photon-initiated processes.

In $\tilde{\ell}\tilde{\ell}^*$ production, tree-level Feynman
diagrams are very similar to the case of $e^+e^-$ production. The
$q\bar q$ initial state proceeds via (non-resonant) Drell-Yan
production, and the $\gamma \gamma$ initial state presents $t-$ and
$u-$channel diagrams. The $\ell^+ \ell^-$ initial state proceeds via
both Drell-Yan production and $t$-channel neutralino exchange. Also,
at variance with the $q\bar q$ and $\gamma \gamma$ initial states, at
LO the $\ell^+ \ell^-$ initial state can produce
$\tilde{\ell}_1\tilde{\ell}_2^*$ where both the chirality and the
flavour of $\tilde{\ell}_1$ and $\tilde{\ell}_2^*$ are different.
  
In order to minimise the dependence on the MSSM parameters, we have
chosen to consider only the cases $\ell=e,\mu$. In this way we avoid
the dependence on the left-right mixing parameters in the stau
sector. Moreover, all the particles besides the selectrons, smuons and
the lightest neutralino have been decoupled by setting their masses to
5 TeV. The neutralino mixing matrix and the physical masses have been
calculated at the scale of the slepton mass $m_{\tilde{\ell}}$, with
the help of {\sc\small Suspect} \cite{Djouadi:2002ze}.

In Tab.~\ref{tab:sleptons} we list the results for the total cross
sections of $\tilde{\ell}\tilde{\ell}^*$ production at the LHC and
FCC-hh, where the soft masses for right-handed and left-handed leptons
of the first two families have been set to $m_{\tilde{\ell}}$=200,
500, 1200 GeV and the mass of the lightest neutralino to
$m_{\tilde \chi^0_1}$=100 GeV. Physical masses are slightly different
from soft masses, however, the small differences are completely
irrelevant for the study pursued here, so we do not report them in the
text. The numbers in Tab.~\ref{tab:sleptons} refer to the sum of all
the possible chirality and flavour configurations. Also for this
process, the contribution due to leptons in the initial state is in
general negligible. Even setting $m_{\tilde \chi^0_1}$= 1 GeV, the
lepton-lepton contribution results slightly enhanced, but the
qualitative picture does not change. Conversely, photon-induced
processes are relevant and cannot be neglected at the LHC, especially
for large masses.

\begin{table}
    \begin{center}
    \begin{tabular}{c|ccc}
        \multicolumn{4}{l}{LHC, $\sigma(pp\to\tilde \ell\tilde \ell^*)$[pb]}\\\hline
            Initial state   &  $m_{\tilde \ell}=200\gev$  &  $m_{\tilde \ell}=500\gev$  &  $m_{\tilde \ell}=1200\gev$  \\\hline
            Total           &  $5.24  \cdot 10^{-2} \pm   4\%$  &  $1.31  \cdot 10^{-3} \pm  10\%$  &  $1.19  \cdot 10^{-5} \pm  38\%$ \\
            $q\bar q$       &  $4.98  \cdot 10^{-2} \pm   1\%$  &  $1.18  \cdot 10^{-3} \pm   2\%$  &  $7.52  \cdot 10^{-6} \pm   6\%$ \\
            $\gamma \gamma$ &  $2.62  \cdot 10^{-3} \pm  72\%$  &  $1.32  \cdot 10^{-4} \pm  84\%$  &  $4.38  \cdot 10^{-6} \pm  95\%$ \\
            $\ell^+\ell^-$            &  $6.70  \cdot 10^{-7} \pm  92\%$  &  $4.92  \cdot 10^{-8} \pm  97\%$  &  $1.68  \cdot 10^{-9} \pm  99\%$ 
    \end{tabular}
    \\[10pt]
    \begin{tabular}{c|ccc}
        \multicolumn{4}{l}{FCC-hh, $\sigma(pp\to\tilde \ell\tilde \ell^*)$[pb]}\\\hline
            Initial state   &  $m_{\tilde \ell}=200\gev$  &  $m_{\tilde \ell}=500\gev$  &  $m_{\tilde \ell}=1200\gev$  \\\hline
            Total           &  $8.42  \cdot 10^{-1} \pm   2\%$  &  $4.88  \cdot 10^{-2} \pm   3\%$  &  $2.35  \cdot 10^{-3} \pm   5\%$ \\
            $q\bar q$       &  $8.13  \cdot 10^{-1} \pm   1\%$  &  $4.64  \cdot 10^{-2} \pm   1\%$  &  $2.19  \cdot 10^{-3} \pm   1\%$ \\
            $\gamma \gamma$ &  $2.94  \cdot 10^{-2} \pm  46\%$  &  $2.32  \cdot 10^{-3} \pm  52\%$  &  $1.58  \cdot 10^{-4} \pm  62\%$ \\
            $\ell^+\ell^-$            &  $6.22  \cdot 10^{-6} \pm  74\%$  &  $8.09  \cdot 10^{-7} \pm  80\%$  &  $7.86  \cdot 10^{-8} \pm  87\%$ 
    \end{tabular}
    \caption{\label{tab:sleptons} Total rates for slepton-pair
      production at the LHC and at the FCC-hh. Quoted uncertainties
      refer to PDF uncertainties at the 68\% CL.}
    \end{center}
\end{table}

\section{Conclusions and outlook}\label{sec:outlook}

In this paper we have presented the first-ever estimate of the lepton
content of the proton. By implementing the complete LO QED corrections
to the DGLAP evolution equations in the {\tt APFEL} program and by
means of a model for the lepton PDFs at the initial scale, we have
produced PDF sets in the {\tt LHAPDF6} format containing, besides the
usual PDFs for quarks and gluons and possibly photons, distributions
for all charged leptons. The main result of this work are the sets
produced according to the ansatz in Eq.~(\ref{eq:ansatz}), {\it i.e.}
by assuming that leptons are produced in pairs by photon splitting
at the respective mass scales.

By studying the feature of the PDFs, we found that our results
are consistent with the expectations. Lepton PDFs are
strongly correlated to the photon PDF and suppressed by a factor
$\alpha$ w.r.t.~it. As a consequence of their small size, their
phenomenological impact at hadron colliders is likely to be
limited. Also, the uncertainties of lepton PDFs can be very large, as
in the case of the photon PDF.  We explicitly checked the
phenomenological impact of lepton-induced partonic channels for
several relevant SM processes at the LHC at 13 TeV and at the 100 TeV
FCC-hh. With reasonable experimental cuts, the contributions of lepton
PDFs is usually extremely suppressed. On the contrary, the photon PDF
contribution is not negligible and has to be carefully studied for
precise predictions at the LHC and FCC-hh. We also presented
representative results for BSM processes (slepton-antislepton
production in the MSSM) and rare SM processes such as same-sign and/or
different-flavour dilepton production. More detailed studies,
including higher-order effects as well as realistic cuts and background simulations, are envisaged,
since the role of the lepton PDFs could be relevant in this context.

\medskip

The combined QCD+QED evolution with lepton PDFs has been implemented
in {\tt APFEL} version 2.4.0 and later, publicly available at:
\begin{center}
  \url{https://github.com/scarrazza/apfel}
\end{center}

The sets of PDFs used in this work, produced with {\tt APFEL}, are
also publicly available in the {\tt LHAPDF6} library format from the
{\tt APFEL} webpage\footnote{\url{http://apfel.hepforge.org/}}:
\begin{flushleft}
  \small
  \tt apfel\_nn23nlo0118\_lep0, apfel\_nn23nnlo0118\_lep0,\\
  \tt apfel\_nn23qedlo0118\_lept0, apfel\_nn23qednlo0118\_lept0, apfel\_nn23qednnlo0118\_lept0, \\
  \tt apfel\_nn23qedlo0118\_lept, apfel\_nn23qednlo0118\_lept,
  apfel\_nn23qednnlo0118\_lept, \\
  apfel\_mrst04qed\_lept0, apfel\_mrst04qed\_lept
\end{flushleft}

\acknowledgments 

We are grateful to Roberto Salerno, Lorenzo Calibbi, Stefano Frixione, 
Fabio Maltoni,
Michelangelo Mangano, Juan Rojo, Stefano Forte and the {\tt
  MG5\_aMC@NLO} Collaboration members for fruitful discussions and
suggestions during the development of this project.

This work is partially supported by an Italian PRIN2010 grant (S.C.),
by a European Investment Bank EIBURS grant (S.C.) and by the ERC grant
291377, LHCtheory: \emph{Theoretical predictions and analyses of LHC
  physics: advancing the precision frontier} (V.B., S.C. and D.P.).
The work of M.Z. is supported by the ERC grant “Higgs@LHC” and
partially by the ILP LABEX (ANR-10-LABX-63), in turn supported by
French state funds managed by the ANR within the ``Investissements
d'Avenir'' programme under reference ANR- 11-IDEX-0004-02.

\appendix

\section{The combined QCD+QED evolution in the presence of lepton PDFs}\label{sec:appendix}

The DGLAP evolution equations employed in this work for the unified
QCD+QED evolution in the presence of lepton PDFs consist of a system
of coupled differential equations which satisfy both QCD and QED
evolution equations. In order to simultaneously diagonalise as much as
possible the respective evolution matrices, avoiding unnecessary
couplings between parton distributions, we have adopted the following
evolution basis:
\begin{equation}\label{EvolBasis}
\begin{array}{ll}
\mbox{\texttt{ 1} : }g & \\
\mbox{\texttt{ 2} : }\gamma & \\
\mbox{\texttt{ 3} : }\displaystyle \Sigma = \Sigma_u + \Sigma_d & \quad
\mbox{\texttt{12} : }\displaystyle V =V_u +  V_d\\
\mbox{\texttt{ 4} : } \displaystyle \Delta_\Sigma = \Sigma_u - \Sigma_d& \quad\displaystyle 
\mbox{\texttt{13} : } \Delta_V = V_u - V_d\\
\mbox{\texttt{ 5} : }\displaystyle \Sigma_\ell 
& \quad \mbox{\texttt{14} : }\displaystyle V_\ell\\
\mbox{\texttt{ 6} : }T_1^u = u^+ - c^+ &\quad \mbox{\texttt{15} : }V_1^u = u^- - c^- \\
\mbox{\texttt{ 7} : }T_2^u = u^+ + c^+ - 2t^+ &\quad \mbox{\texttt{16} : }V_2^u = u^- + c^- - 2t^-\\
\mbox{\texttt{ 8} : }T_1^d = d^+ - s^+ &\quad \mbox{\texttt{17} : }V_1^d = d^- - s^- \\
\mbox{\texttt{ 9} : }T_2^d = d^+ + s^+ - 2b^+ &\quad \mbox{\texttt{18}
                                               : }V_2^d = d^- + s^- -
                                               2b^-\\
\mbox{\texttt{10} : }T_{3}^{\ell} = l_e^+ - l_\mu^+&\quad
                                                          \mbox{\texttt{19} : }V_{3}^{\ell} = l_e^- - l_\mu^-\\
\mbox{\texttt{11} : }T_{8}^{\ell} = l_e^+ + l_\mu^+-2l_\tau^+&\quad
                                                          \mbox{\texttt{20} : }
                                                                       V_{8}^{\ell}
                                                                       = l_e^- + l_\mu^--2l_\tau^-\,,
\end{array}
\end{equation}
where we have defined:
\begin{equation}
\begin{array}{ll}
q^\pm = q\pm\overline{q}&\quad\mbox{with }q = u,d,s,c,b,t,\\
\\
l_\ell^\pm = \ell^-\pm\ell^+&\quad\mbox{with }\ell = e,\mu,\tau,
\end{array}
\end{equation}
and:
\begin{equation}
\begin{array}{ll}
\Sigma_u = u^++c^++t^+, &\quad V_u = u^-+c^-+t^-,\\
\\
\Sigma_d = d^++s^++b^+,&\quad V_d = d^-+s^-+b^-,\\
\\
\Sigma_\ell = l_e^++l_\mu^++l_\tau^+,&\quad V_\ell = l_e^-+l_\mu^-+l_\tau^-.
\end{array}
\end{equation}

When considering LO QED corrections to the DGLAP equations, the
evolution equations in the basis given in Eq.~(\ref{EvolBasis})
decouple into different sectors that evolve independently as shown
below. It should also be noticed that considering only LO QED
corrections allows one to completely separate QCD and QED
corrections. Indeed, defining $\widetilde{P}$ the QCD splitting
functions (up to any of the known orders in
$\alpha_s$~\cite{Altarelli:1977zs,Curci:1980uw,Furmanski:1980cm,Moch:2004pa,Vogt:2004mw})
and $\bar{P}=(\alpha/4\pi){P}^{(0)}$ the LO QED splitting functions,
with ${P}^{(0)}$ given in Eq.~(\ref{QEDLOsplittingFunctions}), we
have:
\begin{itemize}
\item the singlet sector:
\begin{equation}\label{APFELsysLept}
\begin{array}{rcl}
\displaystyle\mu^2\frac{\partial}{\partial \mu^2}
\begin{pmatrix}
g\\
\gamma\\
\Sigma\\
\Delta_\Sigma\\
\Sigma_\ell
\end{pmatrix} &=& \displaystyle \left[
\begin{pmatrix}
\widetilde{P}_{gg} & 0 & \widetilde{P}_{gq} & 0 & 0\\
0 & 0 & 0 & 0 & 0\\
2n_f\widetilde{P}_{qg} & 0 & \widetilde{P}_{qq} & 0 & 0\\
\frac{n_u-n_d}{n_f} 2n_f\widetilde{P}_{qg} & 0 & \frac{n_u-n_d}{n_f}(\widetilde{P}_{qq}-\widetilde{P}^+) &
\widetilde{P}^+ & 0\\
0 & 0 & 0 & 0 & 0\\
\end{pmatrix}\right.
\\
\\
&+&\left.\begin{pmatrix}
0 & 0 & 0 & 0 & 0\\
0 & e_\Sigma^2 \bar{P}_{\gamma\gamma} & \eta^+\bar{P}_{\gamma q} &\eta^-\bar{P}_{\gamma q} & \bar{P}_{\gamma q}\\
0 & 2 e_\Sigma^2 \bar{P}_{q\gamma} & \eta^+\bar{P}_{qq} &
\eta^-\bar{P}_{qq} & 0\\
0 & 2 \delta_e^2 \bar{P}_{q\gamma}
&\eta^-\bar{P}_{qq}
&\eta^+\bar{P}_{qq} & 0\\
0 & 2n_\ell\bar{P}_{q\gamma} & 0 & 0 & \bar{P}_{qq}\\
\end{pmatrix}\right]\otimes
\begin{pmatrix}
g\\
\gamma\\
\Sigma\\
\Delta_\Sigma\\
\Sigma_\ell
\end{pmatrix}\,,
\end{array}
\end{equation}
\item the quark valence sector:
\begin{equation}
\displaystyle\mu^2\frac{\partial}{\partial \mu^2}
\begin{pmatrix}
V\\
\Delta_V
\end{pmatrix} = 
\left[
\begin{pmatrix}
\widetilde{P}^V & 0 \\
\frac{n_u-n_d}{n_f}(\widetilde{P}^V-\widetilde{P}^-)  & \widetilde{P}^- 
\end{pmatrix}
+
\begin{pmatrix}
\eta^+\bar{P}_{qq} & \eta^-\bar{P}_{qq} \\
\eta^-\bar{P}_{qq} & \eta^+\bar{P}_{qq}
\end{pmatrix}
\right]\otimes
\begin{pmatrix}
V\\
\Delta_V
\end{pmatrix}\,,
\end{equation}
\item the lepton valence sector:
\begin{equation}
\begin{array}{l}
\displaystyle  \mu^2\frac{\partial V_\ell}{\partial \mu^2} =
\bar{P}_{qq}\otimes  V_\ell\,,
\end{array}
\end{equation}
\item the quark non-singlet sector:
\begin{equation}
\begin{array}{l}
\begin{array}{rcl}
\\
\displaystyle \mu^2\frac{\partial T^u_{1,2}}{\partial \mu^2} &=&
\displaystyle (\widetilde{P}^+ + e_u^2\bar{P}_{qq}) \otimes T^u_{1,2}\,,\\
\\
\displaystyle \mu^2\frac{\partial T^d_{1,2}}{\partial \mu^2} &=&
\displaystyle (\widetilde{P}^+ + e_d^2\bar{P}_{qq}) \otimes T^d_{1,2}\,,
\end{array}
\\
\\
\begin{array}{rcl}
\displaystyle \mu^2\frac{\partial V^u_{1,2}}{\partial \mu^2} &=&
\displaystyle (\widetilde{P}^- + e_u^2\bar{P}_{qq}) \otimes V^u_{1,2}\,,\\
\\
\displaystyle \mu^2\frac{\partial V^d_{1,2}}{\partial \mu^2} &=&
\displaystyle (\widetilde{P}^- + e_d^2\bar{P}_{qq}) \otimes V^d_{1,2}\,,
\end{array}
\end{array}
\end{equation}
\item and the lepton non-singlet sector:
\begin{equation}
\begin{array}{l}
\displaystyle  \mu^2\frac{\partial T_{3,8}^{\ell}}{\partial \mu^2} =
\bar{P}_{qq}  \otimes T_{3,8}^{\ell}\,,\\
\\
\displaystyle  \mu^2\frac{\partial V_{3,8}^{\ell}}{\partial \mu^2} =
\bar{P}_{qq}  \otimes V_{3,8}^{\ell}\,,
\end{array}
\end{equation}
\end{itemize}
where $e_u$ and $e_d$ are the electric charges of the up- and
down-type quarks, $n_f$ is the number of of active quark flavours,
$n_u$ and $n_d$ are the number of up- and down-type active quark
flavours (such that $n_f=n_u+n_d$) and $n_\ell$ is the number of
active leptons. Moreover, we have defined:
\begin{equation}
\begin{array}{rcl}
e_{\Sigma}^{2}& \equiv &\displaystyle
N_c(n_ue_{u}^{2}+n_de_{d}^{2})\,,\\
\\
\delta_e^2 & \equiv &\displaystyle N_c(n_u e_u^2 -n_d e_d^2)\,,\\
\\
\eta^{\pm} & \equiv & \displaystyle \frac{1}{2}\left(e_{u}^{2}\pm
  e_{d}^{2}\right)\,,\\
\end{array}
\end{equation}
being $N_c$ the number of colours.

The basis presented here is implemented in \texttt{APFEL} and
accessible through the \texttt{QUniD} solution.

\bibliography{leptonpdf}

\end{document}